\documentclass[%
 reprint,
 onecolumn,
nofootinbib,
 aps
]{revtex4-2}
\usepackage{graphicx}% Include figure files
\usepackage{dcolumn}% Align table columns on decimal point
\usepackage{bm}% bold math
\usepackage{float}
\usepackage{amsmath,amssymb,amsfonts}
\usepackage{todonotes}
\usepackage{multirow}

\def\XXint#1#2#3{{\setbox0=\hbox{$#1{#2#3}{\int}$}
\vcenter{\hbox{$#2#3$}}\kern-.5\wd0}}
\def\Xiint#1{\mathchoice
	{\XXiint\displaystyle\textstyle{#1}}%
	{\XXiint\textstyle\scriptstyle{#1}}%
	{\XXiint\scriptstyle\scriptscriptstyle{#1}}%
	{\XXiint\scriptscriptstyle\scriptscriptstyle{#1}}%
	\!\iint}
	
\def\XXiint#1#2#3{{\setbox0=\hbox{$#1{#2{\mathrm{#3\!\! #3}}}{\iint}$}
\vcenter{\hbox{$#2 {\mathrm{#3\!\! #3}}$}}\kern-.5\wd0}}

\usepackage{hyperref}
\usepackage{color}  
\definecolor{darkGreen}{rgb}{0,0.45,0}
\definecolor{darkBlue}{rgb}{0,0,0.7}
\definecolor{darkRed}{rgb}{0.76, 0.13, 0.28}
\hypersetup{
	colorlinks=true, 
	linktoc=all,    
	linkcolor=darkBlue, 
	citecolor=darkGreen,
	urlcolor = darkRed,
}
\usepackage{comment}

\renewcommand{\bf}[1]{\mathbf{#1}}
\renewcommand{\d}{{\mathrm{d}}}

\newcommand{\nn}{\noindent}

\usepackage[margin=2.3cm]{geometry}
\begin{document}

\title{Membrane flutter in three-dimensional inviscid flow}
\author{Christiana Mavroyiakoumou}%
\email[Electronic address: ]{chrismav@umich.edu}
\author{Silas Alben}
\email[Electronic address: ]{alben@umich.edu}
\affiliation{\vspace{.1cm}Department of Mathematics, University of Michigan, Ann Arbor, MI 48109, USA\vspace{.1cm}}

\date{\today}

\begin{abstract}
We develop a model and numerical method to study the large-amplitude flutter of rectangular membranes (of zero bending rigidity) that shed a trailing vortex-sheet wake in a three-dimensional (3D) inviscid fluid flow. 
We apply small initial perturbations and track their decay or growth to large-amplitude steady state motions. For
12 combinations of boundary conditions at the membrane edges we compute the stability thresholds and the subsequent large-amplitude dynamics across the  three-parameter space of membrane mass density, pretension, and stretching rigidity. With free side edges we find good agreement with previous 2D results that used different discretization methods. We find that the 3D dynamics in the 12 cases naturally form four groups based on the conditions at the leading and trailing edges. The deflection amplitudes and oscillation frequencies have scalings similar to those in the 2D case. The conditions at the side edges, though generally less important, may have small or large qualitative effects on the membrane dynamics---e.g. steady versus unsteady, periodic versus chaotic, or the variety of spanwise curvature distributions---depending on the group and the physical parameter values. 
\end{abstract}

\maketitle

%%%%%%%%
\section{Introduction\label{sec:intro}}

Interactions between flexible structures and high-Reynolds-number flows are ubiquitous in nature and engineering applications. 
The physical mechanisms that govern these fluid-structure interactions provide us with important insights into the fluid dynamics of biolocomotion. For example, fish can exploit energy from surrounding vortices and move efficiently by undulating their bodies~\cite{triantafyllou2000hydrodynamics,gazzola2014scaling}, or synchronize their motions with the oncoming vortices~\cite{liao2003fish}. These observations have inspired engineers to design and manufacture continuously deformable robots that exhibit these behaviors~\cite{lauder2007fish,rus2015design,li2017fast}. Birds and other flying animals also take advantage of such mechanisms to achieve efficient locomotion. In particular, bats---one of nature's most agile fliers~\cite{swartz1996mechanical,swartz2007wing,song2008aeromechanics,shyy1999flapping,cheney2015wrinkle}---can adapt to the surrounding flow conditions by deforming their thin, compliant membrane wings. 

An extensible membrane is a soft material that undergoes significant stretching in a fluid flow and has negligible bending rigidity. When it is aligned with a fluid flow, the surrounding fluid forces can cause it to flutter and become unstable. Being able to predict the onset of membrane instability across parameter space, either by flutter, divergence, or a
combination of the two, is fundamental to a wide range of applications. Stable membranes can be used in a variety of configurations in aircraft and shape-morphing airfoils~\cite{abdulrahim2005flight,lian2005numerical,hu2008flexible,stanford2008fixed,jaworski2012high,piquee2018aerodynamic,schomberg2018transition,tzezana2019thrust}, sails~\cite{colgate1996fundamentals,kimball2009physics} parachutes~\cite{pepper1971aerodynamic,stein2000parachute}, and micro-air vehicles~\cite{lian2005numerical}. When a rigid wing moves through a flow its upper surface may experience flow separation and significant reductions in aerodynamic efficiency in both steady and unsteady flows, thus limiting the aircraft’s maneuverability and performance. However, a flexible membrane is able to adapt quickly to unsteady airflow conditions by assuming a deformed shape that can inhibit flow separation and enhance aircraft maneuverability. Recent developments in membrane aerodynamics are reviewed in~\cite{lian2003membrane,tiomkin2021review}. 

The regions in parameter space where membrane flutter occurs have been predicted by linear models that mostly assume an infinite membrane span and two-dimensional (2D) flow. In this work we use a three-dimensional (3D) inviscid flow model based on the vortex lattice method to study membrane stability in the linear regime of small deflections as well as large-amplitude nonlinear membrane dynamics. We find qualitative changes in flutter behavior in some cases due to three-dimensionality.

% Most studies of vortex sheets in 2D flow, including our previous work~\cite{mavroyiakoumou2020large,mavroyiakoumou2021eigenmode,mavroyiakoumou2021dynamics}, are based on formulating the problem in complex variables but such a formulation does not generalize naturally to vortex sheets in 3D flows.

Only a small number of studies have considered fully-coupled interactions of flexible bodies and 3D inviscid flows. The formulation of the mechanical force balance laws and the numerical methods are more complicated in 3D than in 2D and
the computational expense is much higher. Inviscid flow models such as the vortex lattice method are widely used to model viscous high-Reynolds-number flows for aquatic and aerodynamic propulsion \cite{kagemoto2000force,katz2001low,shukla2007inviscid,michelin2009resonance,willis2014multiple,ayancik2022scaling}
because the computational cost is generally much lower than for direct solvers (e.g. in the case of complex and deforming body geometries, immersed-boundary, lattice-Boltzmann, and deforming-mesh methods~\cite{kim2007penalty,zhu2011immersed,willis2007computational,hoover2021neuromechanical}). 
In the inviscid models computational elements are distributed along surfaces rather than throughout the flow volume. However, traditional inviscid approximations of flow separation work well only in certain cases such as low-angle-of-attack airfoils where trailing edge separation is dominant. Recently leading-edge separation has been included in such models~\cite{pan2012boundary,ramesh2014discrete,li2021flow}.

Several recent works have studied the effect of various dimensionless control parameters such as Reynolds number, density ratio, shear modulus, and aspect ratio on the
flutter of one or more thin plates or flags with bending rigidity in three-dimensional viscous, high-Reynolds-number flows~\cite{banerjee2015three,tian2012onset,huang2010three,yu2012numerical,dong2016coupling,chen2020flapping}. Immersed boundary methods were used by~\cite{tian2012onset,huang2010three}, while~\cite{yu2012numerical} used a fictitious domain method and~\cite{banerjee2015three} used a coordinate transformation method.

Immersed boundary methods have also been used to study a number of 3D swimming problems, such as the self-propulsion of flapping flexible plates~\cite{masoud2010resonance,tang2016self}, the role of active muscle contraction, passive body elasticity and fluid forces in forward
swimming~\cite{mittal2008versatile,borazjani2013parallel,hoover2017quantifying,dawoodian2021kinetics,hoover2021neuromechanical}, and the propulsive forces acting on flexible fish bodies~\cite{hoover2020decoding}.

%Similar methods have also been applied for medical purposes; for example~\cite{lee2021bioprosthetic} used immersed boundary methods to study how 3D fluttering of bioprosthetic heart valves depends on the device geometry to improve durability of
%valve replacement.

Boundary element methods have helped
identify the effects of body kinematics on thrust production and efficiency of
3D swimmers with non-deforming~\cite{liu1997propulsive,liu1996time} and deforming wakes~\cite{zhu2002three}. \cite{moored2018unsteady} used a boundary element method to examine the
self-propelled swimming of undulatory fins and manta rays~\cite{fish2016hydrodynamic}, and to derive 
three-dimensional heaving and pitching scaling laws~\cite{ayancik2020three,ayancik2022scaling}.

There are relatively few studies of 3D coupled interactions of inviscid flows and flexible bodies. 
% One recent study
Recently, \cite{hiroaki2021numerical,hiroaki2021three} used nonlinear beam theory and the vortex-lattice method to analyze the limit cycle oscillations of a rectangular plate and plate vibration under harmonic forced excitation~\cite{hiroaki2021theoretical}. Previously~\cite{gibbs2012theory,gibbs2015stability}
studied the 3D linear stability problem for a flexible plate in an inviscid flow with various boundary conditions. \cite{tang1999limit,tang2001effects} used the method to study the stability of delta wings and rectangular plates, and~\cite{murua2012applications} reviewed the vortex lattice method in similar applications. 

To our knowledge the present paper is the first 3D study of large-amplitude dynamics of membranes (of zero bending rigidity) in inviscid flows.  
Here, we use a flat-wake approximation~\cite{kagemoto2000force} and find good agreement between membranes with large aspect ratio in 3D flow and membranes in 2D flow with wake roll-up. Computing wake roll-up with high precision in 3D is challenging as the vortex sheet may undergo complex twisting and shearing deformations that make meshing difficult~\cite{feng2007vortex}. Fast algorithms (e.g. tree codes) for the evolution of vortex sheets in three dimensions have been developed by \cite{kaganovskiy2006adaptive,feng2009azimuthal,lindsay2001particle,willis2007combined,sakajo2001numerical}.

In~\cite{mavroyiakoumou2020large} we studied membrane dynamics in a 2D flow, where the membrane is a 1D curvilinear segment that undergoes large deflections.
We first studied the most common endpoint conditions,
``fixed--fixed," in which the membrane ends were held fixed, as in most previous studies of membrane flutter~\cite{le1999unsteady,sygulski2007stability,tiomkin2017stability,nardini2018reduced}. We found surprisingly that all unstable membranes tended to a steady shape in the large-amplitude regime, except for some cases with unrealistically large deflections. This motivated us to study a second case, ``fixed--free," in which the trailing edge of the membrane is free to move, but only in the direction perpendicular to the oncoming flow. This corresponds to the free-end boundary condition for a string or membrane in classical mechanics~\cite{graff1975wave,farlow1993partial}, where the membrane end has horizontal slope. Physically, this boundary condition means that the end slides without friction perpendicularly to the membrane's flat equilibrium state (for example, in~\cite{farlow1993partial} the end is attached to a frictionless, massless ring). If the membrane end is also free to move in the in-plane direction, compression develops and the problem is ill-posed without a stabilizing effect such as bending rigidity, and difficult to solve computationally \cite{triantafyllou1994dynamic}. Recent work has studied unsteady motions by
membranes with partially free edges. 
In~\cite{hu2008flexible}, the authors study membrane wings with partially free trailing edges and find that trailing-edge flutter can occur when the angle of attack is relatively low. \cite{arbos2013leading} found experimentally that membrane wing flutter can be enhanced by the vibrations of flexible leading and trailing edge supports. Partially free edges exist also in sails with tensioned cables running along the edges~\cite{kimball2009physics}. At sufficiently low tension these edges may flutter~\cite{colgate1996fundamentals}. A related application is to energy harvesting by membranes mounted on tensegrity structures (networks of rigid rods and elastic fibers) and placed in fluid flows~\cite{sunny2014optimal,yang2016modeling}. In such cases the membrane ends have some degrees of freedom akin to the free-end boundary conditions we have defined above.

% For example,~\cite{jaworski2012high} studied numerically a heaving and pitching membrane airfoil in a fluid stream and found that prestress and elastic modulus led to enhanced thrust and propulsive efficiency. Membrane wings were shown to provide greater lift, and improved lift slopes, due to modification of the camber at different angles of attack~\cite{tzezana2019thrust}. In~\cite{tiomkin2017stability} the authors examined the stability of membrane wings as a function of the mass and tension of the membrane. This was also considered in~\cite{mavroyiakoumou2021eigenmode} for different boundary conditions. 
% Elasticity of airfoils is also common to applications such as energy harvesting,
% sails~\cite{colgate1996fundamentals,kimball2009physics}, and more recently in the context of morphing wing sections which change
% their shape continuously~\cite{tiomkin2017stability,tzezana2019thrust,mavroyiakoumou2020large}. These shape-morphing airfoils are currently extensively studied because of their potential to enhance performance of aircraft structures with current approaches including deployable and foldable structures~\cite{thill2008morphing} and energy harvesting systems.

The structure of the paper is as follows. \S\ref{sec:model} describes our large-amplitude membrane-vortex-sheet model. Unlike our previous work~\cite{mavroyiakoumou2020large,mavroyiakoumou2021eigenmode,mavroyiakoumou2021dynamics}, here we consider a membrane surface held in a 3D inviscid flow with 12 different sets of boundary conditions. We solve this nonlinear model using Broyden's method and an unsteady vortex lattice algorithm. \S\ref{sec:numerical} describes the numerical method. We apply a small initial perturbation to the membrane in the direction transverse to the flow and compute the subsequent dynamics, which have large-amplitude deflections if the membrane is unstable. 
In \S\ref{sec:validation} we perform convergence studies and validate our model by comparing the current results with our previous 2D studies of fixed--fixed, fixed--free, and free--free membranes.
\S\ref{sec:results} describes the computed membrane dynamics and how these vary with key parameters such as the membrane mass, pretension, and stretching rigidity. We find that the membrane dynamics naturally form four groups based on the boundary conditions at the leading and trailing edges. \S\ref{sec:conclusions} gives the conclusions.

% %%%%%%%%%%%%%%%%%%%%%%

\section{Large-amplitude membrane-vortex-sheet model}\label{sec:model}

\begin{figure}
    \centering
    \includegraphics[width=\textwidth]{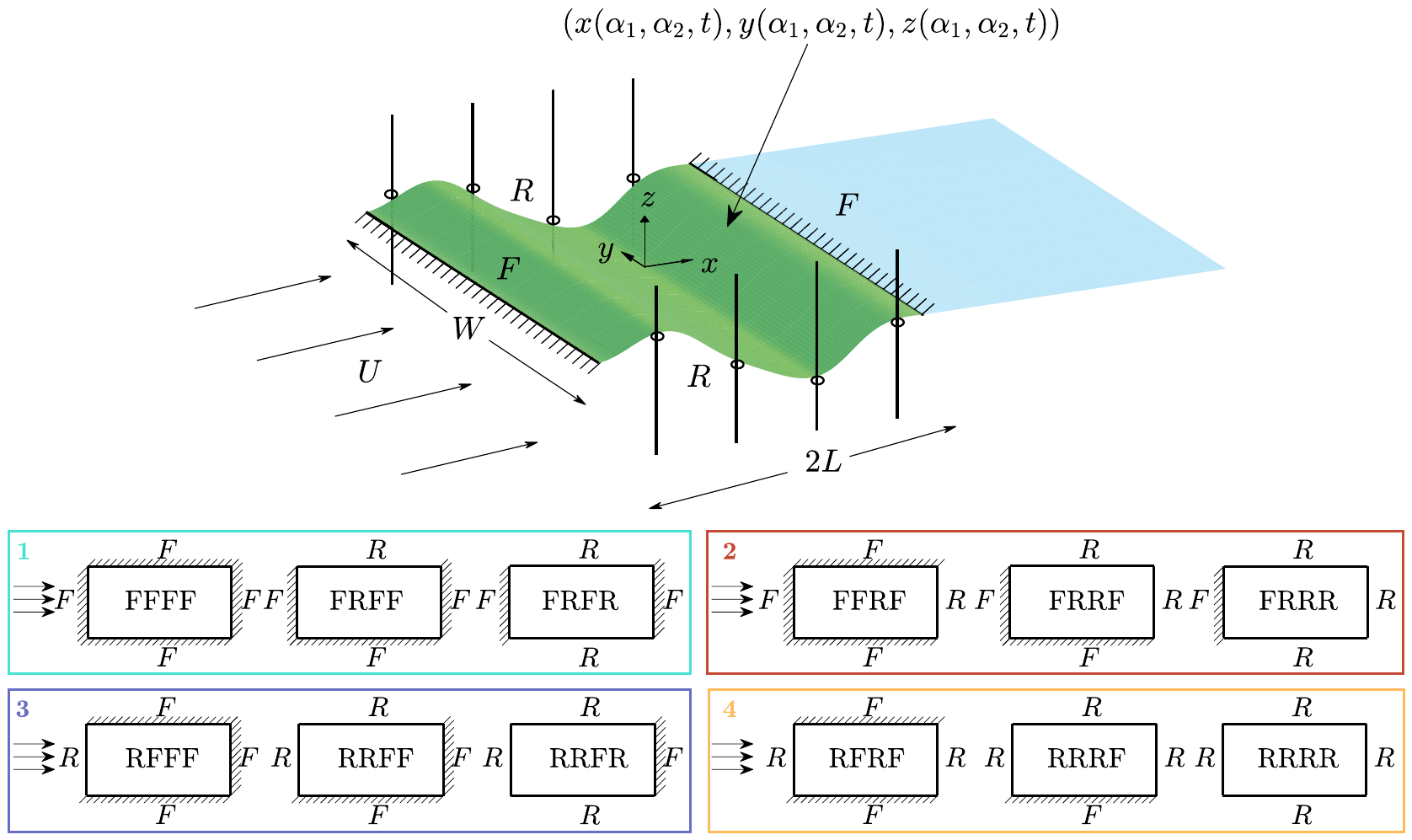}
    \caption{Schematic diagram (in perspective view) showing a three-dimensional membrane (dark green surface) with fixed leading and trailing edges and free side edges. Along a free edge, points are fixed to massless rings that slide without friction along vertical poles. $U\widehat{\mathbf{e}}_x$ is the oncoming flow velocity, $W$ is the membrane's spanwise width, and $2L$ is the membrane's chord. There is also a flat vortex wake (light blue surface) that emanates from the membrane's trailing edge.
In the lower portion of the figure,
we also show schematically (in top view) the 12 distinct boundary conditions explored in the current work. The diagonal marks indicate a fixed (F) boundary and other boundaries
are free (R). The arrows indicate the far-field flow direction which is the same for each configuration.}
    \label{fig:schemBC}
\end{figure}

We consider the motion of an extensible membrane held in a 3D fluid flow with velocity $U\widehat{\mathbf{e}}_x$ in the far field (see figure~\ref{fig:schemBC}). In the undeformed state the membrane is flat and parallel to the flow in the plane $z = 0$. As in~\cite{alben2019semi}, the membrane obeys linear elasticity but undergoes large deflections, so geometrically nonlinear terms enter the force expression. The membrane has a small thickness $h$ and to leading order the deformations do not vary through the thickness, so they can be described by the middle surface, midway through the thickness. The middle surface is a geometric surface, a three-component vector parametrized by two spatial coordinates and time: $\mathbf{r}(\alpha_1,\alpha_2,t)=(x(\alpha_1,\alpha_2,t),y(\alpha_1,\alpha_2,t),z(\alpha_1,\alpha_2,t))\in\mathbb{R}^3$, where the spatial coordinates $\alpha_1\in[-L,L]$ and $\alpha_2\in[-W/2,W/2]$ are the material coordinates, the $x$ and $y$ coordinates of each point in the initially flat state (with a uniform in-plane tension---a ``pretension"---applied by the boundaries).

Each of the four membrane edges is either fixed at zero deflection or free to move in the $z$ direction, perpendicular to the oncoming flow. Then there are $2^4 = 16$ possible boundary conditions but since the problem is symmetrical with respect to reflection in the $x$-$z$ plane, we only need to consider 12 distinct boundary conditions, which can be denoted: FFFF, FRFF/FFFR, FRFR, FFRF, FRRF/FFRR, FRRR, RFFF, RRFF/RFFR, RRFR, RFRF, RRRF/RFRR, and RRRR (with symmetrical pairs identified). Here F stands for a \textit{fixed} edge and R stands for a \textit{free} edge. The first letter in each label is the leading-edge boundary condition type and the following letters are the boundary conditions moving clockwise around the rectangular membrane looking down from larger $z$ values (bottom of figure~\ref{fig:schemBC}), as in~\cite{gibbs2015stability}. Thus, the second and fourth letters are the side-edge boundary conditions and the third letter is the trailing-edge boundary condition. We will see in the results sections that the dynamics in the 12 cases are naturally classified into four groups,
based on whether the leading and trailing edges are fixed or free. Each group is placed in one of the four colored rectangles in the lower portion of figure~\ref{fig:schemBC}. We present the results for each group in the four subsections of~\S\ref{sec:largeResultsMotions}, numbered with the numbers listed at the upper left corner of each group's rectangle in figure~\ref{fig:schemBC}. We list the equations for three examples from the set of 12 boundary conditions (and the others are analogous):
\begin{align}
   \text{FRFR:}&\quad   z(-L,\alpha_2,t)=0,\quad 
    z(L,\alpha_2,t)=0,\quad \partial_{\alpha_2} z(\alpha_1,-W/2,t)=0,\quad \partial_{\alpha_2} z(\alpha_1,W/2,t)=0,\label{eq:bcfrfrn}\\
    \text{FRRR:}&\quad z(-L,\alpha_2,t)=0,\quad \partial_{\alpha_1}
    z(L,\alpha_2,t)=0, \quad \partial_{\alpha_2} z(\alpha_1,-W/2,t)=0,\quad \partial_{\alpha_2} z(\alpha_1,W/2,t)=0,\label{eq:bcfrrrn}\\
     \text{RRRR:} &\quad  \partial_{\alpha_1}z(-L,\alpha_2,t)=0, \quad 
    \partial_{\alpha_1}z(L,\alpha_2,t)=0,\quad \partial_{\alpha_2}z(\alpha_1,-W/2,t)=0,\quad \partial_{\alpha_2}z(\alpha_1,W/2,t)=0. \label{eq:bcrrrrn}
\end{align}
Whether the edge is fixed or free affects only the out-of-plane ($z$-) component of the membrane motion at the edge. In all cases, no in-plane motion of the membrane edges is allowed, i.e. the $x$- and $y$-coordinates of the edges are fixed (see figure \ref{fig:schemBC}):
\begin{align}
\mbox{Upstream/downstream edges:} \quad
\left\{x\left(\pm L,\alpha_2,t\right)=\pm L \;, \; y\left(\pm L,\alpha_2,t\right)=\alpha_2\right\} \; &, -\frac{W}{2} \leq \alpha_2 \leq \frac{W}{2},   \\ \mbox{Side edges:} \quad \left\{x\left(\alpha_1,\pm\frac{W}{2},t\right) = \alpha_1\;, \; y\left(\alpha_1,\pm\frac{W}{2},t\right)= \pm \frac{W}{2}\right\}\; &, -L \leq \alpha_1 \leq L .
\end{align}
%To derive the 3D analogue of the unsteady, nonlinear, extensible elastic equation presented in~\cite{mavroyiakoumou2020large}, 

We start with the stretching energy per unit undeformed area for a thin sheet with isotropic elasticity (described in appendix~\ref{app:moreOnModel} and \cite{efrati2009elastic,alben2019semi}):
% Firstly, we substitute~\eqref{eq:fourtensor} into~\eqref{eq:w2d} and then use the properties of Kronecker delta to obtain
\begin{align}
w_s&=\frac{h}{2}\frac{E}{1+\nu}\left(\frac{1}{1-\nu}(\epsilon_{11}^2+\epsilon_{22}^2)+\frac{2\nu}{1-\nu}\epsilon_{11}\epsilon_{22}+2\epsilon_{12}^2\right),\label{eq:wsnonvar}
\end{align}
where $E$ is Young's modulus, $\nu$ is Poisson's ratio, $h$ is the membrane's thickness, and the strain tensor is
\begin{equation}\label{eq:strainTensor}
    \epsilon_{ij}(\alpha_1,\alpha_2,t)=\overline{e}\delta_{ij}+\frac{1}{2}\left(\partial_{\alpha_i}\mathbf{r}\cdot\partial_{\alpha_j}\mathbf{r}-\delta_{ij}\right).
\end{equation}
Here $\overline{e}$ denotes a constant prestrain corresponding to the pretension applied at the boundaries and $\delta_{ij}$ is the identity tensor, with $i,j\in\{1,2\}$.
%$m,n\in\{1,2\}$
We take the variation of the stretching energy
\begin{equation}\label{eq:Ws}
W_s = \iint w_s \d\alpha_1 \d\alpha_2
\end{equation}
with respect to the position $\mathbf{r}$ to obtain the stretching force per unit material area, i.e. $-\delta w_s/\delta \mathbf{r}$.
%, and use that the variation of the strain tensors is $\delta\epsilon_{11}=\partial_{\alpha_1}\mathbf{r}\cdot\partial_{\alpha_1}(\delta\mathbf{r})$, $\delta \epsilon_{12}=\left[\partial_{\alpha_1}\mathbf{r}\cdot\partial_{\alpha_2}(\delta\mathbf{r})+ \partial_{\alpha_1}(\delta\mathbf{r})\cdot \partial_{\alpha_2}\mathbf{r}\right]/2$, and $\delta\epsilon_{22}=\partial_{\alpha_2}\mathbf{r}\cdot \partial_{\alpha_2}(\delta\mathbf{r})$.

Integrating by parts to move derivatives off of~$\delta\mathbf{r}$ terms, we obtain 
\begin{align} \label{dWs}
\delta W_s=&\frac{Eh}{1+\nu}\oint\left[ \left(\frac{1}{1-\nu}\epsilon_{11}\left(\frac{\partial\mathbf{r}}{\partial \alpha_1}\cdot\delta \mathbf{r}\right)+\frac{\nu}{1-\nu}\epsilon_{22}\left(\frac{\partial\mathbf{r}}{\partial \alpha_1}\cdot\delta \mathbf{r}\right)+\epsilon_{12}\left(\frac{\partial\mathbf{r}}{\partial \alpha_2}\cdot\delta \mathbf{r}\right) \right)\upsilon_1 \right.\nonumber\\
&\hspace{1.1cm}+\left. \left( \frac{1}{1-\nu}\epsilon_{22}\left(\frac{\partial\mathbf{r}}{\partial \alpha_2}\cdot\delta \mathbf{r}\right)+\frac{\nu}{1-\nu}\epsilon_{11}\left(\frac{\partial\mathbf{r}}{\partial \alpha_2}\cdot\delta \mathbf{r}\right)+\epsilon_{12}\left(\frac{\partial\mathbf{r}}{\partial \alpha_1}\cdot\delta \mathbf{r}\right) \right)\upsilon_2 \right]\,\d\sigma \nonumber\\
&-\frac{Eh}{1+\nu} \iint \left[\frac{\partial}{\partial \alpha_1} \left(\frac{1}{1-\nu}\epsilon_{11}\frac{\partial\mathbf{r}}{\partial \alpha_1}+\frac{\nu}{1-\nu}\epsilon_{22}\frac{\partial\mathbf{r}}{\partial \alpha_1}+\epsilon_{12}\frac{\partial\mathbf{r}}{\partial \alpha_2} \right)\right.    \nonumber\\
&\hspace{2.1cm} \left. +\frac{\partial}{\partial \alpha_2} \left(\frac{1}{1-\nu}\epsilon_{22}\frac{\partial\mathbf{r}}{\partial \alpha_2}+\frac{\nu}{1-\nu}\epsilon_{11}\frac{\partial\mathbf{r}}{\partial \alpha_2}+\epsilon_{12}\frac{\partial\mathbf{r}}{\partial \alpha_1} \right)  \right]\cdot\delta\mathbf{r}\,\d \alpha_1\d \alpha_2.
\end{align}
The first integral in (\ref{dWs}) can be used to obtain the free edge boundary conditions. It is a boundary integral with respect to $\d\sigma$, arc length along the boundary in the $\alpha_1$-$\alpha_2$ plane;  $(\upsilon_1,\upsilon_2)$ is the outward normal in this plane. The integrand in the second integral in (\ref{dWs})
gives the stretching force per unit material area,
\begin{align}
\mathbf{f}_s=-\frac{\delta w_s}{\delta\mathbf{r}}=\frac{Eh}{1+\nu}&\left[\frac{\partial}{\partial \alpha_1} \left(\frac{1}{1-\nu}\epsilon_{11}\frac{\partial\mathbf{r}}{\partial \alpha_1}+\frac{\nu}{1-\nu}\epsilon_{22}\frac{\partial\mathbf{r}}{\partial \alpha_1}+\epsilon_{12}\frac{\partial\mathbf{r}}{\partial \alpha_2} \right)\right.\nonumber\\
&\left. + \frac{\partial}{\partial \alpha_2}\left(\frac{1}{1-\nu}\epsilon_{22}\frac{\partial\mathbf{r}}{\partial \alpha_2}+\frac{\nu}{1-\nu}\epsilon_{11}\frac{\partial\mathbf{r}}{\partial \alpha_2}+\epsilon_{12}\frac{\partial\mathbf{r}}{\partial \alpha_1} \right)  \right].\label{eq:3dtension}
\end{align}

% In the 2D case the undeformed coordinate $\alpha_1$ was denoted by $\alpha$ and the position $\mathbf{r}$ was written as $\zeta=x+iy$. Since there are no shear strains in that case $\epsilon_{12}=0$ and the only normal strain is $\epsilon_{11}=T(\alpha,t)/(Eh W)$. We have
% \begin{equation*}
% \mathbf{f}_s=\frac{Eh}{1+\nu}\left[\frac{\partial}{\partial \alpha}\left(\frac{1}{1-\nu}\frac{T(\alpha,t)}{EhW }\frac{\partial \mathbf{r}}{\partial \alpha}\right)\right]=\frac{1}{W(1-\nu^2)}\partial_ \alpha\left(  T(\alpha,t)\partial_\alpha \mathbf{r}\right).
% \end{equation*}

The membrane dynamics are governed by the balance of momentum for a small material element with area $\Delta \alpha_1\Delta \alpha_2$: 
\begin{equation}
\rho_s h\Delta \alpha_1\Delta \alpha_2\partial_{tt}\mathbf{r}=\mathbf{f}_s\Delta \alpha_1\Delta \alpha_2-[p](\alpha_1,\alpha_2,t)\mathbf{\hat{n}}\sqrt{\det[\partial_{\alpha_i}\mathbf{r}\cdot\partial_{\alpha_j}\mathbf{r}]}\Delta \alpha_1\Delta \alpha_2,\label{eq:membrane3d}
\end{equation}
where $\rho_s$ is the mass per unit volume of the membrane, uniform in the undeformed state; $[p](\alpha_1,\alpha_2,t)$ is the fluid pressure; $\mathbf{\hat{n}}$ is the unit normal vector,
\begin{equation} \mathbf{\hat{n}}=(\partial_{\alpha_1}\mathbf{r}\times\partial_{\alpha_2}\mathbf{r})/\|\partial_{\alpha_1}\mathbf{r}\times\partial_{\alpha_2}\mathbf{r}\|,\label{eq:normal}
\end{equation} 
and $\det[\partial_{\alpha_i}\mathbf{r}\cdot\partial_{\alpha_j}\mathbf{r}]$ is the determinant of the metric tensor, so $\sqrt{\det[\partial_{\alpha_i}\mathbf{r}\cdot\partial_{\alpha_j}\mathbf{r}]}\Delta \alpha_1\Delta \alpha_2$ is the area of the material element in physical space.

We nondimensionalize the governing equations by the density of the fluid $\rho_f$, the half-chord $L$, and the imposed fluid flow velocity $U$. The dimensionless time, space, and pressure jump variables (denoted by tildes) are:
\begin{equation}
    \tilde{t}=\frac{t}{L/U}, \quad(\tilde{\mathbf{r}},\widetilde{\alpha_1},\widetilde{\alpha_2})=\frac{(\mathbf{r},\alpha_1,\alpha_2)}{L},\quad\widetilde{[p]}=\frac{[p]}{\rho_fU^2}.
\end{equation}
The membrane equation~\eqref{eq:membrane3d}
becomes
\begin{align}
    \frac{\rho_s hU^2}{L}\widetilde{\partial_{tt}\mathbf{r}}=
\frac{Eh}{L(1+\nu)}\bigg[&\widetilde{\partial_{\alpha_1}} \left(\frac{1}{1-\nu}\widetilde{\epsilon_{11}}\widetilde{\partial_{\alpha_1}\mathbf{r}}+\frac{\nu}{1-\nu}\widetilde{\epsilon_{22}}\widetilde{\partial_{ \alpha_1}\mathbf{r}}+\widetilde{\epsilon_{12}}\widetilde{\partial_{\alpha_2}\mathbf{r}} \right)  \nonumber\\
&\left.+ \widetilde{\partial_{ \alpha_2}}\left(\frac{1}{1-\nu}\widetilde{\epsilon_{22}}\widetilde{\partial_{\alpha_2}\mathbf{r}}+\frac{\nu}{1-\nu}\widetilde{\epsilon_{11}}\widetilde{\partial_{\alpha_2}\mathbf{r}}+\widetilde{\epsilon_{12}}\widetilde{\partial_{\alpha_1}\mathbf{r}} \right)  \right]
    -\rho_fU^2\widetilde{[p]}\mathbf{\hat{n}}\sqrt{\det[\widetilde{\partial_{\alpha_i}\mathbf{r}}\cdot\widetilde{\partial_{\alpha_j}\mathbf{r}}]}.\label{eq:tildeMembraneNonlinear}
\end{align}
Dividing~\eqref{eq:tildeMembraneNonlinear} by $\rho_fU^2$ throughout yields
\begin{align}
     \frac{\rho_s h}{\rho_f L}\widetilde{\partial_{tt}\mathbf{r}}=
\frac{Eh}{\rho_fU^2L(1+\nu)}\bigg[&\widetilde{\partial_{\alpha_1}} \left(\frac{1}{1-\nu}\widetilde{\epsilon_{11}}\widetilde{\partial_{\alpha_1}\mathbf{r}}+\frac{\nu}{1-\nu}\widetilde{\epsilon_{22}}\widetilde{\partial_{ \alpha_1}\mathbf{r}}+\widetilde{\epsilon_{12}}\widetilde{\partial_{\alpha_2}\mathbf{r}} \right)  \nonumber\\
&\left.+ \widetilde{\partial_{ \alpha_2}}\left(\frac{1}{1-\nu}\widetilde{\epsilon_{22}}\widetilde{\partial_{\alpha_2}\mathbf{r}}+\frac{\nu}{1-\nu}\widetilde{\epsilon_{11}}\widetilde{\partial_{\alpha_2}\mathbf{r}}+\widetilde{\epsilon_{12}}\widetilde{\partial_{\alpha_1}\mathbf{r}} \right)  \right]
    -\widetilde{[p]}\mathbf{\hat{n}}\sqrt{\det[\widetilde{\partial_{\alpha_i}\mathbf{r}}\cdot\widetilde{\partial_{\alpha_j}\mathbf{r}}]}.
\end{align}
Thus, the dimensionless membrane equation (dropping tildes) is 
\begin{align}
     R_1{\partial_{tt}\mathbf{r}}-
K_s\left\{{\partial_{\alpha_1}} \left({\epsilon_{11}}{\partial_{\alpha_1}\mathbf{r}}+{\nu}{\epsilon_{22}}{\partial_{ \alpha_1}\mathbf{r}}+(1-\nu){\epsilon_{12}}{\partial_{\alpha_2}\mathbf{r}} \right) + {\partial_{ \alpha_2}}\left({\epsilon_{22}}{\partial_{\alpha_2}\mathbf{r}}+{\nu}{\epsilon_{11}}{\partial_{\alpha_2}\mathbf{r}}+(1-\nu){\epsilon_{12}}{\partial_{\alpha_1}\mathbf{r}} \right)  \right\}\nonumber\\
    =-{[p]}\mathbf{\hat{n}}\sqrt{(\partial_{\alpha_1}\mathbf{r}\cdot \partial_{\alpha_1}\mathbf{r})(\partial_{\alpha_2}\mathbf{r}\cdot \partial_{\alpha_2}\mathbf{r})-(\partial_{\alpha_1}\mathbf{r}\cdot \partial_{\alpha_2}\mathbf{r})^2},\label{eq:nonlinearMembrane}
\end{align}
where $R_1=\rho_s h/(\rho_f L)$ is the dimensionless membrane mass density, $K_s=R_3/(1-\nu^2)$ is a dimensionless stretching stiffness written in terms of $R_3 = Eh/(\rho_fU^2L)$, the dimensionless stretching rigidity, and we have written out the determinant under the square root explicitly in (\ref{eq:nonlinearMembrane}). 
The prestrain $\overline{e}$ in~\eqref{eq:strainTensor} is the strain in a membrane under uniform tension $T_0 = K_s\overline{e}$, the ``pretension," one of the main control parameters here, as in our 2D study~\cite{mavroyiakoumou2020large}. We assume that the thickness ratio $h/L$ is small, but $\rho_s/\rho_f$ may be large, so $R_1$ may assume any non-negative value. As in~\cite{mavroyiakoumou2020large}, we have neglected bending rigidity, denoted as~$R_2$ in~\cite{alben2008flapping}. In the extensible regime studied here, $R_3$ is finite, so $R_2=R_3h^2/(12L^2)\to 0$ in the limit $h/L\to 0$. For simplicity and for ease of comparison with the 2D case we set the Poisson ratio~$\nu$ (the transverse contraction due to axial stretching) to zero.

We solve for the flow using the vortex lattice method, a type of panel method \cite{katz2001low} that solves for the 3D inviscid flow past a thin body by posing a vortex sheet on the body to satisfy the no-flow-through or kinematic condition. The vortex sheet is advected into the fluid at the body's trailing edge, thus avoiding a flow singularity there \cite{katz2001low}. The
velocity $\mathbf{u}$ is a uniform background flow $\widehat{\mathbf{e}}_x$ plus the flow induced by a distribution of vorticity~$\bm{\omega}$, via
the Biot-Savart law \cite{saffman1992vortex}:
\begin{equation}\label{eq:BS}
    \mathbf{u}(\mathbf{x})=\widehat{\mathbf{e}}_x+\frac{1}{4\pi}\iiint_{\mathbb{R}^3}
    \bm{\omega}(\mathbf{x}',t) \times
    (\mathbf{x}-\mathbf{x}')/\|\mathbf{x} -\mathbf{x}'\|^3 \d\mathbf{x}'.
\end{equation}
\nn The vorticity is a vortex sheet on the body and wake. With the body surface parametrized by $\alpha_1$ and $\alpha_2$, we define a local coordinate basis by $\{\mathbf{\widehat{s}}_1, \mathbf{\widehat{s}}_2, \mathbf{\hat{n}}\}$:
\begin{equation}
    \mathbf{\widehat{s}}_1=\frac{\partial_{\alpha_1}\mathbf{r}}{\|\partial_{\alpha_1}\mathbf{r}\|};\qquad \mathbf{\widehat{s}}_2=\frac{\partial_{\alpha_2}\mathbf{r}}{\|\partial_{\alpha_2}\mathbf{r}\|};\qquad \mathbf{\hat{n}}\; \mbox{from (\ref{eq:normal})}.
\end{equation}
\nn Thus $\mathbf{\widehat{s}}_1$ and
$\mathbf{\widehat{s}}_2$ span the body's local tangent plane and $\mathbf{\hat{n}}$ is its normal vector.
$\mathbf{\widehat{s}}_1$ and $\mathbf{\widehat{s}}_2$ are also the tangents to the material lines $\alpha_2=\mathrm{constant}$ and $\alpha_1=\mathrm{constant}$, respectively. When the body experiences in-plane shear, 
$\mathbf{\widehat{s}}_1$ and $\mathbf{\widehat{s}}_2$ are not orthogonal, but they do not become parallel except for singular deformations that we do not consider. 

For the vortex sheet on the body, the vorticity takes the form $\bm{\omega}(\mathbf{x},t) = \bm{\gamma}(\alpha_1,\alpha_2,t)\delta(n) = \gamma_1(\alpha_1,\alpha_2,t)\delta(n)\mathbf{\widehat{s}}_1+\gamma_2(\alpha_1,\alpha_2,t)\delta(n) \mathbf{\widehat{s}}_2$, with $\delta(n)$ the Dirac delta distribution and $n$ the signed distance from the vortex sheet along the sheet normal. The vorticity is concentrated at the vortex sheet, $n = 0$, and $\bm{\gamma}$ is the jump in the tangential flow velocity across the vortex sheet \cite{saffman1992vortex}. The vorticity can be written similarly in the wake vortex sheet, but a different parametrization is used since $\alpha_1$ and
$\alpha_2$ are only defined on the body. The nonlinear kinematic equation states that the normal component of the body velocity equals that of the flow velocity at the body~\cite{saffman1992vortex}:
\begin{equation}\label{eq:kinematic}
    \mathbf{\hat{n}}\cdot \partial_t\mathbf{r}=\mathbf{\hat{n}}\cdot\left(\widehat{\mathbf{e}}_x+\frac{1}{4\pi}\Xiint{-}_{S_B+S_W}
    \bm{\gamma}(\mathbf{x}',t) \times
    (\mathbf{r}-\mathbf{x}')/\|\mathbf{r} -\mathbf{x}'\|^3 \d S_{\mathbf{x'}}\right)
\end{equation}
where $S_B$ and $S_W$ are the body and wake surfaces respectively. Since $\mathbf{r}$ and $\mathbf{x'}$ lie on $S_B$, the integral in (\ref{eq:kinematic}) is singular, defined as a principal value integral. 

The pressure jump across the membrane $[p](\alpha_1,\alpha_2,t)$ can be written in terms of the vortex sheet strength components $\gamma_1$ and $\gamma_2$ using the unsteady Bernoulli equation written at a fixed material point on the membrane. The formula and its derivation are lengthy so they are given in appendix~\ref{app:pressureJump}, but the form of the pressure jump formula is
\begin{align}
\partial_{\alpha_{1}}[p] = G(\mathbf{r},\gamma_1,\gamma_2,\mu_1,\mu_2,\tau_1,\tau_2,\nu_v). \label{eq:pressureJumpAlpha1}
\end{align}
\nn Here $\mu_1$ and $\mu_2$ are 
the tangential components of the average of the flow velocity on the two sides of the membrane, i.e. the dot products of $\mathbf{\widehat{s}}_1$ and $\mathbf{\widehat{s}}_2$ with the term in parentheses on the right side of (\ref{eq:kinematic}). $\tau_1$, $\tau_2$, and $\nu_v$ are the components of the membrane's velocity in the $\{\mathbf{\widehat{s}}_1, \mathbf{\widehat{s}}_2, \mathbf{\hat{n}}\}$ basis:
\begin{equation}\label{eq:tau1}
\tau_1(\alpha_1,\alpha_2,t)=\partial_t\mathbf{r}\cdot \mathbf{\widehat{s}}_1;\qquad
\tau_2(\alpha_1,\alpha_2,t)=\partial_t\mathbf{r}\cdot \mathbf{\widehat{s}}_2;
\qquad\nu_v(\alpha_1,\alpha_2,t)=\partial_t\mathbf{r}\cdot\mathbf{\hat{n}}.
\end{equation}
(\ref{eq:pressureJumpAlpha1}) generalizes the 2D formula from~\cite[appendix A]{mavroyiakoumou2020large} to the case of an extensible body in a 3D flow. We integrate it from the trailing edge using the Kutta condition
\begin{equation}
    [p](\alpha_1=1,\alpha_2,t)=0,
\end{equation}
to obtain $[p](\alpha_1,\alpha_2,t)$ at all points on the membrane.

In summary, equations (\ref{eq:nonlinearMembrane}), (\ref{eq:kinematic}), and (\ref{eq:pressureJumpAlpha1}) are a coupled system of equations for $\mathbf{r}$, 
$\bm{\gamma}$, and $[p]$ that we can solve with suitable initial and boundary conditions to compute the membrane dynamics.

%where $\gamma_1 = \partial_{s_2}\Gamma_2 =(1/\partial_{\alpha_2}s_2)\partial_{\alpha_2}\Gamma$, $\gamma_2 =- \partial_{s_1}\Gamma_1 =-(1/\partial_{\alpha_1}s_1)\partial_{\alpha_1}\Gamma$ are the vortex sheet strength components (see appendix~\ref{app:correspondenceGammaAndgamma} for a relationship between $\gamma$ in 2D and $\gamma_1$, $\gamma_2$ in 3D), 

%%%%%%%%%%%%%%%%%%%%%%%%%%%%%%
\section{Numerical method}\label{sec:numerical}

We solve equations (\ref{eq:nonlinearMembrane}), (\ref{eq:kinematic}), and (\ref{eq:pressureJumpAlpha1}) using an implicit iterative time-stepping approach.
At the initial time $t=0$ the membrane has a very small uniform slope $\partial_{\alpha_1}z/\partial_{\alpha_1}x$ of $10^{-3}$, or $10^{-5}$ if the small-amplitude regime is the focus of interest, and the free vortex wake has length zero. The background flow speed is increased smoothly from 0 to 1 using a saturating exponential function with a time constant 0.2. 

\begin{figure}[H]
    \centering
    \includegraphics[width=\textwidth]{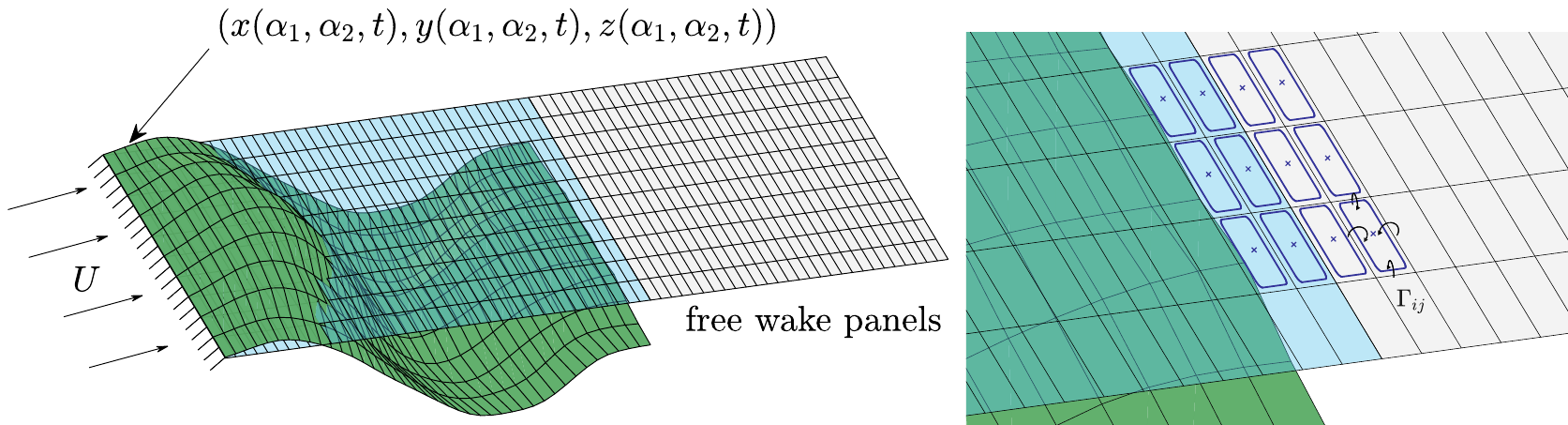}
    \caption{Discretization of the membrane surface into panels with vortex rings. On the left, we show an example of an FRRR deformed membrane (dark green surface) used for computing the inertial and elasticity terms, together with the flat membrane panels (light blue surface) and flat wake panels (light gray surface) used for the kinematic condition. On the right, we show a zoomed-in version of the same membrane with a subset of the vortex rings (blue rounded rectangles) on top of the flat membrane and flat wake panels. The curved arrows illustrate the velocity induced by positive $\Gamma$ according to the right-hand rule.}
    \label{fig:schemNumericalMethod}
\end{figure}

We discretize
the membrane with a uniform rectangular grid of $M+1$ and $N+1$ points in the $\alpha_1$ and $\alpha_2$ directions respectively. The membrane grid becomes deformed in physical space; an example is shown by the black lines on the green membrane surface in figure~\ref{fig:schemNumericalMethod}. The vortex sheet is discretized as a lattice of vortex rings, one per membrane grid cell. Each vortex ring consists of four vortex filaments, one along each side of the grid cell.
This discretization is a type of vortex-lattice method~\cite{katz2001low,hiroaki2021numerical}. To speed up the computations, we approximate the vortex sheet geometries as flat, a common approach in vortex-lattice methods~\cite{katz2001low,hiroaki2021numerical}. This occurs in the kinematic equation~(\ref{eq:kinematic}), which we solve for $\bm{\gamma}$ on the body assuming the body position and velocity (i.e. $\mathbf{r}$, $\partial_t\mathbf{r}$, and $\hat{\mathbf{n}}$) and $\bm{\gamma}$ on the wake are known. The values of $\mathbf{r}$, $\partial_t\mathbf{r}$, and $\hat{\mathbf{n}}$ in~(\ref{eq:kinematic}) are those of the true membrane except in the vortex sheet integrals, where they and $\mathbf{x}'$ correspond to the body in its flat position with uniform pretension in the $z = 0$ plane and points in the wake are those of the body's trailing edge, advected at speed 1 in the $x$-direction. The approximate body and wake vortex sheets lie on the light blue and light gray regions respectively in figure~\ref{fig:schemNumericalMethod}. The vortex sheets are approximated by lattices of rectangular vortex rings that lie on the boundaries of the rectangular grid cells. Each line segment in the grid is the support of two vortex lines, one from the vortex ring of each neighboring grid cell. The inset at the right of figure~\ref{fig:schemNumericalMethod} shows a schematic example of 12 vortex rings in a 3-by-4 portion of the lattice that runs across the membrane's trailing edge. The vortex rings are shown as blue rounded rectangles slightly inside each black rectangular boundary, but they actually run along the black edges.
The circular arrows show the velocity induced by a single vortex ring marked $\Gamma_{ij}$. 

The net circulation along each lattice line segment is the difference of the circulations in the two neighboring vortex rings. The net circulation per unit length in the $\mathbf{\widehat{s}}_1$ direction converges to $\gamma_1$ in the limit of small mesh spacing in that direction, and similarly for $\gamma_2$:
\begin{align} \label{fdGamma}
\gamma_1|_{x_{i+1/2}, y_j} \approx -\frac{\Gamma|_{x_{i+1/2},y_{j+1/2}} - \Gamma|_{x_{i+1/2},y_{j-1/2}}}{y_{j+1/2}-y_{j-1/2}} \quad ; \quad
\gamma_2|_{x_i, y_{j+1/2}} \approx \frac{\Gamma|_{x_{i+1/2},y_{j+1/2}} - \Gamma|_{x_{i-1/2},y_{j+1/2}}}{x_{i+1/2}-x_{i-1/2}}
\end{align}
\nn using centered difference approximations. We linearly interpolate $\gamma_1$ and $\gamma_2$ values at the panel edge midpoints from these expressions to values at the panel corners $\{x_{i}, y_j\}$ and thus obtain $\partial_{\alpha_1}[p]$ in (\ref{eq:pressureJumpAlpha1}).

We solve for the strengths of the vortex rings by imposing (\ref{eq:kinematic}) at a set of $\mathbf{r}$ (in the integral), called the control points or collocation points in vortex-lattice methods, and here located at the centers of each vortex ring (shown as crosses in the inset at the right of figure~\ref{fig:schemNumericalMethod}). \cite{hess1967calculation,yeh1986vortex} also placed the control points at the centers of their vortex rings, quadrilaterals and triangles respectively. Other methods
shift the vortex rings and control points downstream by one-quarter mesh spacing, which reproduces certain formulas for the lift and moment on a flat plate~\cite{james1972remarkable,hough1973remarks,katz2001low,moored2018unsteady,hiroaki2021numerical}. Such methods are considered low-order methods, in contrast to higher-order methods that use curved body panel geometries and nonconstant polynomial representations for the distribution of circulation or velocity potential on each panel~\cite{willis2006quadratic,cole2020unsteady}.
We proceed with the low-order method described here for simplicity and because we find remarkably good agreement with 2D results using a different discretization method based on Chebyshev polynomials and including the non-flat geometries of the body and wake vortex sheets, including wake roll-up \cite{mavroyiakoumou2020large}. We also find good agreement among results with different mesh sizes, although they indicate spatial convergence below first order (see \S\ref{sec:validation}).

The integral in (\ref{eq:kinematic}) is replaced by a sum that gives the velocity induced by the four sides of all the vortex rings. From \cite{katz2001low}, the velocity induced at a point $\mathbf{x}_p$ by a straight vortex line segment with circulation $\Gamma$ that runs from $\mathbf{x}_1$ to $\mathbf{x}_2$ is   
\begin{equation}\label{eq:velocityInduced}
    \mathbf{u}=\frac{\Gamma}{4\pi|\mathbf{r}_1\times \mathbf{r}_2|^2}\left(\frac{(\mathbf{r}_2-\mathbf{r}_1)\cdot \mathbf{r}_1}{\|\mathbf{r}_1\|}-\frac{(\mathbf{r}_2-\mathbf{r}_1)\cdot\mathbf{r}_2}{\|\mathbf{r}_2\|}\right)\cdot(\mathbf{r}_1\times\mathbf{r}_2) \quad ; \quad \mathbf{r}_1 \equiv \mathbf{x}_p - \mathbf{x}_1 \;,\; \mathbf{r}_2 \equiv \mathbf{x}_p - \mathbf{x}_2. 
\end{equation}
When discretized, (\ref{eq:kinematic}) becomes a linear system of equations for the vortex ring strengths. The key advantage of approximating the body and wake vortex sheets as flat is that we can precompute all the matrix entries in the linear system (the ``influence coefficients," i.e. sums of the coefficients of $\Gamma$ in (\ref{eq:velocityInduced})), before the time-stepping iterative solver for the membrane position. For an $M$-by-$N$ lattice of vortex rings on the body and an $M_w$-by-$N$ lattice in the wake (typically $M_w \gg M$), the cost of computing the influence coefficients would be $O(MM_wN^2)$ per iteration within each time step for general membrane and wake shapes, but for flat vortex sheets this can be reduced to a one-time cost of $O(MM_wN)$ (actually less) using repetition in relative positions of control points and vortex rings in the flat wake to change from $N^2$ to $N$. 
We also avoid an even larger cost to evolve the vortex wake position, $O(M_w^2N^2)$, though this could be reduced using fast summation methods \cite{lindsay2001particle}.
The right-hand-side vector for the linear system consists of the remaining terms in (\ref{eq:kinematic}), the body velocity and the background flow, using the actual (non-flat) membrane position, and interpolated to give values at the control points.

%There are different possibilities for where to place the collocation points and singular elements (e.g. vortex rings), and which singularities to use~\cite{murua2012flexible,hess1967calculation,hess1975review,hess1990panel,morino1974subsonic,erickson1990panel,newman1986distributions,voutsinas2006vortex,willis2006quadratic,willis2007combined,ramsey1996boundary}. For instance, one could satisfy the no-penetration condition in terms of the velocity potential instead of the velocity. 
%The uniform grid used in the present work and in~\cite{hou20063d}, may be replaced by a Chebyshev grid~\cite{dejarnette1976arrangement} as in~\cite{mavroyiakoumou2020large,golberg1979solution} for 2D flows, or the vortex rings could be placed according to the quarter-chord method as in~\cite{katz2001low,moored2018unsteady,hiroaki2021numerical}. 
%Thus the acceleration and tension terms in~\eqref{eq:nonlinearMembrane} are both computed at the corner points of the membrane panels but the rest of the terms are computed at the quarter-chord midpoints.
%In~\cite{james1972remarkable} it is found that the quarter-chord method exactly reproduces the lift and moment on a flat plate, unlike the midpoint method. A discussion of this can also be found in~\cite{hough1973remarks}. 
%start here
The algorithm alternates between computing the membrane position at the current time step and updating the vortex wake circulation for the next time step. To compute the membrane position at the current time step, the distribution of circulation in the vortex wake is assumed to be known from the previous steps of the algorithm, and is zero initially. We then use a quasi-Newton iterative method (Broyden's method \cite{ralston2001first}) to solve for the $x$, $y$, and $z$ components of the membrane position on the interior points of the $\alpha_1$-$\alpha_2$ grid, resulting in $3(M-1)(N-1)$ unknowns for the iterative solver. 

Broyden's method solves
a nonlinear system of equations  $\mathbf{f}(\mathbf{x}) = \mathbf{0}$. In our case $\mathbf{x}$ is a vector whose entries are the $x$, $y$, and $z$ components of the membrane position $\mathbf{r}$ on the interior points of the $\alpha_1$-$\alpha_2$ grid.
$\mathbf{f}$ is a vector given by the three components of the membrane equation (\ref{eq:nonlinearMembrane}), discretized at the interior $\alpha_1$-$\alpha_2$ grid points using second-order finite differences for the temporal and spatial derivatives of the membrane position.  One-sided spatial differentiation formulas are used near boundaries and backward temporal differentiation formulas are used, with the given initial values for the membrane position and background flow, and zero initial bound circulation
(see appendix~\ref{app:residualMembraneBroyden} for details). $\mathbf{f}$ requires
the membrane position boundary values, which are either prescribed (for a fixed boundary) or deduced from the guesses for the values at the interior mesh points (for a free boundary). At each time step Broyden's method produces a sequence of iterates that converges to the solution for $\mathbf{x}$ starting from an initial guess, which we choose to be the membrane position at the previous time step. Usually convergence is obtained in just a few iterations. 

Given the free wake position and circulation and the guess for $\mathbf{r}$, we compute $\partial_t \mathbf{r}$ and $\hat{\mathbf{n}}$ and then compute the membrane circulation values by solving the linear system with influence coefficients corresponding to the discretized kinematic equation (\ref{eq:kinematic}). The membrane circulation (and other quantities derived from it and $\mathbf{r}$) are used to compute
$\partial_{\alpha_1}[p]$ in the membrane equation (\ref{eq:nonlinearMembrane}), by integrating (\ref{eq:pressureJumpAlpha1}) from the trailing edge using the trapezoidal rule. 
All quantities on the right hand side of (\ref{eq:pressureJumpAlpha1}) can be evaluated using the guess for $\mathbf{r}$ and the circulations of the vortex rings on the membrane. Some quantities (e.g. $\gamma_1$ and $\gamma_2$) are computed on the panel edge midpoints and others (e.g. $\mu_1$ and $\mu_2$) on the panel centers; these are extrapolated to the corner points of each panel, where $\partial_{\alpha_1}[p]$ is evaluated. Before integrating $\partial_{\alpha_1}[p]$, we decouple it into two parts, as explained at the end of appendix \ref{app:pressureJump}. We integrate one part analytically and the other part numerically using the trapezoidal rule, and apply $[p] = 0$ at the trailing edge. We use this decoupled approach because it gives much better agreement with our previous 2D computations. 

When Broyden's method converges, we obtain the membrane position at the current time step and the circulation values on the membrane. The next step is to update the wake circulation values for the next time step. This is done by moving the vortex rings in the wake and the last row of the vortex rings on the body downstream at speed 1. This is a flat-wake approximation of the more general statement that lines in the wake vortex sheet move at the average of the tangential components of the flow velocity on the two sides of the wake vortex sheet \cite{saffman1992vortex}. This turns out to be equivalent to the condition that the pressure jump is zero across the wake, and
advecting the last row of vortex rings on the membrane into the wake at the trailing-edge flow velocity imposes $[p] = 0$ at the trailing edge, known as the unsteady Kutta condition, which makes flow velocity finite at the trailing edge \cite{saffman1992vortex,katz2001low}.
The time step is set equal to the streamwise grid spacing, so to move downstream at speed 1 each vortex ring in the wake and along the membrane trailing edge in figure \ref{fig:schemNumericalMethod} simply shift to the next panel downstream.

In the next section we study the convergence of this numerical method as the membrane grid is refined, and compare the large-span 3D solutions with 2D membrane solutions computed using the method in~\cite{mavroyiakoumou2020large}.

%%%%%%%%%%%%%%%%%%%%%%%
\section{Validation of the 3D model and algorithm}\label{sec:validation}

We now study the effect of spatial grid refinement on membrane dynamics in test cases using a square membrane at four different combinations of boundary conditions and physical parameters. 
We then compare 3D results in the quasi-2D limit (large span, free side edges) with results at same parameters using our previous 2D algorithm that had several differences: a Chebyshev-Lobatto mesh 
and Chebyshev differentiation matrices were used; the kinematic and Kutta conditions were imposed with a Chebyshev-Galerkin method and non-flat vortex sheets; and the full (non-flat) dynamics of the vortex sheet wake was computed~\cite{mavroyiakoumou2020large}. Three cases are compared: (i) fixed--fixed (2D) and FRFR (3D), (ii) fixed--free (2D) and FRRR (3D), (iii) free--free (2D) and RRRR (3D).  

%%%%%%%%%%%%
\subsection{Spatial convergence for a square membrane}

Ours and most other implementations of the vortex-lattice method are considered low-order methods, as flat rather than curved body panel geometries are used, as well as vortex rings rather than nonconstant polynomial distributions of circulation on each panel \cite{willis2006quadratic,murua2012applications,cole2020unsteady}.
First-order spatial accuracy was found for certain quantities by~\cite{margason1985subsonic,grozdanov2017transonic} using similar versions of the vortex-lattice method. We study the spatial convergence of our method in four test cases with the following boundary conditions and $(R_1,T_0,R_3)$ values, one from each of the colored boxes in figure \ref{fig:schemBC}:
an FFFF membrane with $(R_1,T_0,R_3)=(10^{-0.5},10^{-0.5},10^0)$, an FFRF membrane with $(10^{0.5},10^{-0.75},10^1)$, an RFFF membrane with $(10^{0.5},10^{-0.5
},10^1)$, and an RFRF membrane with $(10^{0},10^{-0.75},10^1)$. All membranes have aspect ratio one ($W/2L = 1$). We compute the dynamics of the membrane up to a time $t_p \in [20,30]$ at which large-amplitude oscillatory motion occurs, and record the
maximum of $|z(\alpha_1,\alpha_2,t)|$ over the membrane surface and time.

In table~\ref{table:dxconvergence} we present the maximum $|z|$ values for four different streamwise mesh spacings $\d x$ together with the change in $\max|z|$ between two successive $\d x$ values, and the estimated order of convergence using three successive $\d x$ values:
\begin{equation}\label{eq:errorAndOrder}
    \Delta\max|z| \equiv \left|\max|z|_{\mathrm{d}x}-\max |z|_{\mathrm{d}x/2}\right|,\qquad \mathrm{Order}\equiv\log_{2}\frac{\left|\max|z|_{\mathrm{d}x}-\max |z|_{\mathrm{d}x/2}\right|}{\left|\max|z|_{\mathrm{d}x/2}-\max |z|_{\mathrm{d}x/4}\right|}.
\end{equation}
We fix the spanwise mesh spacing $\mathrm{d}y=2/10$ (so ten panels cover the span), and recall that
$\d t=\d x$ in all cases, so the time step and streamwise mesh spacing are refined together.

\begin{table}[H]
\caption{$\mathrm{d}x$-convergence with fixed $\mathrm{d}y=2/10$ for four cases with boundary conditions and $t_p$ listed at the top, and $(R_1,T_0,R_3)$ below, for membranes with aspect ratio 1. Recall that $R_1$ is the dimensionless membrane mass, $T_0$ is the dimensionless pretension, and $R_3$ is the dimensionless stretching rigidity. The change in $\max|z|$ between successive $\mathrm{d}x$ values and the estimated order of convergence are defined in~\eqref{eq:errorAndOrder}.}\label{table:dxconvergence}
\centering
\vspace{.1cm}
\begin{tabular}{lccccccccccccccc}
\hline\hline
&    \multicolumn{3}{c}{FFFF, $t_p= 30$}   &  & \multicolumn{3}{c}{FFRF,  $t_p= 22$} &  &    \multicolumn{3}{c}{RFFF, $t_p= 22$} & & \multicolumn{3}{c}{RFRF, $t_p= 20$}  \\
&   \multicolumn{3}{c}{$(10^{-0.5},10^{-0.5},10^0)$}    &   & \multicolumn{3}{c}{$(10^{0.5},10^{-0.75},10^1)$}  &  &  \multicolumn{3}{c}{$(10^{0.5},10^{-0.5},10^1)$}  & & \multicolumn{3}{c}{$(10^{0},10^{-0.75},10^1)$} \\
      \cline{2-4} \cline{6-8} \cline{10-12} \cline{14-16}
$\mathrm{d}x$    & $\max|z|$ & $\Delta\max|z|$ & Order &  & $\max|z|$ & $\Delta\max|z|$ & Order &  & $\max|z|$ & $\Delta\max|z|$ & Order &  & $\max|z|$ & $\Delta\max|z|$ & Order \\
\hline
2/40  & 0.5596    &  0.0171 &     1.89  &     &      0.2169  &  $7.71\times10^{-3}$ &   1.55    &    &  0.2417   & $4.72\times 10^{-3}$ &     ---   &    &   0.2891  & $9.85\times 10^{-3}$ & 0.57  \\
2/80  &   0.5425  & $4.60\times10^{-3}$  & 1.46  &   &   0.2092    & $2.64\times10^{-3}$  & 0.63       &    &   0.2369    & --- &  ---      &           &   0.2989  & $6.64\times 10^{-3}$ & 0.38  \\
2/160  &  0.5379     &$1.67\times10^{-3}$  & $\dots$ &   &  0.2066 & $1.71\times10^{-3}$ &  $\dots$  &    &     ---  & --- &     $\dots$   &       &  0.3056   & $5.09\times 10^{-3}$ & $\dots$ \\
2/320 &  0.5363    & $\dots$ & $\dots$    &     & 0.2048   &  $\dots$ & $\dots$   &     &  ---     & $\dots$ &    $\dots$   &    &  0.3107 & $\dots$ & $\dots$  \\
\hline\hline
\end{tabular}
\end{table}

In general RFFF membranes take a long time to reach the large-amplitude regime (across $R_1$), so the computations with 
$\d x = 2/160$ and 2/320 were omitted because they were very expensive to compute (due to the large wake size needed). Although the order of convergence is below one in many cases, the $\Delta\max|z|$ values are quite small. We guess that $\Delta\max|z|$ is a reasonable estimate of the error in each case, the difference between the computed solution and the solution in the limit $\d x \to 0$. As a compromise between computational effort and accuracy, we use $\d x = 2/40$ ($M$ = 40) in most cases (i.e. throughout \S\ref{sec:results}), but we present a few comparisons between
$\d x$ = 2/40, 2/80, and 2/160 in the next subsection and find that the results are qualitatively similar.

\begin{table}[H]
\caption{$\mathrm{d}y$-convergence with fixed $\mathrm{d}x=2/40$ for a membrane with aspect ratio 1. The error and order of convergence shown are computed using~\eqref{eq:errorAndOrder} but varying $\d y$ instead of $\d x$. The $(R_1,T_0,R_3)$ values are the same as in table~\ref{table:dxconvergence}.}\label{table:dyconvergence}
\centering
\vspace{.1cm}
\begin{tabular}{lccccccccccccccc}
\hline\hline
&    \multicolumn{3}{c}{FFFF, $t_p= 30$}   &  & \multicolumn{3}{c}{FFRF,  $t_p= 22$} &  &    \multicolumn{3}{c}{RFFF, $t_p=22$} & & \multicolumn{3}{c}{RFRF, $t_p= 20$}  \\
 & \multicolumn{3}{c}{$(10^{-0.5},10^{-0.5},10^0)$}    &   & \multicolumn{3}{c}{$(10^{0.5},10^{-0.75},10^1)$}  &  &  \multicolumn{3}{c}{$(10^{0.5},10^{-0.5},10^1)$}  & & \multicolumn{3}{c}{$(10^{0},10^{-0.75},10^1)$}  \\
      \cline{2-4} \cline{6-8} \cline{10-12} \cline{14-16}
$\mathrm{d}y$    & $\max|z|$ & $\Delta\max|z|$ & Order &  & $\max|z|$ & $\Delta\max|z|$ & Order &  & $\max|z|$ & $\Delta\max|z|$ & Order &  & $\max|z|$ & $\Delta\max|z|$ & Order \\
\hline
2/10  & 0.5596    &  0.0119 &  2.05  &     &      0.2169  & $7.56\times10^{-3}$ &  2.29  &    &   0.2417    & $2.14\times 10^{-3}$ & 0.08  &            & 0.2891  &  $7.59\times 10^{-3}$  & 1.70    \\
2/20  &  0.5715  & $2.87\times10^{-3}$  & 6.13  &   &   0.2093  & $1.54\times10^{-3}$ &  8.62   &   &   0.2438   & $2.02\times 10^{-3}$  & 5.54  &   &  0.2815   &   $2.34\times 10^{-3}$   & 1.63   \\
2/40  &  0.5744  & $4.11\times 10^{-5}$ & $\dots$  &    & 0.2078 &  $3.93\times 10^{-6}$  & $\dots$  &  & 0.2418 & $4.35\times 10^{-5}$ & $\dots$ &    & 0.2792   & $7.58\times 10^{-4}$ & $\dots$   \\
2/80 &  0.5744  & $\dots$ & $\dots$  & & 0.2078 & $\dots$ & $\dots$ &  & 0.2418  & $\dots$  & $\dots$   &  & 0.2784   & $\dots$ & $\dots$  \\
\hline\hline
\end{tabular}
\end{table}
Table~\ref{table:dyconvergence} shows the $\max |z|$ values at four choices of the spanwise mesh spacing $\d y$, with $\d x$ fixed at 2/40 and the other parameters the same as in table~\ref{table:dxconvergence}. The $\d y$ convergence is much faster than the $\d x$ convergence, presumably because $\d x$ sets the resolution near the membrane trailing edge, a particularly sensitive region due to the vortex shedding there \cite{alben2009simulating,alben2010regularizing,hiroaki2021numerical}. 
$\Delta\max|z|$ is uniformly small in table~\ref{table:dyconvergence}, at most $4.4\%$ of the last $\max|z|$ value (with $\d y=2/80$). For computational efficiency with reasonable accuracy we use $\d y=2/10$ ($N=10$) in most cases (i.e. throughout \S\ref{sec:results}).
Other recent vortex lattice works have also found that
a modest number of grid points in the spanwise direction is sufficient for good accuracy (e.g. \cite{long2004object}, \cite[p.~429]{katz2001low}, and \cite[figure~6]{nguyen2016extended}).

%%%%%%%%%%
\subsection{Comparisons of 2D and 3D results with different mesh sizes}\label{sec:comparison2D}

We now compare 3D results with free side edges with results from our 2D algorithm~\cite{mavroyiakoumou2020large} in cases with both small- and large-amplitude dynamics. We begin by comparing the
2D fixed--fixed case with the 3D FRFR case (fixed leading and trailing edges, free side edges). We simulate both cases by starting with a flat membrane at $t=0$, and for $t >  0$ we keep the leading edge fixed at $z = 0$ and move the trailing edge slightly away from and back to $z = 0$:
\begin{equation}\label{eq:perturbationFixedFixed}
    z_{2D}(1,t) = z_{3D}(1,\alpha_2,t) = 2\sin\left(\sigma\left(\frac{t}{\eta}\right)^3\mathrm{e}^{-(t/\eta)^3}\right) \; , \quad -\frac{W}{2L} \leq \alpha_2 \leq \frac{W}{2L},
\end{equation}
where $\sigma$ is a small constant, $10^{-6}$--$10^{-3}$, and $\eta=0.2$.

\begin{figure}
    \centering
    \includegraphics[width=\textwidth]{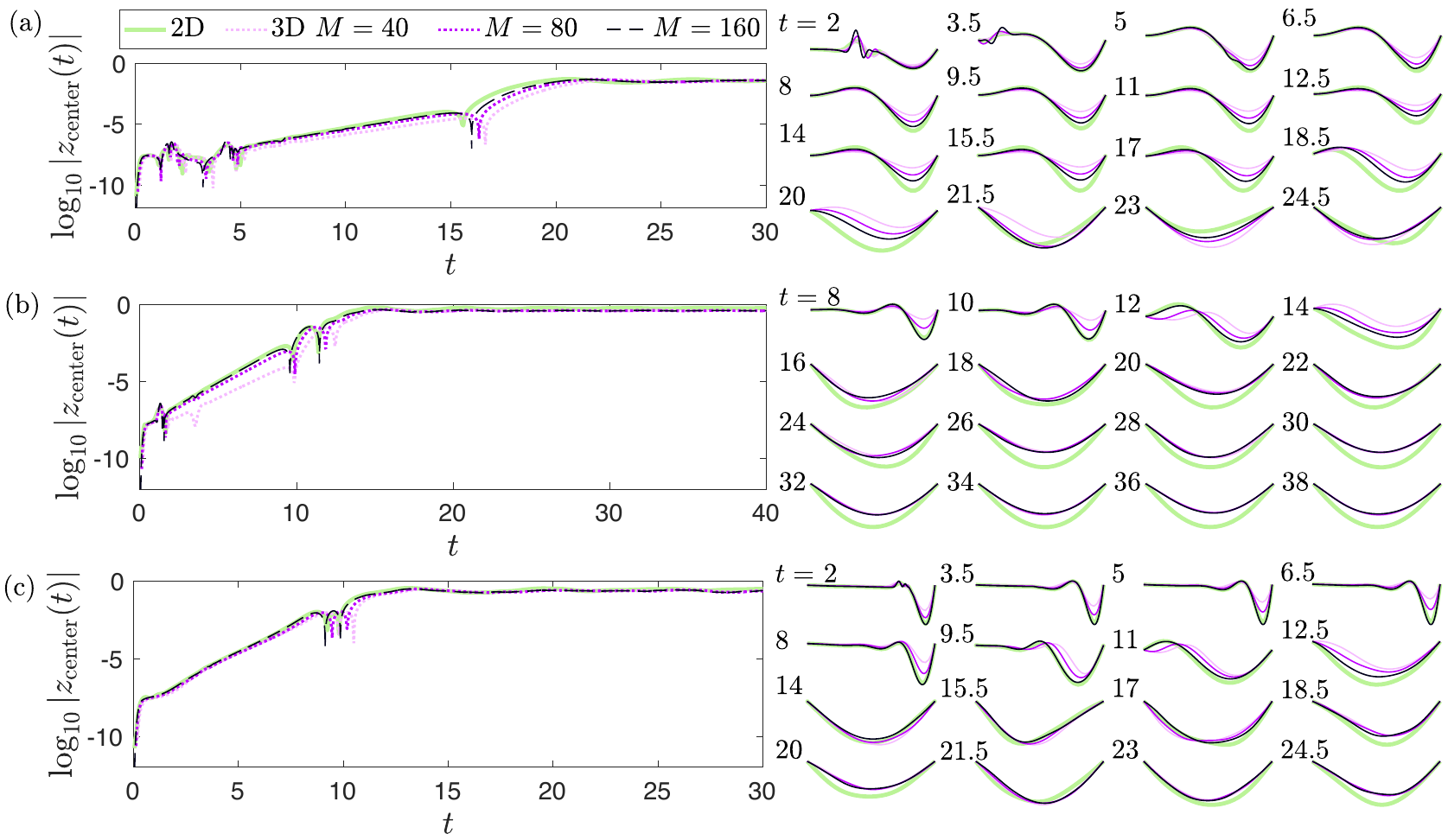}
    \caption{Comparisons of fixed--fixed (2D) and FRFR (3D) computations at three choices of $(R_1,T_0,R_3)$: (a) $(10^{0},10^{-0.25},10^{3})$, (b) $(10^{-0.5},10^{-0.5},10^{1})$, and (c) $(10^{0.25},10^{-0.75},10^{1.5})$. Recall that $R_1$ is the dimensionless membrane mass, $T_0$ is the dimensionless pretension, and $R_3$ is the dimensionless stretching rigidity. The panels at left show plots of $\log_{10}|z_{\mathrm{center}}|$ versus time.
    At right, snapshots of membranes at equally spaced times (labeled at top left) are shown.
        The plots and snapshots are green for 2D membranes and light pink ($M = 40$), dark pink ($M = 80$), or black ($M = 160$) for 3D membranes. Here the aspect ratio $W/2L = 4$, $N=10$, and $\sigma=10^{-6}$ in \eqref{eq:perturbationFixedFixed}.}
    \label{fig:FRFRcomparisonWith2D}
\end{figure}

On the left side of figure~\ref{fig:FRFRcomparisonWith2D}, the panels plot the $z$-deflection of the membrane center (in 3D) or midpoint (in 2D) versus time at three choices of $(R_1,T_0,R_3)$ (listed in the caption) in panels (a)--(c). The 2D membrane has a Chebyshev-Lobatto mesh with 41 points.
In 3D, the number of panels in the streamwise direction ($M$) is 40, 80, or 160, with the aspect ratio $W/2L$ = 4, and the number of panels in the spanwise direction ($N$) is 10 in each case. The perturbation size, $\sigma$ in~\eqref{eq:perturbationFixedFixed}, is $10^{-6}$. In panels (a)--(c), the 3D graphs become closer to the 2D graph as $M$ increases.  

To the right of each set of time plots, membrane snapshots are shown, equally spaced in time, for 2D (green lines) and 3D (with the same colors as in the time plots, for each $M$), where the line shows the midspan profile.
In all three examples, the membranes tend to a steady single-hump shape at large times, as shown for the 2D case in \cite{mavroyiakoumou2020large}.
In figure \ref{fig:FRFRcomparisonWith2D}
there is good qualitative agreement in all the membranes' motions, and good quantitative agreement in most cases, particularly between $M=160$ (black lines) and 2D (green lines) in the small-amplitude (growth) regime. At later times, the 3D snapshots tend to converge, while the 2D snapshot is noticeably displaced from them, particularly in panel (b). The differences between 2D and 3D may be due in part to algorithmic differences, and to differences in the flows due to the finite span in 3D.

\begin{figure}
    \centering
    \includegraphics[width=\textwidth]{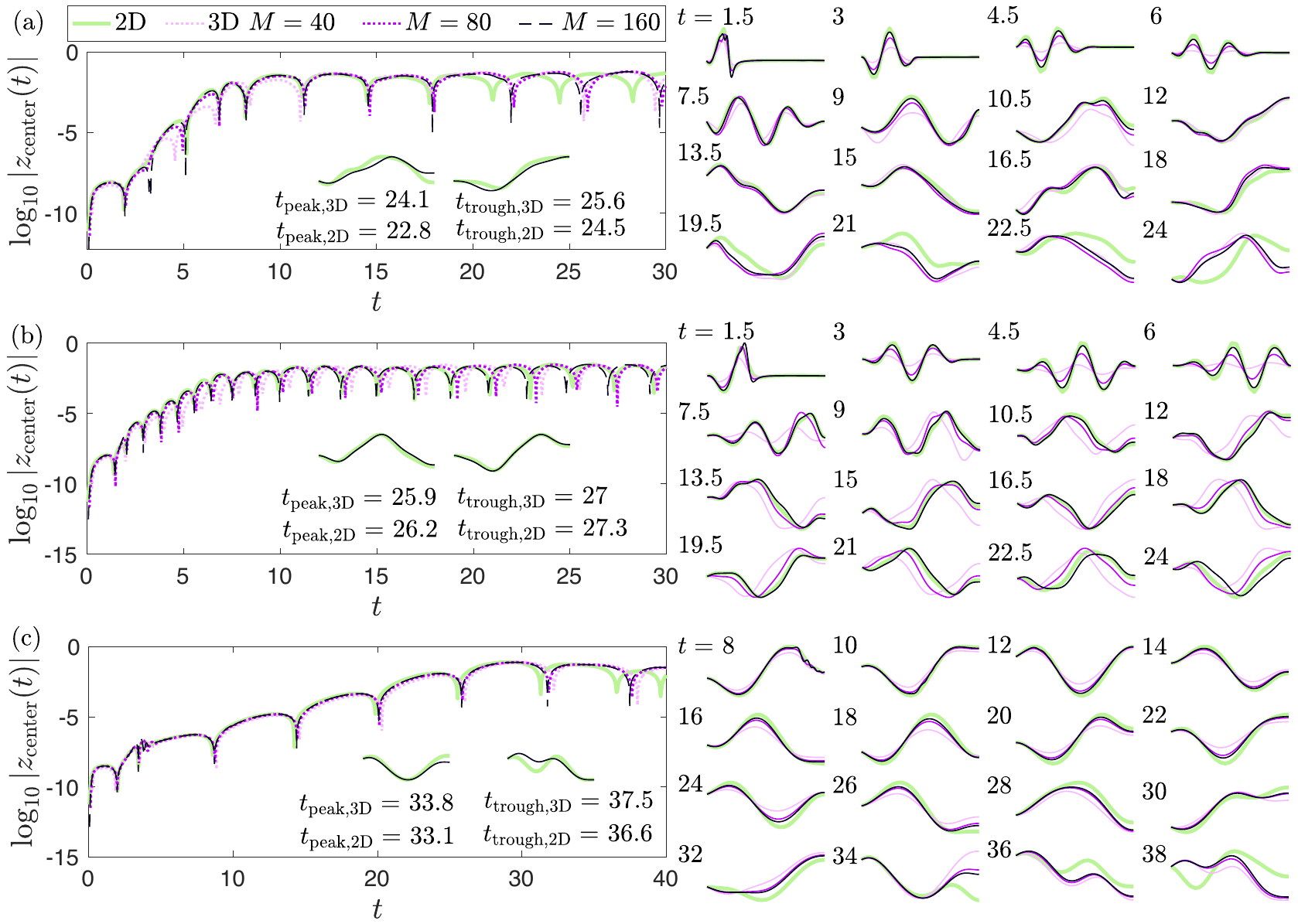}
    \caption{Comparisons of fixed--free (2D) and FRRR (3D) computations at three choices of $(R_1,T_0,R_3)$:
    (a) $(10^0,10^{-1},10^2)$, (b) $(10^{-0.5},10^{-1},10^{2})$, and (c) $(10^{0.5},10^{-0.5},10^{2})$.
    The panels at left show plots of $\log_{10}|z_{\mathrm{center}}|$ versus time.
    At right, snapshots of membranes at equally spaced times (labeled at top left) are shown.
        The plots and snapshots are green for 2D membranes and light pink ($M = 40$), dark pink ($M = 80$), or black ($M = 160$) for 3D membranes. Here the aspect ratio $W/2L = 4$, $N=10$, and $\sigma=10^{-6}$ in~\eqref{eq:perturbationFixedFree}.}
    \label{fig:FRRRcomparisonWith2D}
\end{figure}

Next we compare the 2D fixed--free case with the 3D FRRR case. Now the trailing edge is free, and a small transient perturbation is applied at the leading edge:
\begin{equation}\label{eq:perturbationFixedFree}
    z_{2D}(-1,t) = z_{3D}(-1,\alpha_2,t) = \sin\left(\sigma\left(\frac{t}{\eta}\right)^3\mathrm{e}^{-(t/\eta)^3}\right) \; , \quad -\frac{W}{2L} \leq \alpha_2 \leq \frac{W}{2L}.
\end{equation}
Figure~\ref{fig:FRRRcomparisonWith2D} compares the membranes' center-point deflections and snapshots similarly to figure~\ref{fig:FRFRcomparisonWith2D} but at three different choices of $(R_1,T_0,R_3)$, given in the figure~\ref{fig:FRRRcomparisonWith2D} caption. In the fixed--free case, oscillatory motions occur at large amplitude \cite{mavroyiakoumou2020large}, and they tend to be periodic rather than chaotic at the $O(1)$ values of $R_1$ used in figure~\ref{fig:FRRRcomparisonWith2D}. Again, there is good qualitative agreement in all cases, and good quantitative agreement particularly in the linear growth regime. 
At large times in panels (a) and (c) the three 3D cases are clustered together and are somewhat separate from the 2D case. In panel (b) the $M = 160$ and 2D case are relatively close, even at large times. Within each panel the graphs have somewhat different periods of oscillation at large amplitude, including some abrupt changes in the period, e.g. the 2D graph near the final time in panel (c). However, if we compare the snapshots at qualitatively similar times, i.e. at nearby peaks and troughs of the $\log_{10}|z_{\mathrm{center}}(t)|$ plots, instead of at the same times, then we find better agreement. Comparisons of the 2D case and the 3D case with $M = 160$ at nearby peaks and troughs are shown below the time plots, and show better agreement than in the late-time comparisons on the right side of the figure.

\begin{figure}
    \centering
    \includegraphics[width=\textwidth]{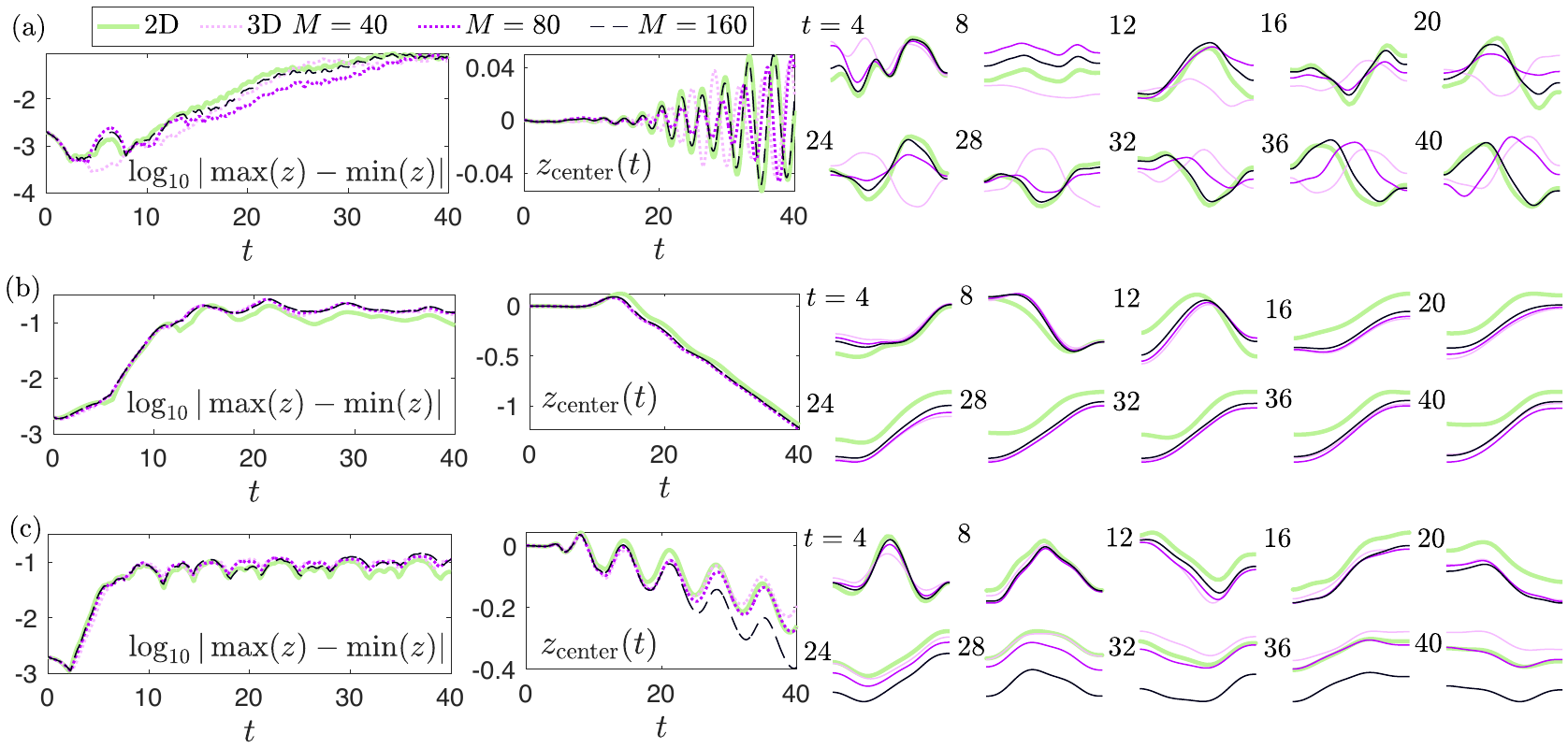}
    \caption{Comparisons of free--free (2D) and RRRR (3D) computations at three choices of $(R_1,T_0,R_3)$:
    (a) $(10^{-0.5},10^{-0.75},10^1)$, (b) $(10^{0.5},10^{-0.5},10^2)$, and (c) $(10^{-0.25},10^{-0.8},10^2)$.
    The panels at left show plots of $\log_{10}|\max (z)-\min (z)|$ and $z_{\mathrm{center}}$ versus time.
    At right, snapshots of membranes at equally spaced times (labeled at top left) are shown.
        The plots and snapshots are green for 2D membranes and light pink ($M = 40$), dark pink ($M = 80$), or black ($M = 160$) for 3D membranes. Here the aspect ratio $W/2L = 4$, $N=10$, and $\sigma=10^{-3}$ in~\eqref{eq:perturbationFreeFree}.}
    \label{fig:RRRRcomparisonWith2D}
\end{figure}

Finally, we compare the free--free 2D case with 3D RRRR cases. Now the membrane starts with a small nonzero slope:
\begin{equation}\label{eq:perturbationFreeFree}
    z_{2D}(\alpha_1,0) = z_{3D}(\alpha_1,\alpha_2,0) = \sigma \alpha_1\; , \quad -\frac{W}{2L} \leq \alpha_2 \leq \frac{W}{2L},
\end{equation}
for $\sigma=10^{-3}$. In figure~\ref{fig:RRRRcomparisonWith2D} we again compare the membranes' deflections and snapshots for three different $(R_1,T_0,R_3)$ combinations
(in the figure caption)
with $R_1 = O(1)$ that have qualitatively different dynamics. With all edges free, the membrane can move uniformly upward or downward yet keep the same shape, so there is a translational $z$ motion to consider along with the evolution of the membrane shape.
Therefore, on the left side we plot the membrane deflection versus time, defined here as $|\max(z)-\min(z)|$, with max and min taken over $\alpha_1$ (and $\alpha_2$ for 3D), and we also plot $z_{\mathrm{center}}(t)$, which shows the net translational motion.
Panel (a) shows an oscillatory motion of $z_{\mathrm{center}}$, (b) shows a steady translation, and (c) shows a combination of the two. As in the fixed--fixed and fixed--free cases, in each panel the agreement in the deflection $|\max (z)-\min (z)|$ is generally best at early times, and at later times there is a more noticeable difference, particularly between the 2D and 3D plots in panels (b) and (c). The snapshots at the right show that in many cases the shapes agree well but there is a translational shift, particularly at $t = 8$ in (a) (where the shift is small, $\approx 10^{-3}$) and at larger times in (c). As in the fixed--free case, there are noticeable phase shifts among the motions at later times for the oscillatory case in panel (a).

In summary, the comparisons of 3D motions at $M$ = 40, 80, and 160 and the 2D motion have shown good qualitative agreement in the types of dynamics---steady or oscillatory, with or without translation---and good quantitative agreement as well in the magnitudes of the deflections, the frequencies of oscillations, and many detailed aspects of the dynamics. 

\begin{figure}
    \centering
    \includegraphics[width=\textwidth]{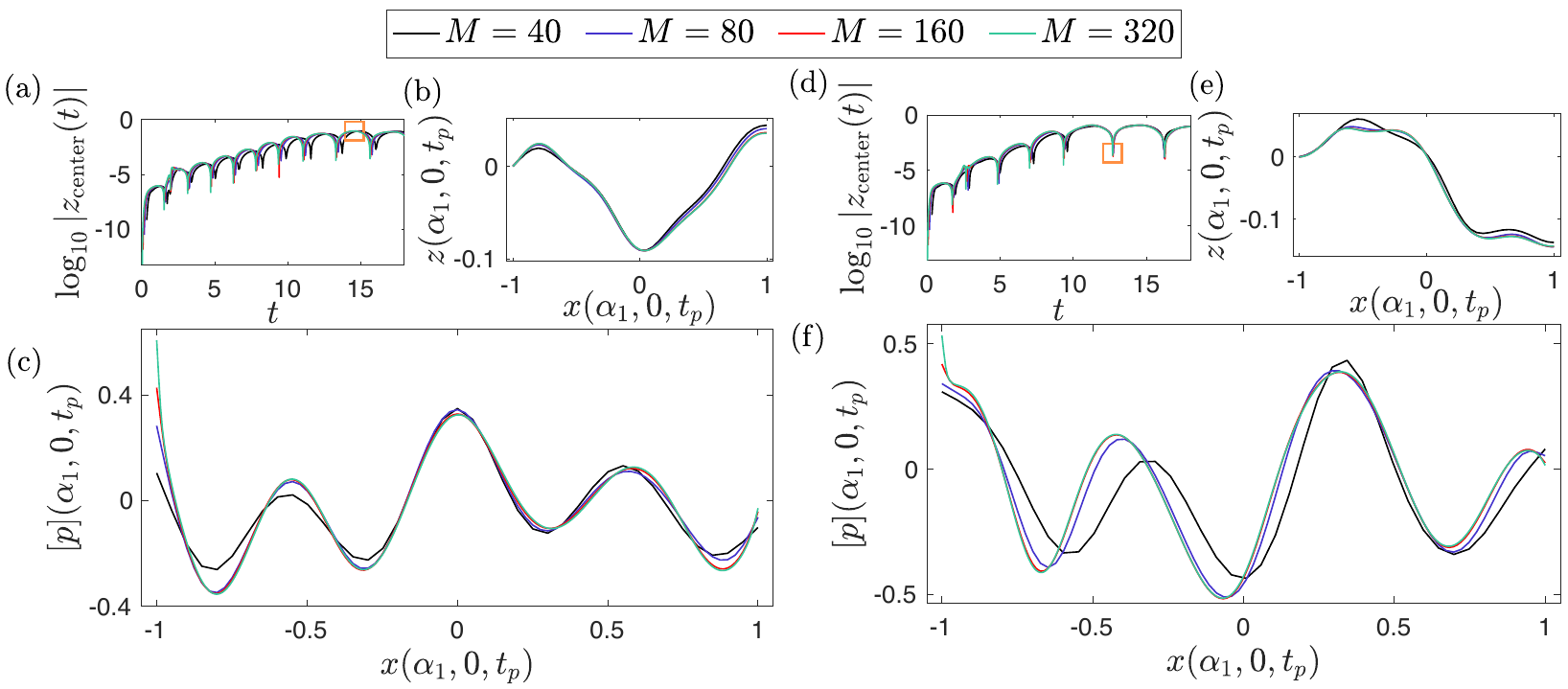}
    \caption{Two examples of how the pressure distribution changes with increasing numbers of streamwise panels. Panels (a)--(c) show plots of  $\log_{10}|z_{\mathrm{center}}|$ versus time and $z(\alpha_1,0,t_p)$ and $[p](\alpha_1,0,t_p)$ versus $x(\alpha_1,0,t_p)$, respectively, for an FRRR membrane with $R_1=10^{-0.25}$, $T_0=10^{-0.75}$, and $R_3=10^1$ at times $t_p$ where $|z_{\mathrm{center}}(t)|$ has a peak that is closest to 14.5 (enclosed in the orange square), for each $M$. Panels (d)--(f) show the same quantities for an FRRR membrane with $R_1=10^{0}$, $T_0=10^{-0.75}$, and $R_3=10^1$, for the troughs in $|z_{\mathrm{center}}(t)|$ nearest to $t_p\approx 12.6$. The plots are black for $M=40$, blue for $M=80$, red for $M=160$, and green for $M=320$. Here the aspect ratio $W/2L=4$, $N=10$, and $\sigma=10^{-4}$ in~\eqref{eq:perturbationFixedFree}.}
    \label{fig:PressureJumpDistribution}
\end{figure}

Inviscid simulations of airfoils and flapping plates in 2D and 3D have shown that the largest computational errors tend to occur in flow quantities near the trailing edge \cite{katz2001low,alben2008optimal,alben2010regularizing,hiroaki2021numerical}. The Kutta condition makes the flow velocity and pressure jump finite there, but their spatial derivatives across the edge are infinite there in general \cite{alben2010regularizing}, and the numerical solution is sensitive to the mesh near the trailing edge \cite{alben2010regularizing,hiroaki2021numerical}. For the present simulations, this sensitivity can be seen in the pressure jump distribution near the trailing edge. In figure~\ref{fig:PressureJumpDistribution} we plot the pressure jump versus streamwise location along the membrane midspan, $y = 0$, for two different examples of FRRR membranes in the large-amplitude regime, at $R_1$, $T_0$, and $R_3$ given in the figure caption, and for various $M$ listed at the top. Both cases reach oscillatory large-amplitude states, shown by the center deflection versus time in panels (a) and (d) respectively. In panels (b) and (e) we compare the midspan deflections that occur at $|z_{\mathrm{center}}|$ peaks near $t = 14.5$ and troughs near 12.6, respectively---the peaks and troughs that occur in the orange boxes of panels (a) and (d). For each $M$ the times of the peaks/troughs are slightly different due to phase shifts, as in the snapshot comparisons in the left panels of figure \ref{fig:FRRRcomparisonWith2D}. Panels (c) and (f) show the midspan pressure jump distributions at these times. The pressure jump distributions are more oscillatory and have larger deviations at these $M$ than the deflections ((b) and (e)), perhaps because in the membrane equation (\ref{eq:nonlinearMembrane}) $[p]$ is balanced by terms that include second derivatives of deflection, so $z$ is smoother than $[p]$. The pressure jump distributions in (c) and (f) only approximately
reach zero at the trailing edge, more closely for larger $M$, which also resemble the generic square-root behavior there, as well as the generic inverse-square-root behaviors at the leading edge \cite{golberg1990ins,alben2009simulating,alben2010regularizing}. Although the deviations from $M = 40$ to 320 are significant, the overall form of the $[p]$ and $z$ distributions at $M$ = 40 are reasonable approximations to those at $M$ = 320.

\begin{figure}
    \centering
    \includegraphics[width=\textwidth]{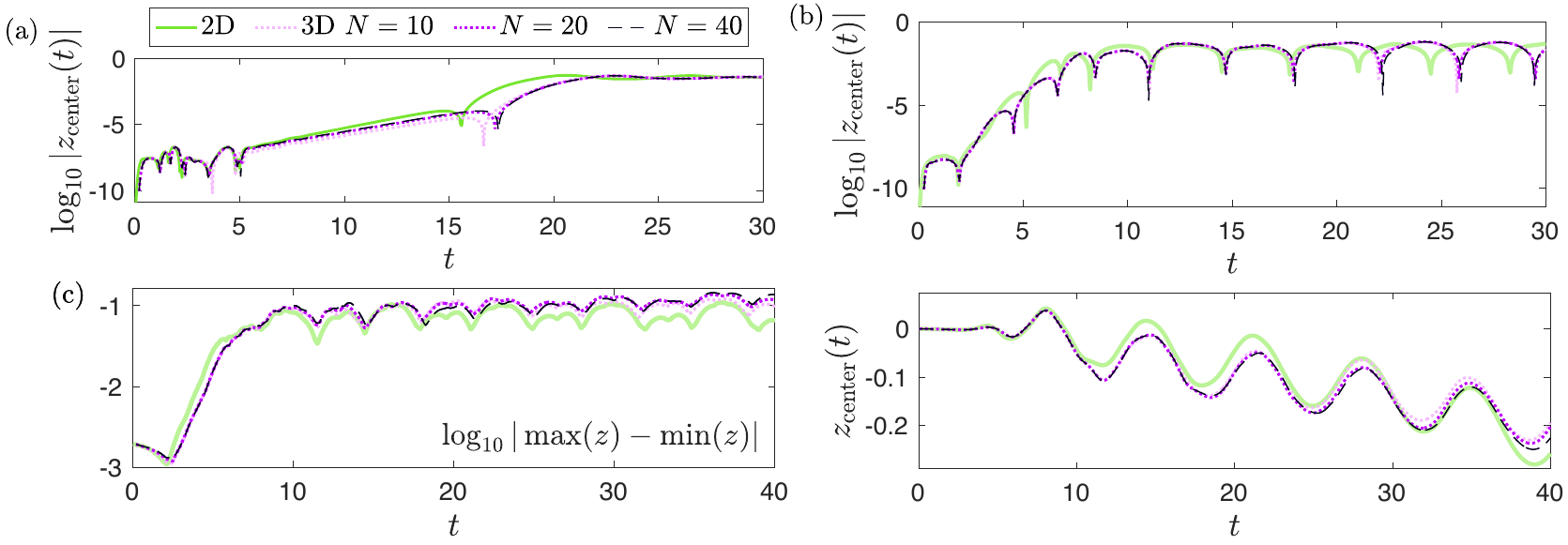}
    \caption{Comparisons of 2D and 3D membrane deflections at three boundary conditions and $(R_1,T_0,R_3)$ values: (a) fixed--fixed versus
    FRFR at $(10^0,10^{-0.25},10^3)$; (b) fixed--free versus
    FRRR at $(10^0,10^{-1},10^2)$ and (c) free--free versus RRRR at
    $(10^{-0.25},10^{-0.8},10^2)$, where both $\log_{10}| \max (z)-\min(z)|$ and $z_{\mathrm{center}}$ are plotted, at left and right respectively. The plots are green for 2D membranes and light pink ($N = 10$), dark pink ($N = 20$), or black ($N = 40$) for 3D membranes. Here 41 Chebyshev-Lobatto points are used in 2D and for the 3D cases $W/2L = 4$, $M=40$, and $\sigma$ = $10^{-6}$, $10^{-6}$, and $10^{-3}$ respectively in (a)--(c).}
    \label{fig:varyN}
\end{figure}

Now we briefly present a comparison of dynamics in 2D and 3D using different numbers of panels across the span for 3D: $N$ = 10, 20, and 40. Figure~\ref{fig:varyN} shows comparisons of (a) fixed--fixed, (b) fixed--free, and (c) free--free 2D cases with the corresponding 3D cases at $M$ = 40 and aspect ratio 4. The 3D plots converge more rapidly here with increasing $N$ than with increasing $M$ in the previous figures, similarly to tables \ref{table:dxconvergence} and \ref{table:dyconvergence}. In the remainder of this work we use $N=10$.

We have compared the small- and large-amplitude dynamics in 2D and 3D at several $(R_1,T_0,R_3)$ values, and shown the effect of varying the numbers of streamwise and spanwise panels in 3D, $M$ and $N$, at aspect ratio~4. Next, we compare the dynamics across a much larger set of $R_1$ and $T_0$ values, but restricted to the small-amplitude regime (where $R_3$ has a negligible effect). We also show the effect of increasing the 3D membrane aspect ratio from 1 to 8, approaching the 2D limit.

%%%%%%%%%%%%%%%%%%%%%%%%
\subsection{Comparison of stability boundaries and mode shapes in 2D and 3D}

The small-amplitude membrane dynamics can be organized in terms of the stability boundary in $R_1$--$T_0$ space. We again apply the
small perturbations \eqref{eq:perturbationFixedFixed}--\eqref{eq:perturbationFreeFree} to the FRFR, FRRR, and RRRR membranes, across a grid of values in $R_1$--$T_0$ space. At each $R_1$ we find exponential decay in membrane deflections above a critical $T_0$ and exponential growth below this $T_0$ (followed by large-amplitude, nonlinear dynamics). The stability boundary is the curve of critical $T_0$ versus $R_1$, and is found by interpolating between neighboring $T_0$ points with positive and negative growth rates.

\begin{figure}
    \centering
    \includegraphics[width=\textwidth]{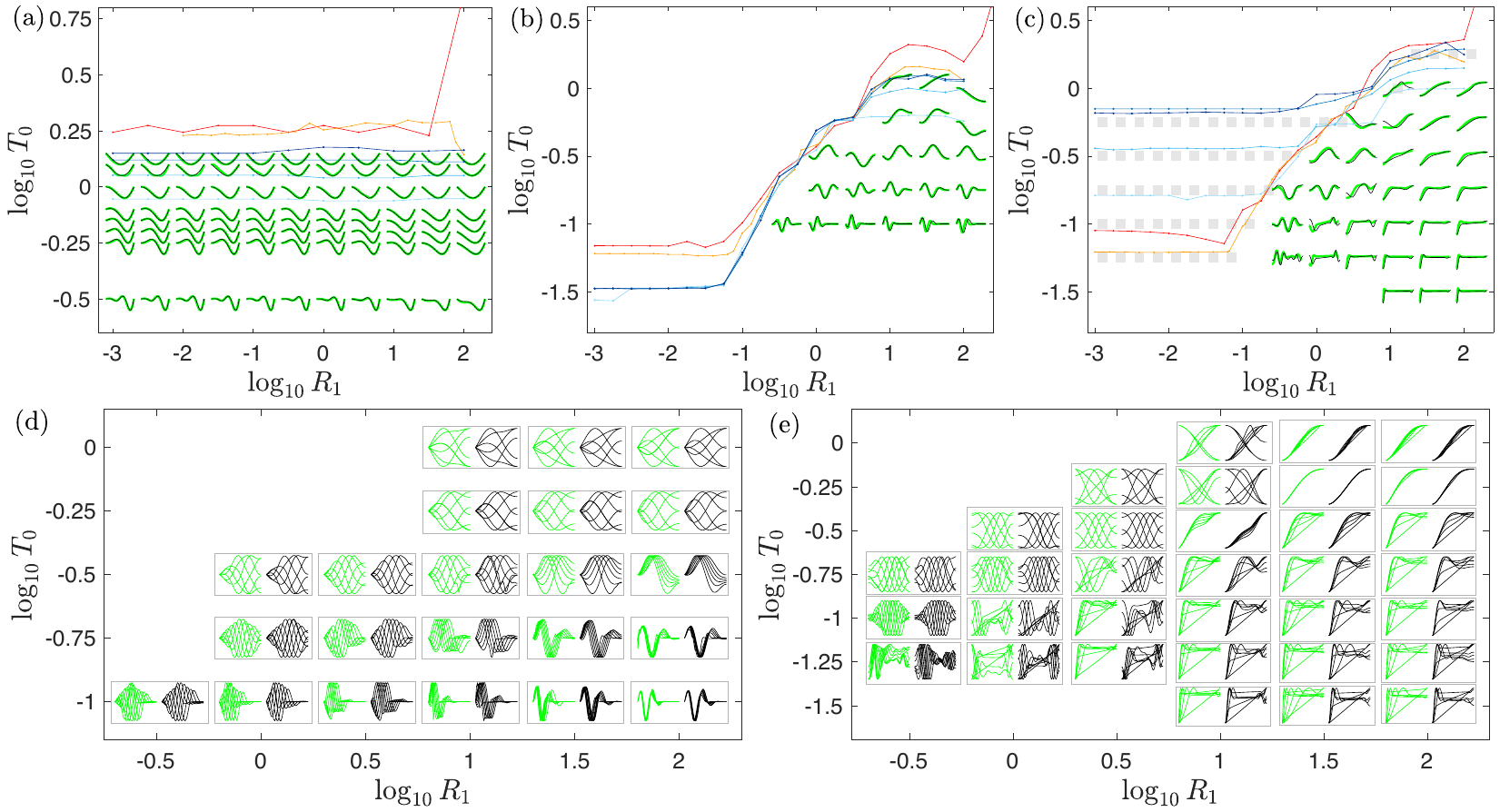}
    \caption{Comparisons of stability boundaries in 2D (orange from \cite{mavroyiakoumou2020large}; red from \cite{mavroyiakoumou2021eigenmode}) and 3D with aspect ratios 1, 2, 4, and 8 (ranging from light blue to dark blue) for (a) fixed--fixed and FRFR, (b) fixed--free and FRRR, and (c) free--free and RRRR cases.
    Below the stability boundary, 3D midspan profiles with aspect ratio 8 (black) are compared with 2D profiles (green) during the linear growth regime. Here $R_3=10^1$, $N=10$, $M=40$ (for FRFR and FRRR membranes) and $M=80$ (for RRRR membranes) for all aspect ratios.}   \label{fig:effectOfSpan}
\end{figure}

In figure \ref{fig:effectOfSpan} we compare the stability boundaries in the 2D fixed--fixed (a), fixed--free (b) and free--free (c) cases with
the 3D FRFR, FRRR, and RRRR cases respectively, with 3D aspect ratios 1, 2, 4, and 8. Two stability boundaries are shown for the 2D case. The orange lines, from \cite{mavroyiakoumou2020large}, are calculated from
the growth rates in time-stepping simulations starting from small perturbations, as in the 3D cases here. The red lines, from \cite{mavroyiakoumou2021eigenmode}, are calculated by linearizing the 2D equations and solving the nonlinear eigenvalue problem; the imaginary parts of the eigenvalues are the growth rates. The eigenvalue method was able to identify very small positive growth rates at $R_1 \geq 10^2$, too small to be distinguishable from no growth or very slow decay in the time-stepping method. Here the red line has a sharply upward slope (similar to \cite{tiomkin2017stability}), unlike the orange line.
The stability boundaries for the 3D cases with aspect ratios 1, 2, 4, and 8 range from light blue to dark blue. In (a) and at large $R_1$ in (b) and (c) the boundaries tend towards the 2D boundaries as aspect ratio increases. As in studies of 3D flexible plates \cite{ESS2007,eloy2008aic,banerjee2015three}, more bodies are unstable at high aspect ratio, i.e. the stability boundary moves up as the aspect ratio increases. In the region of instability for aspect ratio 8, we compare the aspect-ratio-8 midspan snapshots at small amplitude (black lines) with the 2D snapshots (green lines) over a range of $R_1$ and~$T_0$. In (a), all the membranes lose stability by divergence without flutter, so the membranes' deflections grow exponentially without change of shape. The snapshots show the eigenmode shapes, which become more fore-aft asymmetric as $T_0$ decreases, although at large amplitude the membranes tend to single-hump shapes that are nearly fore-aft symmetric as in figure \ref{fig:FRFRcomparisonWith2D}. The agreement is uniformly very good between 2D and 3D in panel~(a). 

Panel (b) shows the same comparison for the 2D fixed--free and 3D FRRR cases. The modes are somewhat underresolved for $T_0 < 10^{-1}$ and
the growth rates in 2D and 3D do not agree well, so we omit the membrane snapshots in this region of the parameter space. Even with a much finer mesh, the fastest growing 2D modes are difficult to resolve in this region using an eigenvalue solver \cite{mavroyiakoumou2021eigenmode}.
This underresolution may affect the apparent lack of convergence of the
3D stability boundaries with increasing aspect ratio to the 2D boundaries at small $R_1$. At moderate $R_1$ ($\in [10^{-0.5},10^{0.5}]$), interestingly, the stability boundaries approximately coincide at all aspect ratios, while at larger $R_1$ the stability boundaries shift upward with increasing aspect ratio similarly to panel~(a). Unlike in panel (a), here the snapshots oscillate and change shape in the small-amplitude growth regime. In panel (d) we show sets of six snapshots for the 2D (green) and 3D (black) cases during time intervals between two consecutive peaks or troughs in the graph of $z_{\mathrm{center}}(t)$. The snapshots are normalized to have uniform maximum deflections. The agreement between 2D and 3D is very good in all cases. The third snapshots in each sequence are overlaid in the instability region in panel (b). In (b) and (d) the values of $\log_{10} R_1$ and $\log_{10} T_0$  for each set of snapshots are the multiples of 0.5 and 0.25 respectively that intercept each set.

Panels (c) and (e) show the same comparison for the 2D free--free and 3D RRRR cases. The stability boundaries and resolution issues are similar to those in panel (b), but here adequate resolution is obtained down to smaller $T_0$. At small $R_1$ values, the 3D stability boundaries actually move away from the 2D boundaries as the aspect ratio increases. This is due to special cases with very small but positive growth rates over a large range of $T_0$ at small $R_1$. Cases with aspect ratio 8 and growth rates $<$ 0.04 are marked in panel (c) by gray boxes.
Such cases only occur close to the stability boundary at large $R_1$, as expected, but occur over a large range of $T_0$ from the dark blue line to the orange (2D) line at small $R_1$. Below the orange line, the 3D growth rates abruptly become much larger, so setting aside the cases with very slow growth, we have better agreement in the stability boundary for 2D and 3D with aspect ratio 8. There is nothing obviously wrong with these slow-growth cases (such as very jagged shapes or nonphysical motions); unlike most of the results in this section, they may be a case where the 3D large-span dynamics are qualitatively different from the 2D dynamics.

The agreement between the snapshots in 2D (green) and 3D (black) in panels (c) and (e) is generally very good, though not quite as good as in (b) and (d), particularly in some cases at $T_0 = 10^{-1}$ and $10^{-1.25}$ that are clearly underresolved. In (c) and (e) higher resolution is used: $M = 80$ for RRRR and 81 Chebyshev-Lobatto points are used in 2D, versus 40 and 41 respectively in (a), (b), and (d). In four cases at the upper right of (c) and (e) $|z_{\mathrm{center}}|$ grows without oscillating, similarly to a divergence instability, so the six snapshots are equally spaced over a large portion of the small-amplitude regime.

We have presented a broad range of evidence of good qualitative agreement between 2D and 3D dynamics, even with fairly modest 3D mesh sizes ($M = 40$ and $N = 10$ in most cases) and other significant computational differences. In 2D we used a Chebyshev-Lobatto mesh, computed wake roll-up, and used a Galerkin method for the kinematic equation and Kutta condition, with a non-flat body and wake in the kinematic condition. In 3D we used a uniform mesh, flat approximate vortex sheets on the body and wake, and a collocation method for the kinematic condition. We now proceed to study and classify the dynamics of 3D membranes with square aspect ratio at all 12 of the boundary conditions in figure \ref{fig:schemBC}.

%%%%%%%%%%%%%%%%%
\section{Membrane dynamics across parameter space for 12 boundary conditions}\label{sec:results}

We now broadly classify membrane dynamics with respect to $R_1$, $T_0$, and $R_3$ at all 12 combinations of fixed and free boundary conditions at the edges. In a few cases the large-amplitude dynamics do not seem to depend very strongly on aspect ratio, so we set it to 1 here, but it would be good to explore this parameter in future work.

To find where the membranes are unstable, we apply the perturbation~\eqref{eq:perturbationFreeFree} with $\sigma=10^{-5}$ for all 12 boundary conditions at a large number of $R_1$ and $T_0$ values, and as before we find the critical $T_0$ at each $R_1$ by interpolation of growth rates.

\begin{figure}[H]
    \centering    \includegraphics[width=.685\textwidth]{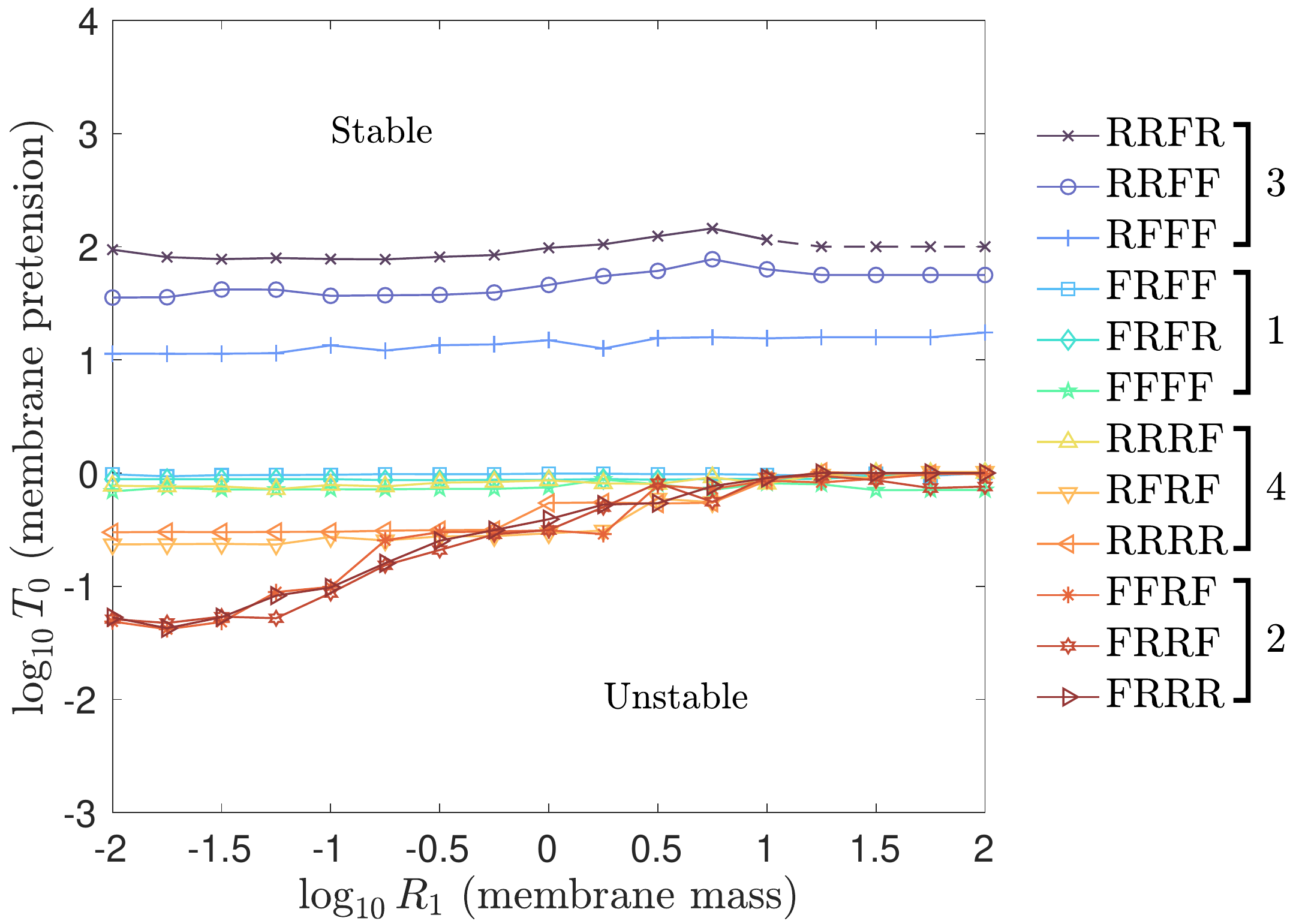}
    \caption{Stability boundaries for the 12 boundary conditions, listed at right, and placed in four groups that depend on only the leading and trailing edge conditions, as in figure \ref{fig:schemBC}.}
    \label{fig:stabilityBoundariesAll}
\end{figure}

Figure \ref{fig:stabilityBoundariesAll} shows the stability boundaries for the 12 boundary conditions, listed at right. Based on the locations of the boundaries and the qualitative features of membranes' large-amplitude dynamics (presented later), the 12 boundary conditions naturally fall into four groups, listed at right in figure \ref{fig:stabilityBoundariesAll} and shown in figure \ref{fig:schemBC}, and determined by only the leading and trailing edge conditions. The conditions at these edges seem to determine the qualitative behaviors of the membranes' dynamics more strongly than the side-edge conditions. However, the side edges can still have important effects, particularly for the large-amplitude results we will present.

In the 2D versus 3D comparisons, we have already shown one stability boundary from three of the four groups: FRFR from group 1, FRRR from group 2, and RRRR from group 4. The other stability boundaries within these groups are similar---nearly flat in group 1, upward sloping in group 2, and either flat or upward sloping in group 4, though the flatness at small $R_1$ for RRRR is affected by the slow-growth cases in figure \ref{fig:effectOfSpan}(c). Perhaps due to these cases, there is more variability in the boundaries of group 4 at small $R_1$, while the boundaries are more uniform across groups 1 and 2. The data for group 3, with the leading edge free and trailing edge fixed, are completely new, and stand out from the other groups in two respects. The group 3 stability boundaries are orders of magnitude higher in $T_0$ than the other groups'.
There is also about an order-of-magnitude variation of the critical $T_0$ within group 3, higher (less stable) with more side edges free. At large $R_1$ it is difficult to distinguish slow growth and slow decay, particularly for RRFR, so the stability boundary is dashed to indicate this lack of certainty.

%%%%%%%%%%%%%%%%%%%%%%%%%
\subsection{Large-amplitude scaling laws}

We now study how key quantities---the typical magnitudes of deflections, and for unsteady cases, frequencies of oscillations---depend on $R_1$, $T_0$, and $R_3$. In our 2D work \cite{mavroyiakoumou2020large,mavroyiakoumou2021eigenmode,mavroyiakoumou2021dynamics} we also studied the typical spatial wave numbers of membrane deformations, but these are somewhat difficult to quantify precisely without higher spatial resolution, which is more feasible computationally in 2D. However, there is sometimes a correlation between wave numbers and frequencies, which are computed here.

\begin{figure}[b]
    \centering
    \includegraphics[width=\textwidth]{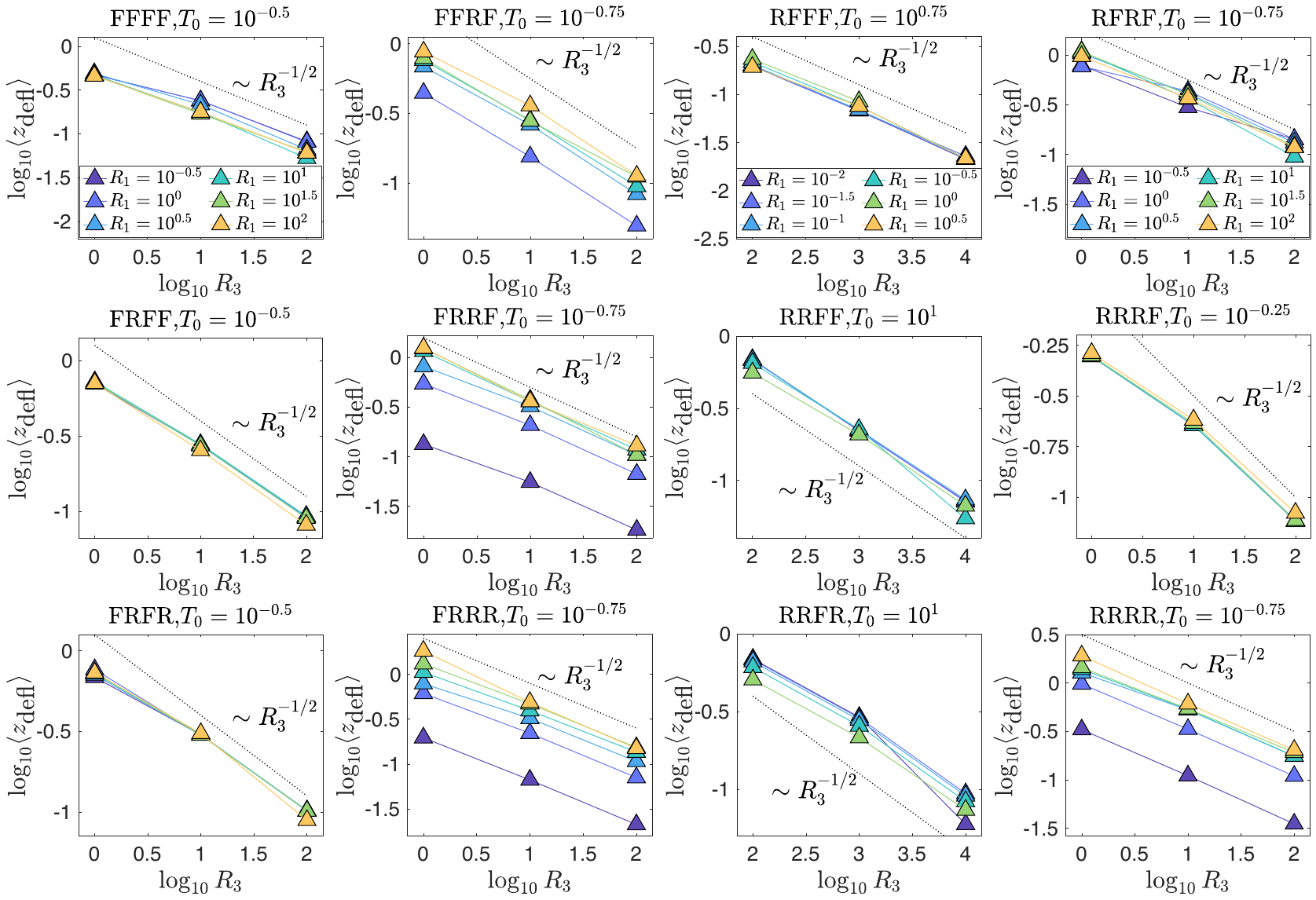}
    \caption{Time-averaged deflections of the membranes (defined by~\eqref{eq:timeAvgDefl}) versus $R_3$ for various $R_1$ (listed in the top panel of each column except column 2 which uses the same values as column 1) and fixed $T_0$ for all 12 boundary conditions. Recall that $R_1$ is the dimensionless membrane mass, $T_0$ is the dimensionless pretension, and $R_3$ is the dimensionless stretching rigidity. Columns 1--4 correspond to groups 1--4 in figures \ref{fig:schemBC} and \ref{fig:stabilityBoundariesAll}. The dotted black line in each subpanel indicates the scaling~$R_3^{-1/2}$.}
    \label{fig:lineDeflR3}
\end{figure}

In figure~\ref{fig:lineDeflR3} we show how the time-averaged deflection depends on $R_3$ at several fixed values of $R_1$ (listed in the legends of the top panels) and $T_0$ for all 12 boundary conditions.
We define the time-averaged deflection of the membrane as the maximum membrane deflection minus the minimum deflection, averaged over time:
\begin{equation}\label{eq:timeAvgDefl}
    \langle z_{\mathrm{defl}}\rangle\equiv \frac{1}{t_2}\int\limits_{t_1}^{t_1+t_2}\left(\max\limits_{-1<\alpha_1,\alpha_2<1}z(\alpha_1,\alpha_2,t)-\min\limits_{-1<\alpha_1,\alpha_2<1}z(\alpha_1,\alpha_2,t)\right)\d t
\end{equation}
where $t_1$ and $t_2$ are sufficiently large that $\langle z_{\mathrm{defl}}\rangle$ changes by less than $1\%$ with further increases in these values. As in the 2D studies \cite{mavroyiakoumou2020large,mavroyiakoumou2021dynamics}, deflection scales as $1/\sqrt{R_3}$ at moderate and large $R_3$ in all 3D cases. The scaling begins at larger $R_3$ for the third column, so the $R_3$ range there is a factor of $10^2$ higher than in the other columns. At $R_3 = 10^2$, the deflection magnitudes are generally much higher in column 3 than in the other columns. Selected $T_0$ values are shown in figure~\ref{fig:lineDeflR3} but the same scaling holds at other $T_0$ values, two of which are shown in appendix~\ref{app:otherT0lineDefl}. We briefly explain the $1/\sqrt{R_3}$ scaling similarly to \cite{mavroyiakoumou2020large}. We expand terms in
the $z$ component of the membrane equation~\eqref{eq:nonlinearMembrane} in the limit of small deflections, by inserting Taylor series for $z$ and its derivatives. $T_0$ and $R_3$ enter through the $\overline{e}$ and $K_s$ terms respectively. $T_0$ multiplies $(\partial^2_{\alpha_1} + \partial^2_{\alpha_2})z$ and $R_3$ multiplies a term that is cubic in derivatives of $z$. Using the kinematic and pressure jump equations, (\ref{eq:kinematic}) and~(\ref{eq:pressureJumpAlpha1}), we find that
$\gamma_2$ is linear in $z$ and therefore so is $[p]$ and the $[p]$ term
in (\ref{eq:nonlinearMembrane}). At steady state, the amplitude is set by a balance of the destabilizing $[p]$ term and the stabilizing $R_3$ term; the $T_0$ term is similar in magnitude to these terms near the stability boundary, and becomes insignificant at smaller $T_0$. Balancing
$R_3 O(|z|^3)$ and the $O(|z|)$ pressure term, we have $|z| \sim R_3^{-1/2}$.

In the third column, smaller $R_1$ values (listed in the top panel) are used compared to the other columns, because the third-column cases did not reach large amplitude within 350 time units at the larger $R_1$. In a few cases data are omitted because membranes were stable (FFRF with $R_1=10^{-0.5}$) or did not attain steady-state motions by $t$ = 350 (RRFF and RRFR with $R_1=10^{0.5}$).

\begin{figure}[b]
    \centering
    \includegraphics[width=\textwidth]{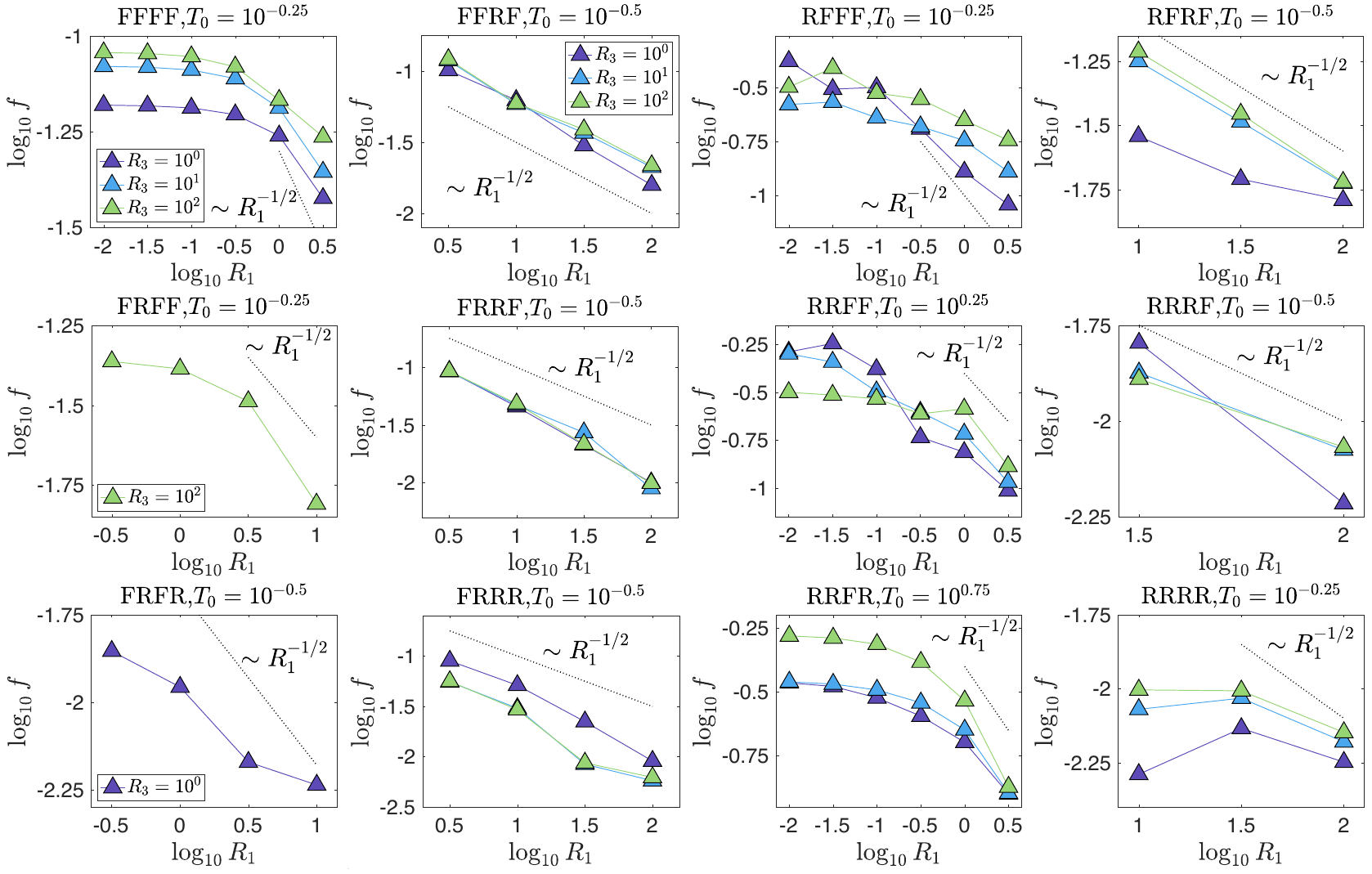}
    \caption{Plots of mean frequency $\log_{10}f$ versus mass density $\log_{10}R_1$ with various $R_3$ and fixed $T_0$ values for all 12 boundary conditions. 
    Columns 1--4 correspond to groups 1--4 in figures \ref{fig:schemBC} and \ref{fig:stabilityBoundariesAll}. The dotted black line in each subpanel indicates the scaling $R_1^{-1/2}$.}
    \label{fig:lineFreqR1}
\end{figure}

The other main quantity we measure is the membranes' frequency of oscillation. This is calculated as the reciprocal of the period of oscillation, the time between successive peaks in $z_{\mathrm{center}}(t)$ near the end of the computation (typically $t = 350$). We plot the frequency data in figure~\ref{fig:lineFreqR1}, similarly to the amplitude data in the previous figure, but now versus $R_1$, the most important parameter for frequency, at various fixed values of $T_0$ and $R_3$ (listed in the titles and in the legends in the first two columns, respectively). Unlike the amplitude, the membranes' shapes and frequencies do not vary much with $R_3$. In most cases with large $R_1$ ($\in [10^1, 10^2]$) and some cases with moderate $R_1$ ($\in [10^0, 10^1]$), the frequency approximately scales as $1/\sqrt{R_1}$, as in the large-amplitude 2D studies \cite{mavroyiakoumou2020large,mavroyiakoumou2021dynamics}. The scaling can be explained in 2D or 3D by balancing the $R_1\partial_{tt}z$ term in the membrane equation with the other terms, which are independent of $R_1$. Assuming a period of oscillation $T$, we have $R_1\partial_{tt}z \sim R_1 z/T^2 \sim O(z)$ for the remaining terms (i.e. the $[p]$ term), so at large $R_1$ the frequency $= 1/T \sim 1/\sqrt{R_1}$.

There is more variability in the frequency dependence on $R_1$ than in the amplitude dependence on $R_3$, for a few main reasons. At large $R_1$, the oscillation period is longer, so longer times are needed to reach a steady state, and the estimates of period and frequency are based on smaller and perhaps noisier data sets. The oscillations are periodic in some cases but chaotic in others, and in the latter cases the frequency is only a rough estimate of the long-time average. Finally, the oscillatory motion may change qualitatively (e.g. the membrane shapes may change) as $R_1$ increases in the ranges considered here, which may cause deviations from the simple $1/\sqrt{R_1}$ scaling law. 

At small $R_1$ the frequency typically plateaus here (in the first and third columns). A different behavior was seen in 2D simulations of tethered membranes in \cite{mavroyiakoumou2021dynamics} which were similar to the 2D free--free case. There the frequencies scaled as $R_1^{-5/6}$ at small $R_1$ and the shapes had very sharp spatial features that required hundreds of points in the streamwise direction to resolve, much higher resolution than we use here in 3D. In the corresponding case here, the fourth column of figure~\ref{fig:lineFreqR1}, we have a limited set of data at small $R_1$ (at different $T_0$ than in the fourth column) that show increasing frequencies as $R_1$ decreases, but it is difficult to characterize the scaling behavior. At small $R_1$ one might expect a plateau if the limit $R_1 \to 0$ is a regular perturbation problem, e.g. if the membrane shape does not change much in the limit $R_1 \to 0$, as seems to occur in the first and third columns. However, if the membrane becomes ever wavier as $R_1$ decreases, the frequency may also diverge, consistent with the 2D tethered membrane data in \cite{mavroyiakoumou2021dynamics} and the closely related problem of a fixed--free flag with bending rigidity \cite{alben2022dynamics}. The fourth and second columns respectively are similar to these cases.

The highest frequencies occur in the third column, and these are much higher than in the first column at the same $R_1$. The third-column frequencies are only moderately higher than those in the second column at the $R_1$ where both are given, $10^{0.5}$.

One interesting difference with 2D that we will discuss further in the next section is that in 3D we find many cases of oscillatory membranes with leading and trailing edges fixed, corresponding to the frequencies in the first column. In 2D, all physically-reasonable fixed--fixed cases assumed steady single-hump shapes. In the first column data are only presented at $R_3 = 10^2$ in the lower two panels, because the membranes did not oscillate at $R_3 = 10^0$ and $10^1$ across the full range of $R_1$ shown at these $T_0$.

%%%%%%%%%%%%%%
\subsection{Large-amplitude dynamics}\label{sec:largeResultsMotions}

We have presented basic scaling relationships for the amplitudes and frequencies of steady-state large-amplitude membrane motions. In this section we characterize the types of motions---e.g. a shape that is steady, or with periodic or chaotic oscillations---across all 12 boundary conditions.
We consider in turn each of the four groups in figure~\ref{fig:schemBC}, determined by the conditions at the leading and trailing edges. We use three values of the stretching rigidity parameter $R_3$: $10^0$, $10^1$, and $10^2$. Somewhat below $10^0$, deflections become unrealistically large in some cases. At $R_3 = 10^2$, the motions generally approximate the large-$R_3$ asymptotic regime. 

We consider values of $T_0$ that are relatively close to the stability boundary, within 1--1.5 orders of magnitude of it. At much smaller $T_0$, the iterative method in our computational method tends to stagnate without converging at an early time step. In some cases, the stagnation occurs after a sharp angular feature appears in the membrane shape. Presumably larger $T_0$ inhibits the formation of such features. By contrast, our 2D algorithm from \cite{mavroyiakoumou2020large,mavroyiakoumou2021dynamics} was able to compute at arbitrarily small $T_0$, where the $T_0$ term became insignificant compared to the $R_3$ term in the total stretching force. There we found only modest changes in the dynamics as $T_0$ increased from zero to the stability boundary, so possibly the 3D results here indicate what would happen at very small $T_0$, but we defer this question to future work.

In each case we consider a wide range of membrane mass $R_1$, from small to large.

%%%%%%%%%%%%%%
\subsubsection{Leading and trailing edges fixed (FFFF, FRFF, FRFR)}\label{sec:FF}

\begin{figure}
    \centering
    \includegraphics[width=\textwidth]{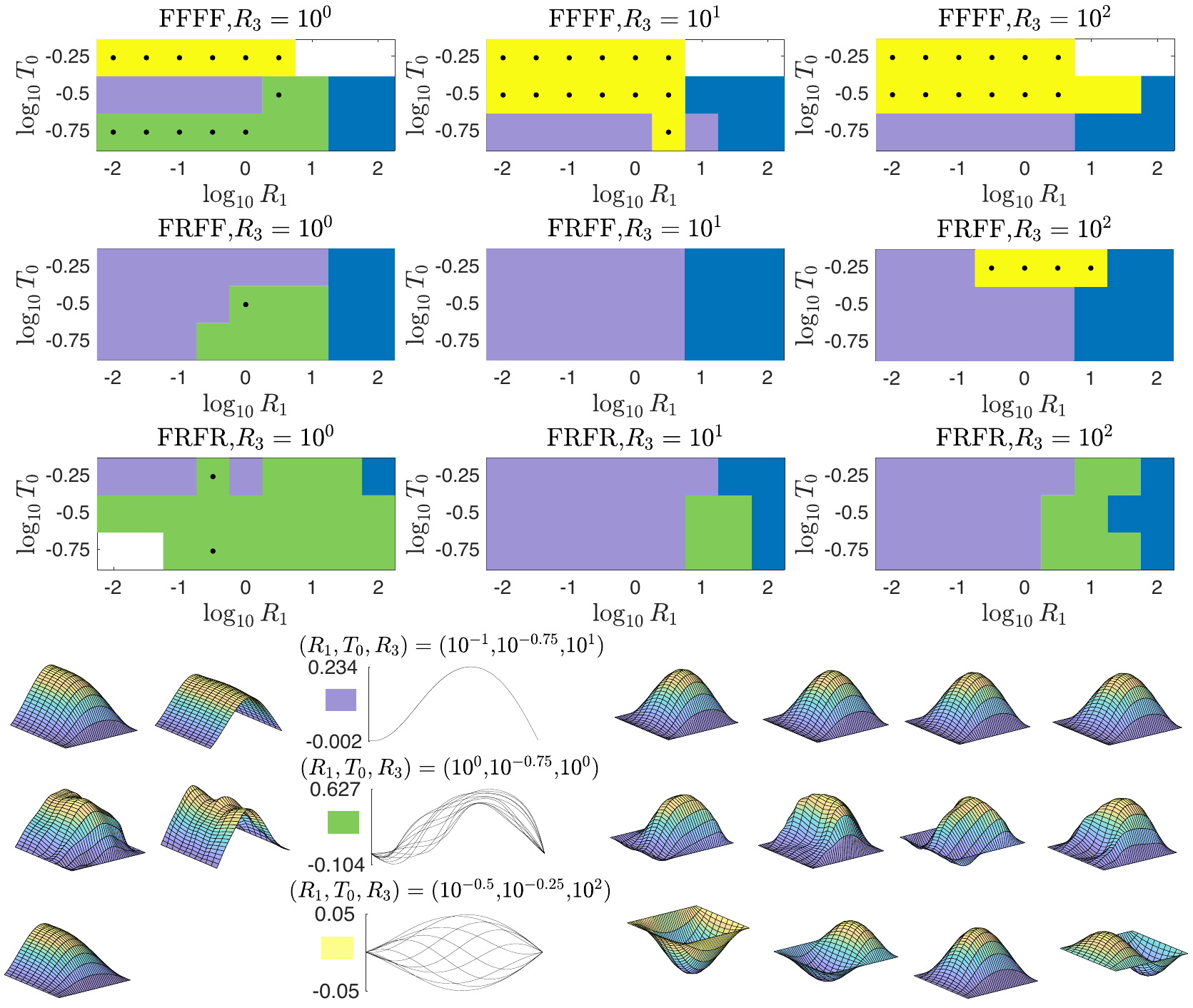}
    %%%%%
    \caption{Typical membrane dynamics with fixed leading and trailing edges. The color plots classify the different membrane dynamics across a grid of $(R_1,T_0)$ values for three values of~$R_3$ ($10^0,10^1,10^2$) with black dots indicating periodic motions. Recall that $R_1$ is the dimensionless membrane mass, $T_0$ is the dimensionless pretension, and $R_3$ is the dimensionless stretching rigidity. Representative examples of motions in the purple, green, and yellow regions are shown in the bottom three rows, identified by small colored rectangles. To the right of the rectangles are overlaid sequences of 11 midspan snapshots in black, for FFFF boundary conditions. To the right of the midspan snapshots are a smaller representative sequence of four 3D snapshots, from left to right with increasing time. To the left of the midspan snapshots are single 3D snapshots representing similar cases with other side-edge conditions (FRFF and FRFR).}
    \label{fig:FFzeroCross}
\end{figure}

We begin with membranes whose leading and trailing edges are fixed at zero deflection (group~1 in figures~\ref{fig:schemBC} and~\ref{fig:stabilityBoundariesAll}). Figure~\ref{fig:FFzeroCross} shows typical motions and where they occur in parameter space. Each row of the top 3-by-3 array of panels corresponds to different side-edge
conditions (FFFF, FRFF, or FRFR), and each column to a different $R_3$ value ($10^0$, $10^1$, or $10^2$). The different colors in each panel indicate where different types of motions occur in $R_1$--$T_0$ space, with representative examples below. Purple is used for steady single-hump shapes, green for small oscillations (periodic or not) about single-hump shapes, and yellow for up-down symmetric oscillations, usually periodic. Small black dots indicate periodic cases. Blue is used for cases that had slowly decaying oscillations at $t = 350$, and could approach the green or purple states at later times. Such cases occur at large $R_1$ and the oscillations tend to be less periodic than at smaller $R_1$. A few white regions in the upper corners of the FFFF panels indicate motions that were still in the exponential growth regime at $t = 350$, while cases in the white region in the FRFR panel at $R_3 = 10^0$ failed to converge at an early time.
The three rows at the bottom of figure~\ref{fig:FFzeroCross} show representative examples of
the membrane motions in the purple, green, and yellow regions.
Just to the right of small rectangles with these colors, the motions are shown by sequences of 11 midspan snapshots (overlaid), and then further to the right, smaller sequences of four 3D snapshots, arrayed horizontally. All these sequences are for FFFF cases, but single 3D snapshots from the same type of motion with other side-edge conditions are shown at the far left in each row.

As noted previously, only steady single-hump shapes (purple) occur in 2D with fixed leading and trailing edges \cite{mavroyiakoumou2020large}, whereas steady-state oscillations are surprisingly common in 3D, particularly with all four edges fixed. Here up-down symmetric periodic motions (yellow) are much more common than with the other side-edge conditions, which typically have moderately small or no oscillations about a single-hump shape. The yellow motion includes a traveling wave of deflection that moves {\it upstream}, unlike in the 2D fixed--free and free--free cases \cite{mavroyiakoumou2020large}. Compared to the steady purple shape, the green motion has more variability not just in time but also in the spanwise direction, exemplified by the snapshot to the left of the small green rectangle in the second row from the bottom. In experiments, oscillatory cases with fixed leading and trailing edges have been reported mainly at nonzero angle of attack, where leading edge vortex shedding seems particularly significant~\cite[figure~4]{rojratsirikul2010unsteady}.

%The FRFR membrane has a single hump that is fore-aft symmetric, similar to the 2D case in~\cite{mavroyiakoumou2020large}. 

%The snapshot of the FRFR  surface (in row~2 at the lower portion of figure~\ref{fig:FFzeroCross}) reveals that the membrane deflection, in this family of motions, is less uniform along the span than it is in the purple region.  

%%%%%%%%%%%%%%
\subsubsection{Leading edge fixed and trailing edge free (FFRF, FRRF, FRRR)}\label{sec:FR}

\begin{figure}
    \centering
    \includegraphics[width=\textwidth]{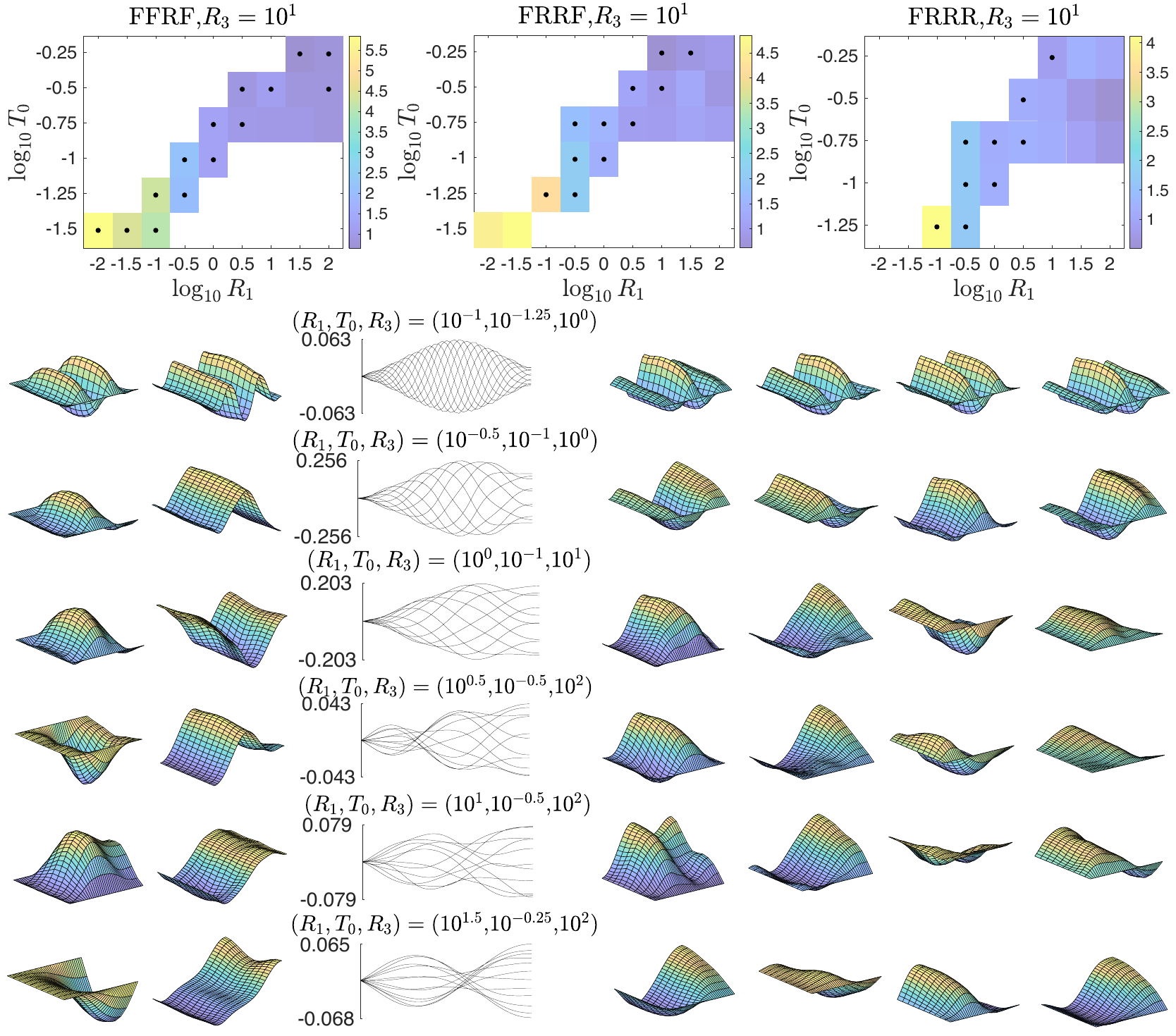}
    \caption{Typical membrane dynamics with fixed leading edges and free trailing edges. The color plots in the top row show the time-averaged number of zero-crossings of the membranes' midspan profiles at various ($R_1$,$T_0$) values for $R_3=10^1$. The black dots indicate periodic motions. 
    Representative examples of motions at a sequence of $R_1$ values from $10^{-1}$ to $10^{1.5}$ are shown in the bottom six rows. Overlaid sequences of 11 midspan snapshots are shown in black, for FRRF boundary conditions. To the right of the midspan snapshots are a smaller representative sequence of four 3D snapshots, from left to right with increasing time. To the left of the midspan snapshots are single 3D snapshots representing similar cases with other side-edge conditions (FFRF and FRRR).}
    \label{fig:FRzeroCross}
\end{figure}

The typical motions are qualitatively different in the second group (group 2 in figures~\ref{fig:schemBC} and~\ref{fig:stabilityBoundariesAll}), with the leading edge fixed and the trailing edge free. Hence we classify the motions somewhat differently in figure~\ref{fig:FRzeroCross},
with colors denoting the number of times the midspan profile crosses $z=0$ (not including the leading edge), time-averaged. This classifies the spatial waviness of the profiles. Here $R_3$ is fixed at $10^1$; at other $R_3$ the deflection amplitude varies but the numbers of zero crossings are almost unchanged. The numbers of zero crossings decrease almost monotonically as $R_1$ increases, and chaotic states (without black dots) with larger amplitudes and more up-down asymmetry become more common, similarly to the 2D results in \cite{mavroyiakoumou2020large}. In the white regions at the upper left of each panel, membranes are stable. In the white regions at the lower right, the computations did not converge when the membranes reached large amplitude but before they attained a steady-state motion, so results are not reported.

Typical membrane motions are shown in the rows at the bottom, one for each $R_1$ in a sequence from $10^{-1}$ to $10^{1.5}$. Here the motions resemble different modes of a flapping flag \cite{connell2007flapping,alben2008ffi,alben2022dynamics}, with traveling waves that pass downstream, unlike for group 1. Another difference here is that the side-edge conditions do not change the type of motion. As in \cite{mavroyiakoumou2020large}, at small $R_1$ the deflection envelope has small amplitude near the trailing edge. At large $R_1$ (e.g. $10^{1.5}$) the traveling waves become more like standing waves, with fixed nodes and antinodes.

%%%%%%%%%%%%%%%%%%%%%%%%
\subsubsection{Leading edge free and trailing edge fixed (RFFF, RRFF, RRFR)}\label{sec:RF}

Next, we consider membranes whose leading edge is free and trailing edge is fixed (group~3 in figures~\ref{fig:schemBC} and~\ref{fig:stabilityBoundariesAll}). We showed in figure~\ref{fig:stabilityBoundariesAll} that here the critical pretensions for instability are much larger than in the other cases. The space of motions is qualitatively different than for groups 1 and 2. We identify just three types of motions, with approximately periodic, up-down symmetric oscillations in most cases, characterized by three typical midspan profiles shown in the bottom three rows: a low-mode shape with at most a single interior $z$ extremum (light purple), a higher mode shape with one or two interior $z$ extrema (yellow), and a low-mode shape with a multi-valued $z$ profile near the trailing edge at certain times (dark purple). This last motion is more common at smaller $R_1$ and does not appear with both side edges fixed or at the largest $R_3$ (right column), where the first motion dominates. White regions are again cases in which the computations did not converge when the membranes reached large amplitude but before they attained a steady-state motion. Another special feature of group 3 is that the small-amplitude growth rates are generally much less than in the other groups, so data are omitted for $R_1 > 10^0$ because the amplitudes were growing and had not yet reached a steady state at $t = 350$. In some cases at $R_1 = 10^{1.5}$ and $10^2$ (not shown) the amplitude neither grows nor decays noticeably after the initial perturbations, as also occurred in some 2D cases at large $R_1$ \cite{mavroyiakoumou2020large}.

\begin{figure}[h]
    \centering
    \includegraphics[width=\textwidth]{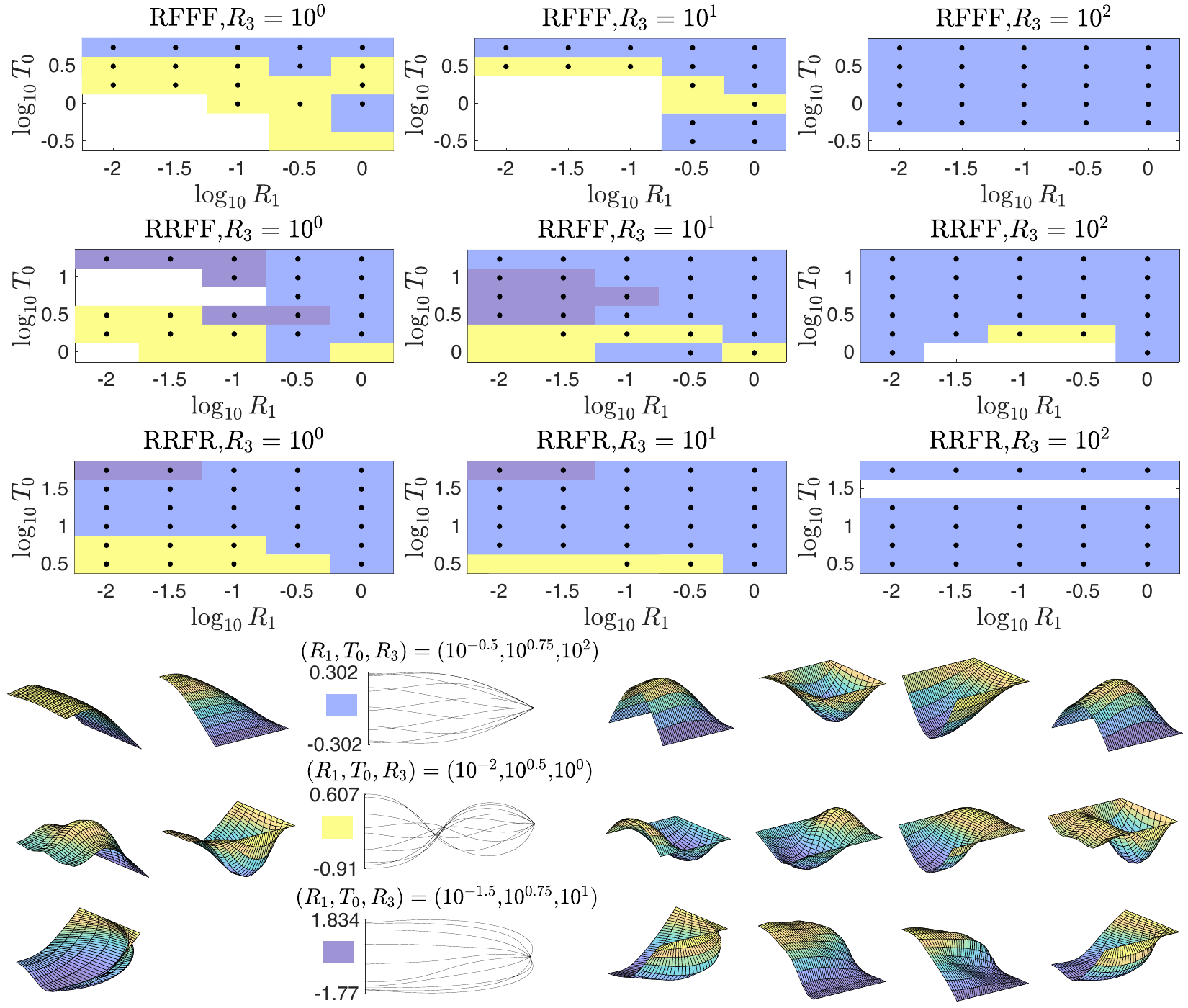}
    \caption{Typical membrane dynamics with free leading edges and fixed trailing edges. The color plots classify the different membrane dynamics across a grid of $(R_1,T_0)$ values for three values of~$R_3$ ($10^0,10^1,10^2$) with black dots indicating periodic motions. Representative examples of motions in the light purple, yellow, and dark purple regions are shown in the bottom three rows, identified by small colored rectangles. To the right of the rectangles are overlaid sequences of 11 midspan snapshots in black, for RFFF boundary conditions (top two rows) and
    RRFF boundary conditions (bottom row). To the right of the midspan snapshots are a smaller representative sequence of four 3D snapshots, from left to right with increasing time. To the left of the midspan snapshots are single 3D snapshots representing similar cases with other side-edge conditions (RRFR and RRFF).}
    \label{fig:RFzeroCross}
\end{figure}

The motions of group 3 seem more likely to violate the physical assumptions of the model than the other groups. The oscillation amplitude is highest at the leading edge, so leading-edge vortex shedding is probably important unless the oscillation amplitudes are very small (i.e. at $R_3  = 10^3$ and $10^4$, where the motions are similar to those in the $R_3 = 10^2$ column, light purple generally). Vortex shedding would probably occur in multiple locations upstream of the trailing edge for the dark purple motions, due to large deflection amplitudes and slopes. 

%start here
%%%%%%%%%%%%%%
\subsubsection{Leading and trailing edges free (RFRF, RRRF, RRRR)}\label{sec:RR}

\begin{figure}
    \centering
  \includegraphics[width=.875\textwidth]{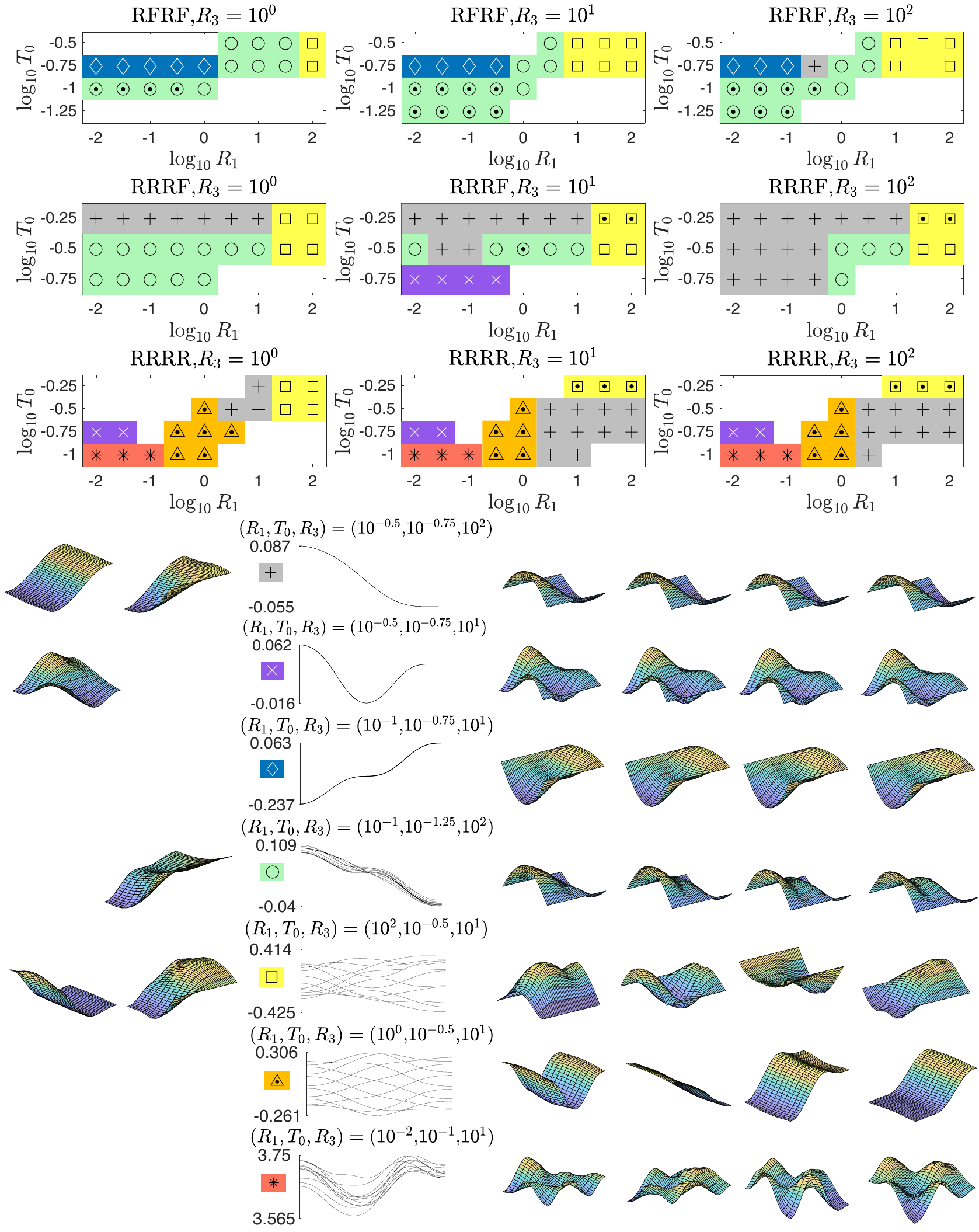}
    \caption{Typical membrane dynamics with free leading and trailing edges. The color plots classify the different membrane dynamics across a grid of $(R_1,T_0)$ values for three values of~$R_3$ ($10^0,10^1,10^2$) with black dots indicating periodic motions. Representative examples of motions in all the colored regions are shown in the bottom seven rows, identified by small colored rectangles with symbols. To the right of the rectangles are overlaid sequences of 11 midspan snapshots in black, for RFRF (rows 1, 3, 4, and 5), RRRF (row 2), and RRRR boundary conditions (rows 6 and 7). To the right of the midspan snapshots are a smaller representative sequence of four 3D snapshots, from left to right with increasing time. To the left of the midspan snapshots are single 3D snapshots representing similar cases with other side-edge conditions (RRRF and RRRR).}
    \label{fig:RRzeroCross}
\end{figure}

Group 4 consists of membranes with the leading and trailing edges free (figures~\ref{fig:schemBC} and~\ref{fig:stabilityBoundariesAll}). With fewer of the edges fixed, we find a wider variety of dynamics. 
In figure~\ref{fig:RRzeroCross} we use seven categories to classify the motions, with representative examples in the seven rows at the bottom, each labeled using a small rectangle with a certain color and symbol that is repeated at data points in this category in the 3-by-3 array of color plots above. The first three rows represent categories of steady shapes with one, two, or three internal inflection points, respectively. The fourth row represents a category of motions with small oscillations about one of the steady shapes. To identify these cases, we decompose the midspan deflection $z(\alpha_1,0,t)$ as the sum of its time-average and the remainder, the unsteady part. Motions are in the fourth category if the maximum minus the minimum of the unsteady deflection is nonzero but much less than the maximum minus the minimum of the time-averaged deflection. The fifth row represents unsteady motions whose unsteady part is comparable to or larger than the steady part (in the same sense as for the fourth category). The sixth row is for a special type of unsteady motion with all edges free (RRRR) that is time-periodic. These motions have translational and deformational dynamics with $O(1)$ frequencies, and their translational motions are more precisely time-periodic than those at large $R_1$. The seventh row is for another special type of unsteady RRRR motion that has both large spanwise and large chordwise curvatures.

The sixth and seventh categories occur only for the RRRR case, which seems to have a wider variety of motions because it is free to translate in $z$. As in the 2D case \cite{mavroyiakoumou2020large}, the translational motion is distinct from but coupled to the shape deformation dynamics, and the combination of the two results in more types of dynamics. Other categories of motions that occur only with certain edge conditions are the second and third, the steady states with two or three interior inflection points respectively. The second occurs for RRRF and RRRR and the third for RFRF, both mainly at small $R_1$. The fourth category, small oscillations about a steady shape, occurs only with one or both side edges fixed (RFRF and RRRF). The first category, a steady shape with a single inflection point, occurs with all side edge conditions, but in the RRRR case it translates either up or down in $z$ at almost constant speed, with occasional reversals in direction (as in~\cite[figure 20]{mavroyiakoumou2020large}).
The fifth category, of general unsteady motions, occurs with all side edge conditions at large $R_1$. 
Here motions tend to be more asymmetrical and irregular and have larger oscillation amplitudes. The large body mass may allow the body to maintain its momentum against resisting fluid forces for longer times.

%%%%%%%%%%%%%%%%%%

\section{Conclusions} \label{sec:conclusions}

We have developed a model and numerical method to compute small- and large-amplitude dynamics of thin membranes in 3D inviscid flows. We provided numerical evidence that indicates convergence of our method with respect to spatial grid refinement. The 3D solutions agree remarkably well with those computed in 2D flows with different discretization techniques, particularly when the aspect ratio in 3D is large.  

With fixed or free boundary conditions at each of the four edges, we have 16 combinations of boundary conditions for rectangular membranes, reduced to 12 when symmetric cases are accounted for. We computed the dynamics in all 12 cases with various values of the three key physical parameters: membrane mass ($R_1$), pretension ($T_0$), and stretching rigidity ($R_3$). The 12 cases fit naturally into four groups that are determined by the leading- and trailing-edge boundary conditions. Within each group there are strong similarities in the shapes and locations of the stability boundaries, the typical magnitudes of membranes' deflections and their oscillation frequencies in unsteady cases, and the typical shapes of the membranes at the midspan location. The side-edge boundary conditions can cause small or large qualitative differences in the dynamics within each group. The differences are relatively small for the fixed-free case (group 2); the midspan shape is almost unchanged at a free side edge or gradually scaled to flat at a fixed side edge. The differences are larger for the fixed-fixed and free-fixed cases (groups 1 and 3); changes in side-edge conditions lead to the appearance or disappearance of certain types of oscillation modes. In group 4 (free-free) the side edges have an even larger effect, for example by allowing or preventing translational motions that lead to a wider variety of dynamics. 

As in 2D, the deflection magnitudes scale as $R_3^{-1/2}$ and the oscillation frequencies scale as $R_1^{-1/2}$ at large values of $R_3$ and $R_1$ respectively. The stability boundaries and dynamics in groups 1, 2, and 4 have many resemblances to the 2D cases with same leading and trailing edge conditions studied in \cite{mavroyiakoumou2020large}. The fixed-free case in 3D is very similar to the 2D case in terms of increasing waviness with decreasing $R_1$, a more limited effect of $T_0$ on the shapes, and $R_3$ mainly setting the amplitudes of oscillation but not the shapes. The free-free case in 3D has some of the same types of states as in 2D: periodic oscillatory states and vertical translational states that may be nearly steady with a non-zero mean slope, or undergo periodic or chaotic oscillations.

However, there are important qualitative differences in 3D. Unlike the 2D fixed-fixed case where all membranes converge to steady single-hump shapes, the 3D version (group 1) has cases with small and large unsteady oscillations, particularly when both side edges are fixed. The 3D results contradicted our intuition from the 2D study that more fixed edges leads to fewer unsteady states.
The 3D free-free case (group 4) has complicated spanwise curvature distributions, particularly with one or both side edges free. There are more types of motions here than in the 2D free-free case. We did not study the 2D analog of group 3, free-fixed, but the variations with side-edge conditions here show that 3D effects can be significant. Taken together these results validate the use of 2D models for a good qualitative understanding of membrane motions and scaling with parameters, but also show that certain important phenomena can only be seen with the 3D model.

Future work could investigate several topics, for example: the effect of aspect ratio; a nonzero Poisson ratio; including non-flat vortex sheets in the kinematic condition and computing wake roll-up (e.g. using a treecode \cite{lindsay2001particle}) and consequences on membrane dynamics, especially at large mass density ratio and small stretching rigidity~\cite[figures~17 and 23]{mavroyiakoumou2020large}; the types of motions at smaller $T_0$, which would require altering the algorithm to be able to compute steady state motions in this regime; mixed fixed and free boundary conditions on a single edge; and different membrane shapes.

%For instance, one could replace the uniform grid with a Chebyshev grid (as in~\cite{mavroyiakoumou2020large}) or with a quarter-chord (as in~\cite{katz2001low}) and determine whether the choice of grid affects convergence.

%%%%%%%%%%%%%
\subsection*{Acknowledgments}
We acknowledge support from a Rackham Predoctoral Fellowship Award (University of Michigan) and  the NSF Mathematical Biology program under award number DMS-1811889.
%%%%%%%%%%%%%

%==========
\appendix
%==========

%%%%%%%%%%

\section{Thin-membrane elasticity}\label{app:moreOnModel}

We compute the stretching energy of the membrane using the position $\mathbf{r}(\alpha_1,\alpha_2,t)$. The membrane has thickness $h\ll L,W$, the lateral dimensions. We assume the stretching strain is constant through the thickness, accurate to leading order in $h$ \cite{efrati2009elastic}.

We denote the flat prestrained configuration of the membrane central surface by $\bm{\alpha} \equiv (\alpha_1,\alpha_2,0) = \mathbf{r}(\alpha_1,\alpha_2,0)$. A small line of material connecting two material points $\bm{\alpha}$ and $\bm{\tilde{\alpha}}$ in the flat prestrained configuration is $\d\bm{\alpha} = \bm{\alpha} - \bm{\tilde{\alpha}}$.
We assume that the prestrained state is obtained from the zero-energy state by applying a uniform prestrain $\overline{e}$.
Thus in the zero-energy state the small line of material is $\d\bm{\alpha}/(1+\overline{e})$, the prestrain having been removed by dividing by $(1+\overline{e})$.

One of the most common measures of deformation in nonlinear elasticity is the difference between the squared length of a material line in the deformed and zero-energy configurations \cite{landau:1986a,tadmor2012continuum}:
\begin{equation}\label{eq:2eps}
\|\d\mathbf{r}\|^2-\frac{1}{(1+\overline{e})^2}\|\d\bm{\alpha}\|^2=2\epsilon_{ij}\,\d \alpha_i\d \alpha_j\;,\quad \epsilon_{ij}=\frac{1}{2}\left(a_{ij}-\frac{1}{(1+\overline{e})^2}\delta_{ij}\right)
\end{equation}
where $\d\mathbf{r} = \mathbf{r}(\alpha_1,\alpha_2,0) - \mathbf{r}(\tilde{\alpha}_1,\tilde{\alpha}_2,0)$ and $\epsilon_{ij}$ is the (Green-Lagrange) strain tensor, written in terms of the metric tensor
\begin{equation}
a_{ij}=\partial_{\alpha_i}\mathbf{r}\cdot\partial_{\alpha_j}\mathbf{r}=\frac{\partial r_k}{\partial \alpha_i}\frac{\partial r_k}{\partial \alpha_j},\quad i,j=1,2.\label{eq:gtensor}
\end{equation}
and the identity tensor $\delta_{ij}$.

For an isotropic membrane with Young's modulus $E$, Poisson ratio $\nu$, and thickness $h$, the elastic energy per unit midsurface area~\cite{alben2019semi} is
\begin{equation}\label{eq:wAepsilon}
    w_s=\frac{h}{2}\bar{A}^{mnop}\epsilon_{mn}\epsilon_{op} \;, \quad \bar{A}^{mnop}=\frac{E}{1+\nu}\left(\frac{\nu}{1-\nu}\delta_{mn}\delta_{op}+\delta_{mo}\delta_{np}\right)
\end{equation}
where $\bar{A}^{mnop}$ is the elasticity tensor for an isotropic material \cite{landau:1986a}. For small-to-moderate prestrains, the focus of this work, the strain tensor in (\ref{eq:2eps}) is approximately
\begin{equation}
    \epsilon_{ij}(\alpha_1,\alpha_2,t)=\overline{e}\delta_{ij}+\frac{1}{2}\left(\partial_{\alpha_i}\mathbf{r}\cdot\partial_{\alpha_j}\mathbf{r}-\delta_{ij}\right), \quad i,j=1,2.
\end{equation}

%%%%%%%%%%%%%%%%%%%%
\section{Derivation of  pressure jump equation in 3D flow}\label{app:pressureJump}

In this appendix we derive an expression for the pressure jump $[p](\alpha_1,\alpha_2,t)$ across the membrane in terms of the membrane vortex sheet strength and other quantities, as a generalization of~\cite[appendix~A]{mavroyiakoumou2020large}. 

The Euler momentum equation given by 
\begin{equation}\label{euler}
\partial_t \boldsymbol{u}(\boldsymbol{x},t)+ \boldsymbol{u}(\boldsymbol{x},t)\cdot\nabla\boldsymbol{u}(\boldsymbol{x},t)=-\nabla p(\boldsymbol{x},t)
\end{equation}
couples the fluid velocity $\boldsymbol{u}(\boldsymbol{x},t)$ and the pressure $p(\boldsymbol{x}, t)$. 
We calculate the fluid pressure at a point that is adjacent to and follows a material point $\boldsymbol{r}(\alpha_1,\alpha_2,t)$ on the membrane. The rate of change of fluid velocity at such a point is 
\begin{equation}\label{material}
\frac{\d}{\d t}\boldsymbol{u}(\boldsymbol{r}(\alpha_1,\alpha_2,t),t)=\partial_t\boldsymbol{u}(\boldsymbol{x},t)|_{\boldsymbol{x}=\boldsymbol{r}(\alpha_1,\alpha_2,t)}+\left(\partial_t\boldsymbol{r}(\alpha_1,\alpha_2,t)\cdot\nabla\right)\boldsymbol{u}(\boldsymbol{x},t)|_{\boldsymbol{x}=\boldsymbol{r}(\alpha_1,\alpha_2,t)}.
\end{equation}
We use (\ref{material}) to replace the first term in (\ref{euler}) and write the pressure gradient at a point that moves with $\boldsymbol{r}(\alpha_1,\alpha_2,t)$.
To obtain the jump in fluid pressure at a material point on the membrane, we write (\ref{euler}) (modified by (\ref{material})) separately at points in the fluid that approach the membrane from either side:
\begin{equation}\label{bothSidespm}
\frac{\d}{\d t}\boldsymbol{u}(\boldsymbol{r}(\alpha_1,\alpha_2,t),t)^{\pm}+\left((\boldsymbol{u}(\boldsymbol{x},t)-\partial_t\boldsymbol{r})\cdot\nabla\boldsymbol{u}(\boldsymbol{x},t)\bigg|_{\boldsymbol{x}=\boldsymbol{r}(\alpha_1,\alpha_2,t)}\right)^{\pm}=-(\nabla p(\boldsymbol{x},t)|_{\boldsymbol{x}=\boldsymbol{r}(\alpha_1,\alpha_2,t)})^{\pm},
\end{equation}
using $+$ for the side toward which the membrane normal $\boldsymbol{\mathbf{\hat{n}}}$ is directed and $-$ for the other side.

Next, we decompose the fluid velocity into components tangential and normal to the membrane. The normal component matches that of the membrane, $\nu_v$.
The tangential component of the fluid velocity may be written in terms of its jump across the membrane, using the vortex sheet strength components $\gamma_1$, $\gamma_2$~\cite{saffman1992vortex},
% \begin{equation}\label{eq:gammancrossu}
% \left(\boldsymbol{u}^+-\boldsymbol{u}^-\right)\cdot \mathbf{\widehat{s}}_2=-\gamma_1;\quad
% \left(\boldsymbol{u}^+-\boldsymbol{u}^-\right)\cdot \mathbf{\widehat{s}}_1=\gamma_2,
% \end{equation}
and the average of the tangential components of the fluid velocity on the two sides of the membrane, denoted $\mu_1$ and $\mu_2$. The subscripts $\{1,2\}$ denote
the components in the $\mathbf{\widehat{s}}_1$ and $\mathbf{\widehat{s}}_2$ directions, respectively.
% To obtain Eqs.~\eqref{eq:gammancrossu} we use $\boldsymbol{\gamma}=\boldsymbol{\hat{n}}\times [\boldsymbol{u}]$.
The fluid velocity can be written
\begin{equation}\label{fluidvel}
\boldsymbol{u}^{\pm}=\left(\mu_1\pm\frac{\gamma_2}{2}\right)\boldsymbol{\mathbf{\widehat{s}}}_1+\left(\mu_2\mp\frac{\gamma_1}{2}\right)\boldsymbol{\mathbf{\widehat{s}}}_2+\nu_v\boldsymbol{\mathbf{\hat{n}}}.
\end{equation}

We take the difference of (\ref{bothSidespm}) on the $+$ and $-$ sides:
\begin{align}\label{diff}
\frac{\d}{\d t}&\left(\boldsymbol{u}^+(\boldsymbol{r}(\alpha_1,\alpha_2,t),t)-\boldsymbol{u}^-(\boldsymbol{r}(\alpha_1,\alpha_2,t),t)\right)
+\left((\boldsymbol{u}^+(\boldsymbol{x},t)-\partial_t\boldsymbol{r})\cdot\nabla\boldsymbol{u}^+(\boldsymbol{x},t)\bigg|_{\boldsymbol{x}=\boldsymbol{r}(\alpha_1,\alpha_2,t)}\right)\nonumber\\&-  
\left((\boldsymbol{u}^-(\boldsymbol{x},t)-\partial_t\boldsymbol{r})\cdot\nabla\boldsymbol{u}^-(\boldsymbol{x},t)\bigg|_{\boldsymbol{x}=\boldsymbol{r}(\alpha_1,\alpha_2,t)}\right)
=-(\nabla p(\boldsymbol{x},t)^+-\nabla p(\boldsymbol{x},t)^-) |_{\boldsymbol{x}=\boldsymbol{r}(\alpha_1,\alpha_2,t)}.
\end{align}
We then take the $\mathbf{\widehat{s}}_1$ components of (\ref{diff}),
term by term. The $\mathbf{\widehat{s}}_1$ component of the right hand side of (\ref{diff}) is $-\partial_{s_{1}}[p]_-^+$ which we will ultimately integrate to obtain $[p]_-^+$. $-\partial_{s_{1}}[p]_-^+$ is equal to the $\mathbf{\widehat{s}}_1$ components of the terms on the left hand side of (\ref{diff}), which we now compute.
Using (\ref{fluidvel}), the $\mathbf{\widehat{s}}_1$ component of the first term on the left-hand side of (\ref{diff}) is
\begin{align}
\mathbf{\widehat{s}}_{1}\cdot\frac{\d}{\d t}(\boldsymbol{u}^+(\boldsymbol{r}(\alpha_1,\alpha_2,t),t)-\boldsymbol{u}^-(\boldsymbol{r}(\alpha_1,\alpha_2,t),t)) &= \mathbf{\widehat{s}}_{1}\cdot \partial_t(-\gamma_{1} \mathbf{\widehat{s}}_{2}+\gamma_{2} \mathbf{\widehat{s}}_{1})\nonumber\\
&=\partial_t\gamma_2-\partial_t\gamma_1(\mathbf{\widehat{s}}_{1}\cdot\mathbf{\widehat{s}}_{2})-\gamma_1(\mathbf{\widehat{s}}_{1}\cdot\partial_t\mathbf{\widehat{s}}_{2}),\label{eq:dtdiffs1}
\end{align}
% and
% \begin{align}
% \mathbf{\widehat{s}}_{2}\cdot\frac{\d}{\d t}(\boldsymbol{u}^+(\boldsymbol{r}(\alpha_1,\alpha_2,t),t)-\boldsymbol{u}^-(\boldsymbol{r}(\alpha_1,\alpha_2,t),t)) &= \mathbf{\widehat{s}}_{2}\cdot \partial_t(\gamma_{1} \mathbf{\widehat{s}}_{2}-\gamma_{2} \mathbf{\widehat{s}}_{1})\nonumber\\
% &=\partial_t\gamma_1-\partial_t\gamma_2(\mathbf{\widehat{s}}_{2}\cdot \mathbf{\widehat{s}}_{1})-\gamma_2(\mathbf{\widehat{s}}_{2}\cdot\partial_t\mathbf{\widehat{s}}_{1}),\label{eq:dtdiffs2}
% \end{align}
where we use $\mathbf{\widehat{s}}_{i}\cdot \mathbf{\widehat{s}}_{i}=1$  and $\mathbf{\widehat{s}}_{i}\cdot \partial_t\mathbf{\widehat{s}}_{i}=0$ for $i=1,2$. 

For $\boldsymbol{\mathbf{\widehat{s}}}_i\cdot[(\boldsymbol{u}^{\pm}(\boldsymbol{x},t)-\partial_t\boldsymbol{r})\cdot\nabla\boldsymbol{u}^{\pm}(\boldsymbol{x},t)]$ where $i=1,2$, we first write $\boldsymbol{u}^{\pm}-\partial_t\boldsymbol{r}$ as $A\boldsymbol{\mathbf{\widehat{s}}}_1+B\boldsymbol{\mathbf{\widehat{s}}}_2$. Then we compute the dot product between that and $\nabla \boldsymbol{u}^{\pm}$ and obtain $A\partial_{s_1}\boldsymbol{u}^{\pm}+B\partial_{s_2}\boldsymbol{u}^{\pm}$. If we substitute $\boldsymbol{u}^{\pm}$, we get an expression of the form $C\boldsymbol{\mathbf{\widehat{s}}}_1+D\boldsymbol{\mathbf{\widehat{s}}}_2$, which we can finally dot with $\boldsymbol{\mathbf{\widehat{s}}}_1$ and $\boldsymbol{\mathbf{\widehat{s}}}_2$.

Using
\begin{equation}\label{membranevel}
\partial_t\boldsymbol{r}=\tau_1\boldsymbol{\mathbf{\widehat{s}}}_1+\tau_2\boldsymbol{\mathbf{\widehat{s}}}_2+\nu_v{\mathbf{\hat{n}}},
\end{equation}
and (\ref{fluidvel}), we find that the $\mathbf{\widehat{s}}_1$ components of the second and third terms on the left-hand side of (\ref{diff}) are
% \begin{align}
%   \mathbf{\widehat{s}}_{1}\cdot \left[(\boldsymbol{u}^\pm(\boldsymbol{x},t)-\partial_t\boldsymbol{r})\cdot\nabla\boldsymbol{u}^\pm(\boldsymbol{x},t)\right]\bigg|_{\boldsymbol{x}=\boldsymbol{r}(\alpha_1,\alpha_2,t)} = \left(\mu_{1} \pm \frac{\gamma_{2}}{2} -\tau_{1}\right)\left[\partial_{s_{1}}\left(\mu_{1}\pm\frac{\gamma_{2}}{2}\right) -\nu_v\kappa_1\right],\label{advpm1}\\
%   \mathbf{\widehat{s}}_{2}\cdot \left[(\boldsymbol{u}^\pm(\boldsymbol{x},t)-\partial_t\boldsymbol{r})\cdot\nabla\boldsymbol{u}^\pm(\boldsymbol{x},t)\right]\bigg|_{\boldsymbol{x}=\boldsymbol{r}(\alpha_1,\alpha_2,t)} = \left(\mu_{2} \mp \frac{\gamma_{1}}{2} -\tau_{2}\right)\left[\partial_{s_{2}}\left(\mu_{2}\mp\frac{\gamma_{1}}{2}\right) -\nu_v\kappa_2\right].\label{advpm2}
% \end{align}
\begin{align}
   \mathbf{\widehat{s}}_{1}\cdot \left[(\boldsymbol{u}^\pm(\boldsymbol{x},t)-\partial_t\boldsymbol{r})\cdot\nabla\boldsymbol{u}^\pm(\boldsymbol{x},t)\right]\bigg|_{\boldsymbol{x}
   =\boldsymbol{r}(\alpha_1,\alpha_2,t)} \nonumber\\= \left(\mu_{1} \pm \frac{\gamma_{2}}{2} -\tau_{1}\right)\left[\partial_{s_1}\left(\mu_{1}\pm\frac{\gamma_{2}}{2}\right) +\left(\partial_{s_1}\left(\mu_2\mp\frac{\gamma_1}{2}\right)\right)\mathbf{\widehat{s}}_{1}\cdot\mathbf{\widehat{s}}_{2}+\left(\mu_2\mp\frac{\gamma_1}{2}\right)\mathbf{\widehat{s}}_{1}\cdot\partial_{s_1}\mathbf{\widehat{s}}_{2}+\nu_v\mathbf{\widehat{s}}_{1}\cdot\partial_{s_1}\mathbf{\hat{n}}\right]\nonumber\\
   +\left(\mu_{2} \mp \frac{\gamma_{1}}{2} -\tau_{2}\right)\left[\partial_{s_2}\left(\mu_{1}\pm\frac{\gamma_{2}}{2}\right) +\left(\partial_{s_2}\left(\mu_2\mp\frac{\gamma_1}{2}\right)\right)\mathbf{\widehat{s}}_{1}\cdot\mathbf{\widehat{s}}_{2}+\left(\mu_2\mp\frac{\gamma_1}{2}\right)\mathbf{\widehat{s}}_{1}\cdot\partial_{s_2}\mathbf{\widehat{s}}_{2}+\nu_v\mathbf{\widehat{s}}_{1}\cdot\partial_{s_2}\mathbf{\hat{n}}\right.\nonumber\\
   +\left.\left(\mu_1\pm \frac{\gamma_2}{2}\right)\mathbf{\widehat{s}}_{1}\cdot\partial_{s_2}\mathbf{\widehat{s}}_{1} \right].\label{advpm1}
   \end{align}
%   and
%   \begin{align}
%   \mathbf{\widehat{s}}_{2}\cdot \left[(\boldsymbol{u}^\pm(\boldsymbol{x},t)-\partial_t\boldsymbol{r})\cdot\nabla\boldsymbol{u}^\pm(\boldsymbol{x},t)\right]\bigg|_{\boldsymbol{x}=\boldsymbol{r}(\alpha_1,\alpha_2,t)} 
% \nonumber\\
% =\left(\mu_{1} \mp \frac{\gamma_{2}}{2} -\tau_{1}\right)
% \left[\partial_{s_1}\left(\mu_{2}\pm\frac{\gamma_{1}}{2}\right) +\left(\partial_{s_1}\left(\mu_1\mp\frac{\gamma_2}{2}\right)\right)\mathbf{\widehat{s}}_{1}\cdot\mathbf{\widehat{s}}_{2}+\left(\mu_2\pm\frac{\gamma_1}{2}\right)\mathbf{\widehat{s}}_{2}\cdot\partial_{s_1}\mathbf{\widehat{s}}_{2}+\nu_v\mathbf{\widehat{s}}_{2}\cdot\partial_{s_1}\mathbf{\hat{n}}\right.\nonumber\\
% +\left.\left(\mu_1\mp\frac{\gamma_2}{2}\right) \mathbf{\widehat{s}}_{2}\cdot\partial_{s_1}\mathbf{\widehat{s}}_1\right] \nonumber\\
%   +\left(\mu_{2} \pm \frac{\gamma_{1}}{2} -\tau_{2}\right)\left[\partial_{s_2}\left(\mu_{2}\pm\frac{\gamma_{1}}{2}\right) +\left(\partial_{s_2}\left(\mu_1\mp\frac{\gamma_2}{2}\right)\right)\mathbf{\widehat{s}}_{1}\cdot\mathbf{\widehat{s}}_{2}+\left(\mu_1\mp\frac{\gamma_2}{2}\right)\mathbf{\widehat{s}}_{2}\cdot\partial_{s_2}\mathbf{\widehat{s}}_{1}+\nu_v\mathbf{\widehat{s}}_{2}\cdot\partial_{s_2}\mathbf{\hat{n}}\right].\label{advpm2}
% \end{align}
The difference of the $+$ and $-$ terms on the right-hand side of (\ref{advpm1}) is
\begin{align}
(\mathbf{\widehat{s}}_{1}\cdot\mathbf{\widehat{s}}_{2})(-\mu_1\partial_{s_1}\gamma_1+\gamma_2\partial_{s_1}\mu_2+\tau_1\partial_{s_1}\gamma_1-\mu_2\partial_{s_2}\gamma_1-\gamma_1\partial_{s_2}\mu_2+\tau_2\partial_{s_2}\gamma_1)\nonumber\\
+(\mathbf{\widehat{s}}_{1}\cdot\partial_{s_1}
\mathbf{\widehat{s}}_{2})(-\mu_1\gamma_1+\gamma_2\mu_2+\tau_1\gamma_1)+(\mu_1\partial_{s_1}\gamma_2+\gamma_2\partial_{s_1}\mu_1-\tau_1\partial_{s_1}\gamma_2+\mu_2\partial_{s_2}\gamma_2-\gamma_1\partial_{s_2}\mu_1-\tau_2\partial_{s_2}\gamma_2)\nonumber\\
+(\mathbf{\widehat{s}}_{1}\cdot\partial_{s_2}\mathbf{\widehat{s}}_{1})(\mu_2\gamma_2-\gamma_1\mu_1-\tau_2\gamma_2)
+(\mathbf{\widehat{s}}_{1}\cdot\partial_{s_2}\mathbf{\widehat{s}}_{2})(-2\mu_2\gamma_1+\tau_2\gamma_1)
+(\mathbf{\widehat{s}}_{1}\cdot \partial_{s_1}\mathbf{\hat{n}})\gamma_2\nu_v-(\mathbf{\widehat{s}}_{1}\cdot \partial_{s_2}\mathbf{\hat{n}})\gamma_1\nu_v,\label{advdiff1}
\end{align}
using $\mathbf{\widehat{s}}_{i}\cdot\partial_{s_i}\mathbf{\widehat{s}}_{i}=0$ and $\mathbf{\hat{n}}\cdot \mathbf{\widehat{s}}_{i}=0$ for $i=1,2$.
% and
% \begin{align}
% (\mathbf{\widehat{s}}_{1}\cdot\mathbf{\widehat{s}}_{2})(-\mu_1\partial_{s_1}\gamma_2-\gamma_2\partial_{s_1}\mu_1+\tau_1\partial_{s_1}\gamma_2-\mu_2\partial_{s_2}\gamma_2+\gamma_1\partial_{s_2}\mu_1+\tau_2\partial_{s_2}\gamma_2)\nonumber\\
% +(\mathbf{\widehat{s}}_{2}\cdot\partial_{s_1}
% \mathbf{\widehat{s}}_{2})(\mu_1\gamma_1-\gamma_2\mu_2-\tau_1\gamma_1)\nonumber\\
% +(\mu_1\partial_{s_1}\gamma_1-\gamma_2\partial_{s_1}\mu_2-\tau_1\partial_{s_1}\gamma_1+\mu_2\partial_{s_2}\gamma_1+\gamma_1\partial_{s_2}\mu_2-\tau_2\partial_{s_2}\gamma_1)\nonumber\\
% +(\mathbf{\widehat{s}}_{2}\cdot\partial_{s_2}\mathbf{\widehat{s}}_{1})(\mu_1\gamma_1-\gamma_2\mu_2+\tau_2\gamma_2)
% +(\mathbf{\widehat{s}}_{2}\cdot\partial_{s_1}\mathbf{\widehat{s}}_{1})(-2\mu_1\gamma_2+\tau_1\gamma_2)\nonumber\\
% -(\mathbf{\widehat{s}}_{2}\cdot \partial_{s_1}\mathbf{\hat{n}})\gamma_2\nu_v+(\mathbf{\widehat{s}}_{2}\cdot \partial_{s_2}\mathbf{\hat{n}})\gamma_1\nu_v.\label{advdiff2}
% \end{align}

Combining (\ref{eq:dtdiffs1}) and (\ref{advdiff1}),
the $\mathbf{\widehat{s}}_{1}$ component of (\ref{diff}) is
\begin{align}\label{ds1p}
\partial_t\gamma_2-\partial_t\gamma_1(\mathbf{\widehat{s}}_{1}\cdot\mathbf{\widehat{s}}_{2})-\gamma_1(\mathbf{\widehat{s}}_{1}\cdot\partial_t\mathbf{\widehat{s}}_{2})\nonumber\\
+
(\mathbf{\widehat{s}}_{1}\cdot\mathbf{\widehat{s}}_{2})(-\mu_1\partial_{s_1}\gamma_1+\gamma_2\partial_{s_1}\mu_2+\tau_1\partial_{s_1}\gamma_1-\mu_2\partial_{s_2}\gamma_1-\gamma_1\partial_{s_2}\mu_2+\tau_2\partial_{s_2}\gamma_1)\nonumber\\
+(\mathbf{\widehat{s}}_{1}\cdot\partial_{s_1}
\mathbf{\widehat{s}}_{2})(-\mu_1\gamma_1+\gamma_2\mu_2+\tau_1\gamma_1)
+(\mu_1\partial_{s_1}\gamma_2+\gamma_2\partial_{s_1}\mu_1-\tau_1\partial_{s_1}\gamma_2+\mu_2\partial_{s_2}\gamma_2-\gamma_1\partial_{s_2}\mu_1-\tau_2\partial_{s_2}\gamma_2)\nonumber\\
+(\mathbf{\widehat{s}}_{1}\cdot\partial_{s_2}\mathbf{\widehat{s}}_{1})(\mu_2\gamma_2-\gamma_1\mu_1-\tau_2\gamma_2)
+(\mathbf{\widehat{s}}_{1}\cdot\partial_{s_2}\mathbf{\widehat{s}}_{2})(-2\mu_2\gamma_1+\tau_2\gamma_1)
+(\mathbf{\widehat{s}}_{1}\cdot \partial_{s_1}\mathbf{\hat{n}})\gamma_2\nu_v-(\mathbf{\widehat{s}}_{1}\cdot \partial_{s_2}\mathbf{\hat{n}})\gamma_1\nu_v
\nonumber\\
=-\partial_{s_{1}}[p]_-^+.  
\end{align}

We multiply (\ref{ds1p}) through by $\partial_{\alpha_1}s_1$ which converts $\partial_{s_{1}}[p]_-^+$ to $\partial_{\alpha_{1}}[p]_-^+$. We integrate with respect to $\alpha_1$ from the trailing edge, applying $[p]_-^+ = 0$ at the trailing edge, to obtain $[p]_-^+$ at all points on the membrane. 

If we directly integrate $\partial_{\alpha_{1}}[p]_-^+$ numerically (e.g. using the trapezoidal rule), the results disagree significantly with our 2D benchmark results \cite{mavroyiakoumou2020large}. Using a different formulation that agrees with the 2D results in the small-amplitude regime, we have found empirically that the following method agrees well with the 2D results in both the small- and large-amplitude regimes. We first write $\partial_{\alpha_{1}}[p]_-^+$ as a sum of two terms:
\begin{equation}\label{eq:addsubtract}
    \partial_{\alpha_{1}}[p]_-^+ = (\partial_{\alpha_1}[p]_{-}^{+} + \partial_{\alpha_1}s_1\partial_t\gamma_2 +  \partial_{\alpha_1}\gamma_2 ) + (- \partial_{\alpha_1}s_1\partial_t\gamma_2 -  \partial_{\alpha_1}\gamma_2).
\end{equation}
We have added and subtracted $\partial_{\alpha_1}s_1\partial_t\gamma_2 +  \partial_{\alpha_1}\gamma_2$. Its integral with respect to $\alpha_1$ can be written $\partial_t\Gamma + \partial_{s_1}\Gamma$, where $\Gamma$ is the integrated vortex sheet strength \cite{saffman1992vortex}. It is conserved at points of a free vortex sheet that move at the average of the tangential flow velocities on the two sides of the sheet, equal to 1 in this case. Hence $\partial_t\Gamma + \partial_{s_1}\Gamma = 0$ at the trailing edge, as does $[p]_-^+$ by the unsteady Kutta condition \cite{saffman1992vortex,katz2001low}. 

Therefore we integrate the first term in~\eqref{eq:addsubtract} (in parentheses) with respect to $\alpha_1$ using the trapezoidal rule, with the boundary condition that its integral is 0 at the trailing edge. The integral of the remaining terms in~\eqref{eq:addsubtract} is $-\partial_t\Gamma - \partial_{s_1}\Gamma$ evaluated at $\alpha_1$ minus its value at the trailing edge, which is zero as we have just discussed. To evaluate
$-\partial_t\Gamma - \partial_{s_1}\Gamma$ on the ($\alpha_1$,$\alpha_2$) grid, 
we approximate $\Gamma$ at the center point of a membrane vortex panel by the circulation of the vortex ring at that panel. Derivatives of $\Gamma$ with respect to $t$ and $\alpha_1$ are obtained by the usual finite-difference formulas and extrapolation to the ($\alpha_1$,$\alpha_2$) grid points---i.e. the corner points of the panels. 

%%%%%%%%%%%%%%%%%%%%%
\section{Residual membrane equations in Broyden's method}\label{app:residualMembraneBroyden}

In this appendix we write down the discretized system of nonlinear membrane equations that we solve using Broyden's method. Having computed each of the quantities in the membrane equation~\eqref{eq:nonlinearMembrane} we keep iterating until $\mathbf{f}(\mathbf{x})$ drops below a certain tolerance that we set to $10^{-5}$. Here $\mathbf{f}(\mathbf{x})$ is given by:
\begin{align}
    f_j(\bf{x}) = R_1\partial_{tt}x_{j}^{k}-K_s\left\{(D_{\alpha_1}^2\mathbf{r}_{j}^{k}\cdot D_{\alpha_1}^1\mathbf{r}_{j}^{k})D_{\alpha_1}^1x_{j}^{k}+\epsilon_{11}D_{\alpha_1}^2x_{j}^{k} +\nu\left(( D_{\alpha_1\alpha_2}^2\mathbf{r}_{j}^{k}\cdot D_{\alpha_2}^1\mathbf{r}_{j}^{k})D_{\alpha_1}^1x_{j}^{k}+\epsilon_{22}D_{\alpha_1}^2x_{j}^{k} \right)\right.\nonumber\\
 +(1-\nu)\left[((D_{\alpha_1}^2\mathbf{r}_{j}^{k}\cdot D_{\alpha_2}^1\mathbf{r}_{j}^{k}+D_{\alpha_1}^1\mathbf{r}_{j}^{k}\cdot  D_{\alpha_1\alpha_2}^2\mathbf{r}_{j}^{k})/2) D_{\alpha_2}^1x_{j}^{k}+\epsilon_{12} D_{\alpha_1\alpha_2}^2x_{j}^{k}\right] \nonumber\\
  +  (D_{\alpha_2}^2\mathbf{r}_{j}^{k}\cdot  D_{\alpha_2}^1\mathbf{r}_{j}^{k}) D_{\alpha_2}^1x_{j}^{k}+\epsilon_{22}D_{\alpha_2}^2x_{j}^{k}+\nu(( D_{\alpha_1\alpha_2}^2\mathbf{r}_{j}^{k}\cdot D_{\alpha_1}^1\mathbf{r}_{j}^{k}) D_{\alpha_2}^1x_{j}^{k}+\epsilon_{11}D_{\alpha_2}^2x_{j}^{k})\nonumber\\
  + \left.(1-\nu)\left[(( D_{\alpha_1\alpha_2}^2\mathbf{r}_{j}^{k}\cdot D_{\alpha_2}^1\mathbf{r}_{j}^{k}+D_{\alpha_1}^1\mathbf{r}_{j}^{k}\cdot D_{\alpha_2}^2\mathbf{r}_{j}^{k})/2)D_{\alpha_1}^1x_{j}^{k}+\epsilon_{12} D_{\alpha_1\alpha_2}^2x_{j}^{k} \right] \right\}\nonumber\\
    +[p]_{j}^{k}\mathbf{\hat{n}}_{x,j}^{k}\sqrt{(D_{\alpha_1}^1\mathbf{r}_{j}^{k}\cdot D_{\alpha_1}^1\mathbf{r}_{j}^{k})( D_{\alpha_2}^1\mathbf{r}_{j}^{k}\cdot D_{\alpha_2}^1\mathbf{r}_{j}^{k})-(D_{\alpha_1}^1\mathbf{r}_{j}^{k}\cdot D_{\alpha_2}^1\mathbf{r}_{j}^{k})^2},
\end{align}
\begin{align}
    f_{j+(M-1)(N-1)}(\bf{x}) =\nonumber\\ R_1\partial_{tt}y_{j}^{k}-K_s\left\{(D_{\alpha_1}^2\mathbf{r}_{j}^{k}\cdot D_{\alpha_1}^1\mathbf{r}_{j}^{k})D_{\alpha_1}^1y_{j}^{k}+\epsilon_{11}D_{\alpha_1}^2y_{j}^{k} +\nu\left(( D_{\alpha_1\alpha_2}^2\mathbf{r}_{j}^{k}\cdot D_{\alpha_2}^1\mathbf{r}_{j}^{k})D_{\alpha_1}^1y_{j}^{k}+\epsilon_{22}D_{\alpha_1}^2y_{j}^{k} \right)\right.\nonumber\\
 +(1-\nu)\left[((D_{\alpha_1}^2\mathbf{r}_{j}^{k}\cdot D_{\alpha_2}^1\mathbf{r}_{j}^{k}+D_{\alpha_1}^1\mathbf{r}_{j}^{k}\cdot  D_{\alpha_1\alpha_2}^2\mathbf{r}_{j}^{k})/2) D_{\alpha_2}^1y_{j}^{k}+\epsilon_{12} D_{\alpha_1\alpha_2}^2y_{j}^{k}\right] \nonumber\\
  +  (D_{\alpha_2}^2\mathbf{r}_{j}^{k}\cdot  D_{\alpha_2}^1\mathbf{r}_{j}^{k}) D_{\alpha_2}^1y_{j}^{k}+\epsilon_{22}D_{\alpha_2}^2y_{j}^{k}+\nu(( D_{\alpha_1\alpha_2}^2\mathbf{r}_{j}^{k}\cdot D_{\alpha_1}^1\mathbf{r}_{j}^{k}) D_{\alpha_2}^1y_{j}^{k}+\epsilon_{11}D_{\alpha_2}^2y_{j}^{k})\nonumber\\
  + \left.(1-\nu)\left[(( D_{\alpha_1\alpha_2}^2\mathbf{r}_{j}^{k}\cdot D_{\alpha_2}^1\mathbf{r}_{j}^{k}+D_{\alpha_1}^1\mathbf{r}_{j}^{k}\cdot D_{\alpha_2}^2\mathbf{r}_{j}^{k})/2)D_{\alpha_1}^1y_{j}^{k}+\epsilon_{12} D_{\alpha_1\alpha_2}^2y_{j}^{k} \right] \right\}\nonumber\\
    +[p]_{j}^{k}\mathbf{\hat{n}}_{y,j}^{k}\sqrt{(D_{\alpha_1}^1\mathbf{r}_{j}^{k}\cdot D_{\alpha_1}^1\mathbf{r}_{j}^{k})( D_{\alpha_2}^1\mathbf{r}_{j}^{k}\cdot D_{\alpha_2}^1\mathbf{r}_{j}^{k})-(D_{\alpha_1}^1\mathbf{r}_{j}^{k}\cdot D_{\alpha_2}^1\mathbf{r}_{j}^{k})^2},
\end{align}
\begin{align}
    f_{j+2(M-1)(N-1)}(\bf{x}) =\nonumber\\ R_1\partial_{tt}z_{j}^{k}-K_s\left\{(D_{\alpha_1}^2\mathbf{r}_{j}^{k}\cdot D_{\alpha_1}^1\mathbf{r}_{j}^{k})D_{\alpha_1}^1z_{j}^{k}+\epsilon_{11}D_{\alpha_1}^2z_{j}^{k} +\nu\left(( D_{\alpha_1\alpha_2}^2\mathbf{r}_{j}^{k}\cdot D_{\alpha_2}^1\mathbf{r}_{j}^{k})D_{\alpha_1}^1z_{j}^{k}+\epsilon_{22}D_{\alpha_1}^2z_{j}^{k} \right)\right.\nonumber\\
 +(1-\nu)\left[((D_{\alpha_1}^2\mathbf{r}_{j}^{k}\cdot D_{\alpha_2}^1\mathbf{r}_{j}^{k}+D_{\alpha_1}^1\mathbf{r}_{j}^{k}\cdot  D_{\alpha_1\alpha_2}^2\mathbf{r}_{j}^{k})/2) D_{\alpha_2}^1z_{j}^{k}+\epsilon_{12} D_{\alpha_1\alpha_2}^2z_{j}^{k}\right] \nonumber\\
  +  (D_{\alpha_2}^2\mathbf{r}_{j}^{k}\cdot  D_{\alpha_2}^1\mathbf{r}_{j}^{k}) D_{\alpha_2}^1z_{j}^{k}+\epsilon_{22}D_{\alpha_2}^2z_{j}^{k}+\nu(( D_{\alpha_1\alpha_2}^2\mathbf{r}_{j}^{k}\cdot D_{\alpha_1}^1\mathbf{r}_{j}^{k}) D_{\alpha_2}^1z_{j}^{k}+\epsilon_{11}D_{\alpha_2}^2z_{j}^{k})\nonumber\\
  + \left.(1-\nu)\left[(( D_{\alpha_1\alpha_2}^2\mathbf{r}_{j}^{k}\cdot D_{\alpha_2}^1\mathbf{r}_{j}^{k}+D_{\alpha_1}^1\mathbf{r}_{j}^{k}\cdot D_{\alpha_2}^2\mathbf{r}_{j}^{k})/2)D_{\alpha_1}^1z_{j}^{k}+\epsilon_{12} D_{\alpha_1\alpha_2}^2z_{j}^{k} \right] \right\}\nonumber\\
    +[p]_{j}^{k}\mathbf{\hat{n}}_{z,j}^{k}\sqrt{(D_{\alpha_1}^1\mathbf{r}_{j}^{k}\cdot D_{\alpha_1}^1\mathbf{r}_{j}^{k})( D_{\alpha_2}^1\mathbf{r}_{j}^{k}\cdot D_{\alpha_2}^1\mathbf{r}_{j}^{k})-(D_{\alpha_1}^1\mathbf{r}_{j}^{k}\cdot D_{\alpha_2}^1\mathbf{r}_{j}^{k})^2},
\end{align}
where $j=1,\dots, (M-1)(N-1)$ and $[p]$ is the result of integrating~\eqref{eq:addsubtract}.
Here $D_{\alpha_i}^1\approx\partial_{\alpha_i}$, $D_{\alpha_i}^2\approx\partial_{\alpha_i\alpha_i}$ for $i=1,2$ and $D_{\alpha_1\alpha_2}^2\approx\partial_{\alpha_1\alpha_2}$ are second-order accurate finite-difference matrix approximations to first and second-order derivatives (order listed in the superscript) on uniform grid nodes, one-sided at boundaries. $\partial_{tt}\{x_j^k, y_j^k,z_j^k\}$ are the second-order accurate finite difference formulas (backward differentiation formulas) at time step $k$ based on the values of $\{x_j,y_j,z_j\}$ at times steps 
$k-3,\ldots, k$.

%%%%%%%%%%
\section{Membrane deflections versus stretching rigidity at different pretension values}\label{app:otherT0lineDefl}

\begin{figure}[H]
    \centering
    \includegraphics[width=\textwidth]{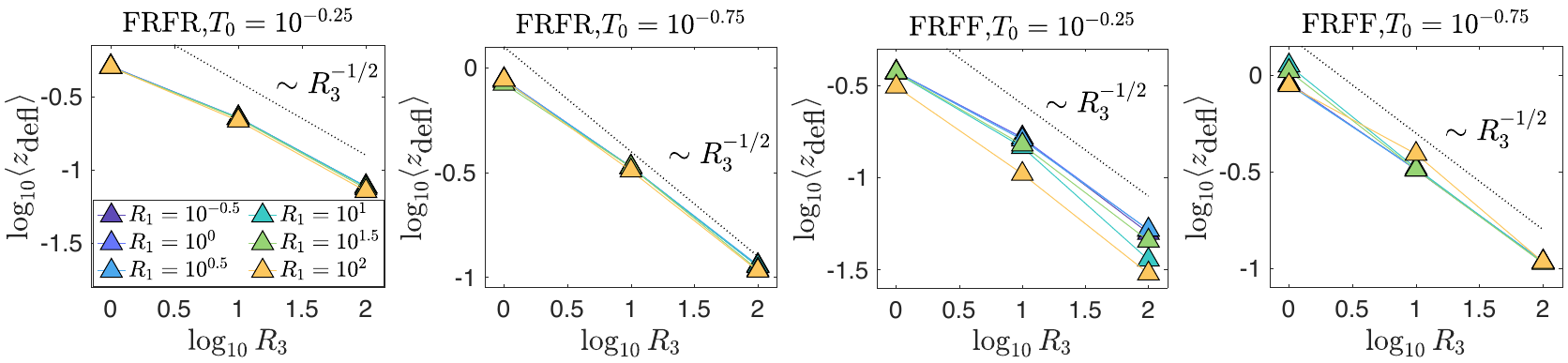}
    \caption{Time-averaged deflections of FRFR and FRFF membranes (defined by~\eqref{eq:timeAvgDefl}) versus $R_3$ for various $R_1$ and fixed $T_0$ ($10^{-0.25}$ and $10^{-0.75}$). The dotted black line indicates the scaling $R_3^{-1/2}$.}
    \label{fig:extraT0lineDeflR3}
\end{figure}

For two of the 12 boundary conditions (FRFF and FRFR) we plot in figure 
\ref{fig:extraT0lineDeflR3} the average membrane deflection at $T_0=10^{-0.25}$ and $10^{-0.75}$, instead of $10^{-0.5}$ in figure~\ref{fig:lineDeflR3}, and see that the deflection values increase at smaller $T_0$ (similarly to \cite{mavroyiakoumou2020large,tiomkin2022unsteady}) but the same $1/\sqrt{R_3}$ scaling holds in all cases.

%%%%%%%%

\bibliographystyle{unsrt}  
\bibliography{biblio.bib} 

\end{document}